\documentclass[twocolumn,amsmath,aps,superscriptaddress,longbibliography]{revtex4-1}
\usepackage{amsmath,amsfonts}
\usepackage{dcolumn}
\usepackage{bm}
\usepackage{color}
\usepackage{multirow}
\usepackage{graphicx}
\usepackage{hyperref}
\usepackage{float} 
\usepackage[utf8x]{inputenc}

\begin{document}


\title{Pressure effects on the electronic structure and superconductivity of (TaNb)$_{0.67}$(HfZrTi)$_{0.33}$ high entropy alloy}



\author{K. Jasiewicz}
\affiliation{Faculty of Physics and Applied Computer Science, AGH University of Science and Technology, Al. Mickiewicza 30, 30-059 Krakow, Poland}

\author{B. Wiendlocha}
\email{wiendlocha@fis.agh.edu.pl}
\affiliation{Faculty of Physics and Applied Computer Science, AGH University of Science and Technology, Al. Mickiewicza 30, 30-059 Krakow, Poland}

\author{K. G{\'o}rnicka}
\affiliation{Faculty of Applied Physics and Mathematics, Gda{\'n}sk University
of Technology, ul. Narutowicza 11/12, 80-233 Gda{\'n}sk, Poland}

\author{K. Gofryk}
\affiliation{Idaho National Laboratory, Idaho Falls, Idaho 83415, USA}	

\author{M. Gazda}
\author{T. Klimczuk}
\affiliation{Faculty of Applied Physics and Mathematics, Gda{\'n}sk University
of Technology, ul. Narutowicza 11/12, 80-233 Gda{\'n}sk, Poland}

\author{J. Tobola}
\affiliation{Faculty of Physics and Applied Computer Science, AGH University of Science and Technology, Al. Mickiewicza 30, 30-059 Krakow, Poland}


\date{\today}

\begin{abstract}
Effects of pressure on the electronic structure, electron-phonon interaction{,} and superconductivity of the high entropy alloy (TaNb)$_{0.67}$(HfZrTi)$_{0.33}$ {are} studied in the {pressure} range 0 - 100 GPa. {The} electronic structure is calculated using the Korringa-Kohn-Rostoker method with the coherent potential approximation. 
Effects of pressure on the lattice dynamics {are} simulated using the Debye-Gr\"{u}neisen model and the Gr\"{u}neisen parameter at ambient conditions{.} {In addition,} the Debye temperature and Sommerfeld electronic heat capacity coefficient were experimentally determined. The electron-phonon coupling parameter $\lambda$ is calculated using the McMillan-Hopfield parameters {and} computed {within} the rigid muffin tin approximation. We find, that the system undergoes the Lifshitz transition, as one of the bands {crosses} the Fermi level at elevated pressures. The electron-phonon coupling parameter $\lambda$ decreases above 10 GPa. {The} calculated superconducting $T_c$ increases up to 40 - 50 GPa and, later, is stabilized at the larger value than for the ambient conditions, in agreement with the experimental findings. Our results show that the experimentally observed evolution of $T_c$ with pressure in (TaNb)$_{0.67}$(HfZrTi)$_{0.33}$ can be well explained by the classical electron-phonon mechanism.
\end{abstract}

\pacs{}
\keywords{superconductivity, electronic structure, high-entropy alloys}

\maketitle



\section{Introduction}	

High-pressure studies of superconducting materials have brought about the latest breakthrough in the field of superconductivity. The record-high $T_c$ of 203~K in H$_3$S~\cite{h2s} and 250 K in LaH$_{10}$~\cite{lah10} at $P >  150$~GPa were recently reported, and theoretical predictions show that even larger values of $T_c$ are possible~\cite{mgh6,durajski-h3s}. As superconductivity in these materials is mediated by the electron-phonon interaction, recent discoveries also turned attention to the effect of extreme pressure on superconductivity in other materials, including bulk conventional superconductors. This includes recent high-pressure studies on superconductivity in Nb-Ti alloy~\cite{nb-ti} and (TaNb)$_{0.67}$(HfZrTi)$_{0.33}$ high-entropy alloy (HEA)~\cite{guo-pressure}, on which we are focusing in the current work. 

High entropy alloys~\cite{yeh_2004,yeh_2007} contain {five} or more elements and, due to stabilization by the configurational entropy, form simple "monoatomic" crystal structures {such as cubic} {\it bcc} or {\it fcc}, with statistical occupation of the single crystal site. The first superconducting HEA, Ta$_{34}$Nb$_{33}$Hf$_{8}$Zr$_{14}$Ti$_{11}$\cite{kozejl_2014} was reported in 2014. It crystallizes in a $Im$-$3m$ {\it bcc}-type {of} structure, with a lattice parameter of 3.36~\AA. {In this system, all atoms randomly occupy (2a) crystal site (on average).} It is a type-II superconductor with the transition temperature of $T_c = 7.3$~K. Experimental data as well as theoretical calculations~\cite{jasiewicz_2016} suggest conventional mechanism of superconductivity with a relatively strong electron phonon coupling parameter $\lambda \sim 1$. Several other examples of superconducting HEAs were later reported~\cite{hea-sup1,hea-sup2,hea-sup3}, however, the TaNbHfZrTi family is still the most investigated one \cite{kozejl_2014,jasiewicz_2016,tnhzt1,tnhzt2}. When the atomic concentration is slightly changed to Ta$_{33.5}$Nb$_{33.5}$Hf$_{11}$Zr$_{11}$Ti$_{11}$~\cite{tnhzt2} [denoted as (TaNb)$_{0.67}$(HfZrTi)$_{0.33}$ or TNHZT in short], superconducting {transition temperature} slightly increases to 7.7~K. This alloy also {hosts} a {cubic} body-centered {crystal} structure, with the lattice {parameter} of 3.34~\AA.

When the external pressure is applied, $T_c$ of (TaNb)$_{0.67}$(HfZrTi)$_{0.33}$ increases up to about 10~K {at} around 50-60 GPa {and then} it remains practically constant up to about 100 GPa. {After that,  it} slightly decreases to 9 K at 190 GPa~\cite{guo-pressure}. In our work we investigate effects of pressure on the electronic structure and superconductivity {in this disordered system to better understand microscopic mechanisms controlling these} interesting $T_c(P)$ characteristics. As the crystal structure was determined experimentally to about 96 GPa~\cite{guo-pressure} we perform our studies in the pressure range from 0 to 100 GPa. {The} electronic structure is calculated using the Korringa-Kohn-Rostoker method with the coherent potential approximation (KKR-CPA)~\cite{stopa_2004,kaprzyk_1990,bansil_1999,soven_1967}. From the KKR-CPA results, by using the rigid muffin tin approximation (RMTA),~\cite{gaspari_1972} the McMillan-Hopfield parameters are calculated. Effect of pressure on the lattice dynamics is simulated using the Debye model and Gr\"{u}neisen parameter $\gamma_G$. To obtain $\gamma_G$ it {becomes} necessary to determine the volume thermal expansion coefficient, thus experimental measurements of the crystal structure evolution with temperature were performed. Additionally, to assure the consistency of the analysis the low-temperature heat capacity was measured {on the same sample} to obtain $T_c$, Debye temperature $\theta_D$, and the Sommerfeld coefficient $\gamma$. As a final result, the pressure evolution of the electron-phonon coupling parameter $\lambda$(P) and the superconducting critical temperature $T_c(P)$ are determined.

\section{Methodology}
\subsection{Synthesis \& X-ray Crystallography}

\begin{figure}[t]
\begin{center}
\includegraphics[width=0.45\textwidth]{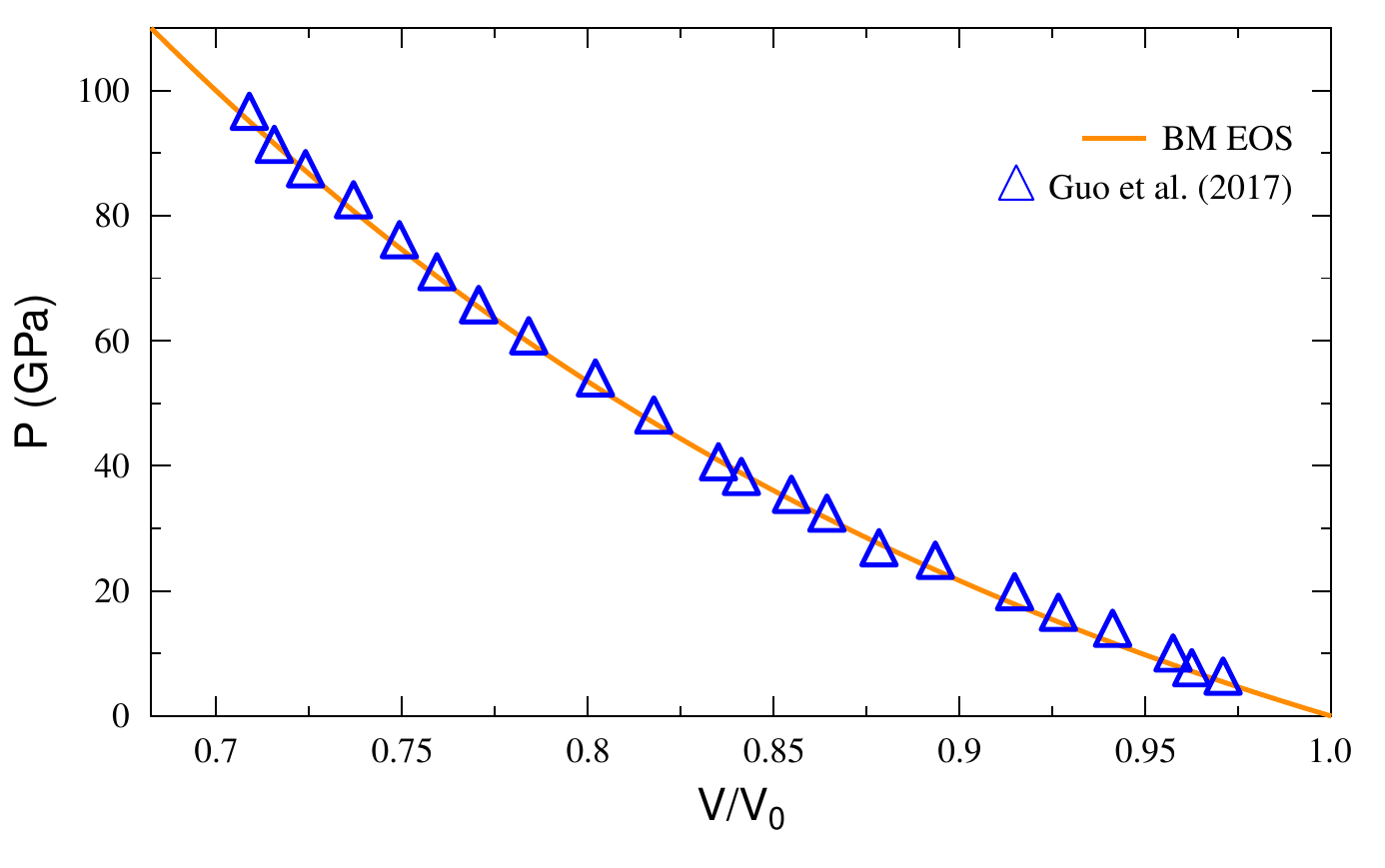}
\end{center}
\caption{Pressure dependence of the unit cell volume. Points correspond to the experimental data~\cite{guo-pressure} and line is determined from the fitted Birch-Murnaghan equation of state.}
\label{murnaghan_eos}
\end{figure}

The Ta$_{0.335}$Nb$_{0.335}$Hf$_{0.11}$Zr$_{0.11}$Ti$_{0.11}$ sample was prepared by melting the required high-purity elements, i.e., tantalum foil (99.9\%), niobium pieces (99.99\%), hafnium pieces (99.99\%), zirconium foil (99.8\%) and titanium pieces (99.99\%). The elemental metals were arc-melted to a single metallic button under an argon atmosphere on a water-chilled copper plate. A piece of zirconium was used as a getter at each melting {steps}. After the initial melt, the sample nugget was turned and remelted three times to ensure the optimal mixing of the constituents. Mass loss during the synthesis was smaller than 1\% and the resulting material was hard and silver in color. 

The phase purity of the obtained material was checked by X-ray diffraction (XRD) using a Philips X'pert Pro MPD with Cu K$_{\alpha}$ radiation. 
The sample exhibited ductility, and therefore could not 
be ground. {Because of that,} for qualitative and quantitative characterization the sample had to be converted into a plate form. In order to prepare the sample for the XRD analysis, the button was cut into smaller piece and then transformed into a plate using hydraulic press. The mechanical handling did not cause any sample contamination. The plate was put on the Al$_2$O$_3$ (corundum) sample holder and mounted in a small furnace inside a diffractometer. Above 400$^\circ$C the sample oxidizes and, hence, the XRD analysis in higher temperatures was not continued. The lattice parameter for TNHZT at different temperatures was estimated from the LeBail fit using a HighScore program.  

\subsection{Heat capacity}
Heat capacity measurements were carried out using a Quantum Design Physical Property Measurement System (PPMS) Evercool-II. The two-$\tau$ {relaxation} method was used to measure the specific heat {without external magnetic field and under} 8 T magnetic field, in the temperature range 1.9 - 10 K. The sample was attached to the measuring stage {by} using Apiezon N grease to ensure good thermal contact.

\subsection{Electronic structure}
Electronic structure calculations were performed using the Korringa-Kohn-Rostoker method with the coherent potential approximation (KKR-CPA)~\cite{stopa_2004,kaprzyk_1990,bansil_1999,soven_1967} to account for {atomic} disorder. Crystal potential of the muffin-tin type was constructed using the local density approximation (LDA), Perdew-Wang parametrization~\cite{perdew_1992}, and in the semi-relativistic approach. Angular momentum cut-off was set to $l_{max}$ = 3. Highly converged results were obtained for about 450 {\bf k}-points grid in the irreducible part of Brillouin zone for self-consistent cycle and 2800 {\bf k}-points for the densities of states (DOS) computations. Muffin-tin radius was set to the largest non-overlapping spheres (i.e. $R_{MT}=a\sqrt{3}/4$) {and the} Fermi level ($E_F$) was accurately determined from the generalized Lloyd formula \cite{kaprzyk_1990}. {It is worth noting, that the KKR-CPA method has already been successfully
applied to study different physical properties} of high entropy alloys~\cite{jasiewicz_2015,jin_2016,Calvo-2017,app-ccfna}. 

Electron-phonon coupling and its evolution under external pressure is studied using the so-called Rigid Muffin Tin Approximation (RMTA). {This method has been} successfully applied to many superconducting materials, mostly containing transition metal elements~\cite{gaspari_1972,gomersall_1974,papa_a15,mazin_1990,wiendlocha_2006,wiendlocha2008,wiendlocha2014} and more recently, to the HEA Ta$_{34}$Nb$_{33}$Hf$_{8}$Zr$_{14}$Ti$_{11}$~\cite{jasiewicz_2016} at ambient pressure. 
In this approach, the electron-phonon interaction is decoupled into electronic and lattice contributions. {The} coupling parameter $\lambda$ is computed as:

\begin{equation}\label{eq:lambda0}
\lambda = \sum_i \frac{\eta_i}{M_i\langle{\omega_i^2}\rangle},
\end{equation}
where $\eta_i$ are the McMillan-Hopfield parameters~\cite{mcmillan_1968,hopfield_1969} computed for each of the atom $i$ in the unit cell, $M_i$ is the atomic mass, and $\langle{\omega_i^2}\rangle$ is the properly defined average square atomic vibration frequency (see the discussion of the frequency moments in the Supplemental Material~\cite{suppl}).
Within RMTA, McMillan-Hopfield parameters are calculated using the band-structure related quantities \cite{gaspari_1972,gomersall_1974,mazin_1990} employing expression:
\begin{equation}\label{eq:eta}
\eta_i =\!\sum_l \frac{(2l + 2)\,n_l(E_F)\,
n_{l+1}(E_F)}{(2l+1)(2l+3)N(E_F)} \left|\int_0^{R_{\mathsf{MT}}}\!\!r^2
R_l\frac{dV}{dr}R_{l+1} \right|^2\!,
\end{equation}
where $V(r)$ is the self-consistent potential at site $i$, $R_\mathsf{MT}$ is the radius of the $i$-th MT sphere, $R_l(r)$ is a regular solution of the radial Schr\"odinger equation (normalized to unity inside the MT sphere), $n_l(E_F)$ is the $l$--th partial DOS per spin at the Fermi level $E_F$, and $N(E_F)$ is the total DOS per primitive cell and per spin.
For a more detailed discussion of the approximations involved in this methodology, see e.g. Refs.~\cite{mazin_1990,wiendlocha_2006} and references therein. 

In the case of a random alloy, where a single crystal site $i$ is occupied by several different atoms {that have} different concentrations, modification of Eq.(\ref{eq:lambda0}) is necessary. In calculations of $\lambda$ for binary alloys {having} similar atomic masses of elements (e.g. Nb-Mo), where one can expect similar denominators in Eq.(\ref{eq:lambda0}) the McMillan-Hopfield parameters obtained from self-consistent KKR-CPA calculations, were simply weighted by atomic concentrations $c_i$~\cite{kaprzyk_1996}, {and were} predicting composition dependence of $\lambda$ reasonably well. {Besides, in the} case of a monoatomic system {that} has a Debye-like phonon spectrum, $\langle{\omega_i^2}\rangle$ may be reasonably well approximated using the experimental Debye temperature~\cite{kaprzyk_1996,papaconstantopoulos_1977,massida_1988} as $\langle{\omega^2}\rangle = \frac{1}{2}\theta_D^2$
(see Supplemental Material~\cite{suppl} for the derivation of this formula). That is especially useful in the present case of a multicomponent HEA, since it allows to estimate $\lambda$ without knowledge of the phonon spectrum. As the Debye temperature represents the characteristic frequency of the whole system, we use it in combination with the concentration-weighted average atomic mass. {In this approach the} denominator in Eq. (\ref{eq:lambda0}) takes the form: 
$M_i \langle{\omega_i^2}\rangle \simeq \langle{M}\rangle{1\over 2}\theta_D^2$, where $\langle{M}\rangle = \sum_i c_iM_i$.
The final formula for the electron-phonon coupling (EPC) parameter $\lambda$ of HEA used in our work becomes:
\begin{equation}\label{eq:lambda}
\lambda = \frac{{\sum_i c_i \eta_i}}{{1\over 2}\langle{M}\rangle\theta_D^2},
\end{equation}
where McMillan-Hopfield parameters $\eta_i$ of each atom in the system are computed in the self-consistent KKR-CPA calculations {and} $c_i$ is the atomic concentration of the element. As mentioned above, this approach was recently applied to the first superconducting high-entropy alloy Ta$_{34}$Nb$_{33}$Hf$_{8}$Zr$_{14}$Ti$_{11}$~\cite{jasiewicz_2016} at ambient pressure ($T_c = 7.3$~K). 
The value of $\lambda = 1.16$ was obtained in good agreement with the value of $\lambda = 0.98$ determined from the renormalization of the electronic heat capacity coefficient $\gamma$. 
Here, the same approach is applied to (TaNb)$_{0.67}$(HfZrTi)$_{0.33}$ system to study evolution of superconducting properties under pressure. 

As far as the crystal structure is concerned, high pressure synchrotron X-ray diffraction measurements were performed in the pressure range from 0 to 96 GPa and {it was shown that}
(TaNb)$_{0.67}$(HfZrTi)$_{0.33}$ {maintains} the bcc structure, since no structural distortion was {observed}{\cite{guo-pressure}}. As there is no information on the crystal structure above this pressure we limit our studies to the pressure range from 0 to 100 GPa. Available experimental data of volume vs. pressure are shown in Fig.~\ref{murnaghan_eos} and were fitted to the third order Birch-Murnaghan equation of state:~\cite{birch}

\begin{equation}\label{bm-eos}
\begin{aligned}
P(V)=\frac{3}{2}B\Bigg[\bigg(\frac{V}{V_0}\bigg)^{-\frac{7}{3}}-\bigg(\frac{V}{V_0}\bigg)^{-\frac{5}{3}}\Bigg]\\ \Bigg\{1+\frac{3}{4}(B'-4)\Big[\bigg(\frac{V}{V_0}\bigg)^{-\frac{2}{3}}-1\Big]\Bigg\},
\end{aligned}
\end{equation}
%
where $V_0$ is the equilibrium volume. Bulk modulus of $B=177.35$ GPa and its derivative $B'=2.87$ were obtained and are used in the subsequent analysis. 

\subsection{Evolution of Debye temperature with pressure}

\begin{figure}[t]
\begin{center}
\includegraphics[width=0.48\textwidth]{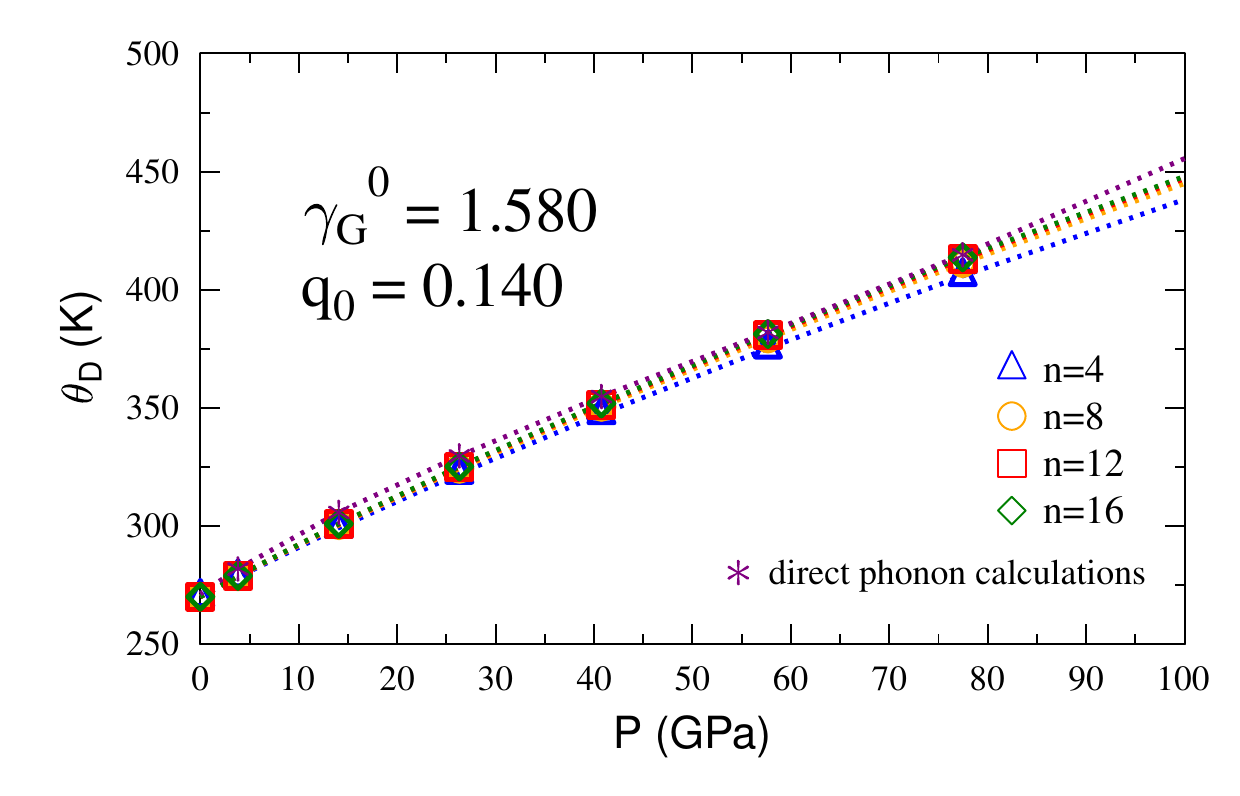}
\end{center}
\caption{\label{debye_nb}Pressure dependence of $\theta_D$ of niobium from the direct phonon calculations and from the Gr\"{u}neisen model for several values of $n$.}
\end{figure}

\begin{figure}[t]
\begin{center}
\includegraphics[width=0.48\textwidth]{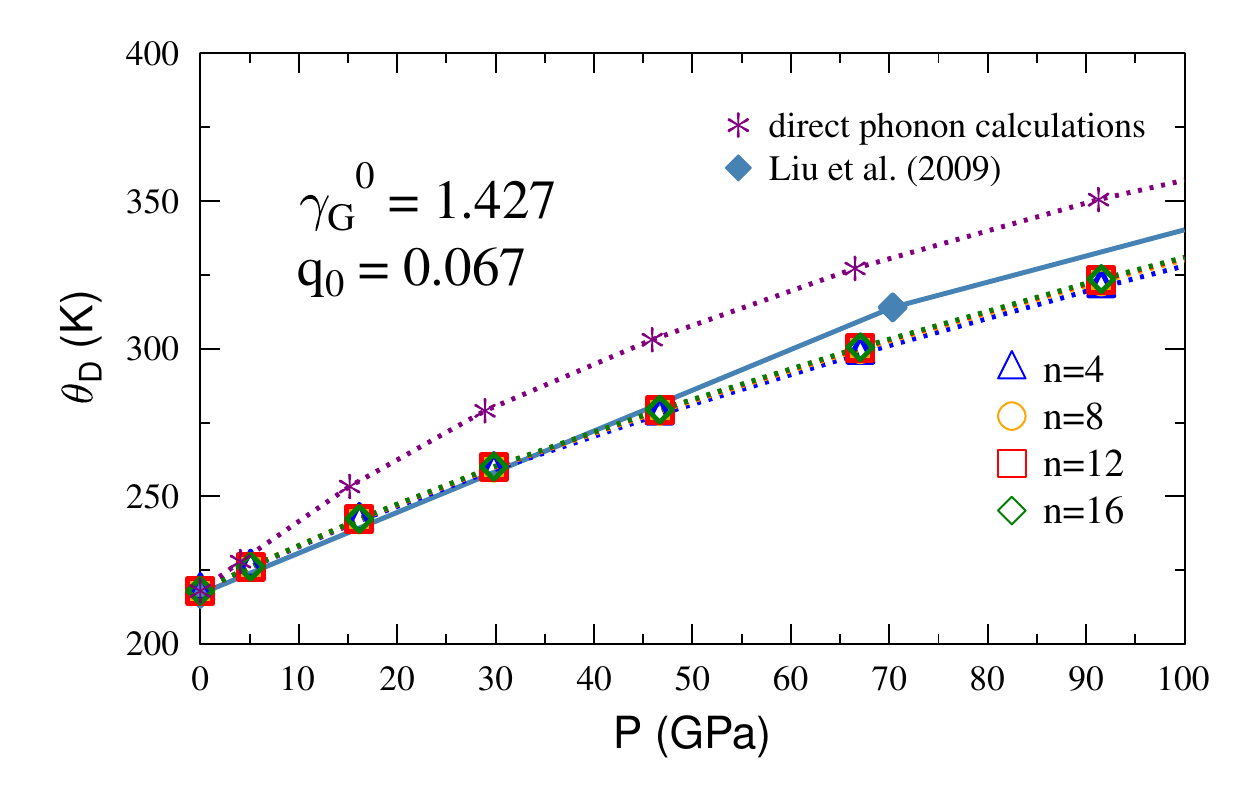}
\end{center}
\caption{\label{debye_ta}Pressure dependence of $\theta_D$ of tantalum from the direct phonon calculations, from the Gr\"{u}neisen model for several values of $n$, and from the quasi-harmonic calculations of Liu {\it et al.}~\cite{liu_2009}.}
\end{figure}

The Debye temperature of TNHZT $\theta_D^0$ was measured only at ambient conditions~\cite{tnhzt2}, {therefore} it was also necessary to simulate its pressure dependence to calculate $\lambda(P)$ and $T_c(P)$ {using} Eq.(\ref{eq:lambda}). This can be {performed} using analytic model based on Gr\"{u}neisen parameter $\gamma_G$ \cite{grimvall} where
\begin{equation}
\gamma_G(V) = -\frac{\partial \ln\theta_D}{\partial \ln V}.
\label{eq:gruneisen1}
\end{equation}
As the volume compression (in our case) reaches 30\% (see, Fig.~\ref{murnaghan_eos}), a variation of Gr\"{u}neisen parameter with pressure (volume) has to be taken into account. This can be done using the so-called second order Gr\"{u}neisen parameter $q$:
\begin{equation}
q(V) =\frac{\partial \ln\gamma_G(V)}{\partial\ln V},
\label{eq:sec_gruneisen}
\end{equation}
which also may be pressure-dependent. Equations (\ref{eq:gruneisen1}) and (\ref{eq:sec_gruneisen}) cannot be solved in a simple way as both $\gamma_G$ and $q$ are volume-dependent parameters. 
Assuming that the next logarithmic derivative is constant:~\cite{jeanloz-gruneisen}
\begin{equation}
q'(V) =\frac{\partial \ln q(V)}{\partial\ln V} = {\rm const}.
\end{equation}
we may write $q(V)$ as a power-law relation 
\begin{equation}
q(V)=q_0\zeta^n,
\label{eq:sec_gruneisen_2}
\end{equation}
where $\zeta=V/V_0$, $n$ is a material-dependent constant parameter, and $q_0 = q(V_0)$ is the value at ambient conditions. 
Such approximation leads to the formula for $\gamma_G(V)$ \cite{nie-gruneisen}:  
\begin{equation}
\gamma_G(V)=\gamma_G^0e^{[\frac{q_0}{n}(\zeta^n-1)]}.
\label{eq:gruneisen_v}
\end{equation}

Once $\gamma_G(V)$ is calculated {the} Debye temperature for a given volume (or equivalently pressure) is computed from:
\begin{equation}\label{eq:thetap}
\theta_D(V) = \theta_D^0\left(\frac{V}{V_0}\right)^{-\gamma_G(V)}.
\end{equation}
Input parameters, required to compute $\theta_D(V)$ are ambient-pressure Debye temperature $\theta_D^0$ and ambient-pressure Gr\"{u}neisen parameter $\gamma_G^0$, which have not been determined for our system yet.
To obtain $\gamma_G^0$ we have performed the volume thermal expansion coefficient $\alpha$ measurements, described in the next section. This allows to calculate the Gr\"{u}neisen parameter at ambient conditions:~\cite{grimvall}
\begin{equation}
\gamma_G^0 = \frac{\alpha BV_0N_A}{C_V},
\label{eq:gruneisen_ambient}
\end{equation}
where $V_0$ is the primitive cell volume, $N_A$ is the Avogadro number, and $C_V$ is the molar constant-volume heat capacity taken as the Dulong-Petit limit of 24.94 (J K$^{-1}$ mol$^{-1}$).
The second order Gr\"{u}neisen parameter is given by the following relation:\cite{anderson1993,anderson-book}
\begin{equation}
q_0=1+\delta_T-B',
\label{eq:q0}
\end{equation}
where $\delta_T$ is the so-called Gr\"{u}neisen-Anderson parameter:\cite{anderson_q}
\begin{equation}
\delta_T \equiv \frac{\partial \ln \alpha}{\partial \ln V}.
\end{equation}
{Using Dugdale and MacDonald \cite{dugdale} work, Chang et al.,}  
%
obtained a simple relation between $\delta_T$ and $\gamma_G^0$:  \cite{chang_gruneisen}
\begin{equation}
\delta_T = 2\gamma_G^0.
\end{equation}
%
Finally, second order Gr\"{u}neisen parameter may be calculated {at ambient conditions} as~\cite{chang_gruneisen}:
\begin{equation}
q_0=1+2\gamma_G^0 -B'.
\label{eq:q0_final}
\end{equation}
Bulk modulus values $B$ and $B'$ were determined above from the $P(V)$ fit, thus the only parameter which remained to be determined is the power-law coefficient $n$ from Eq.(\ref{eq:sec_gruneisen_2}). Unfortunately there are no available literature data to estimate $n$, even for the constituent elements of TNHZT. To overcome this difficulty, first-principles phonon calculations in the pressure range of 0 - 100 GPa were performed for elemental Nb and Ta, which are the main components of our HEA and have the same bcc crystal structure. This allowed us to validate the above-described method of calculating $\theta_D(P)$ as well as to obtain some information about the value of $n$.

\begin{table}[t]
\caption{\label{tab:grun}Computed and experimental values of the Debye temperature $\theta_D$~\cite{mcmillan_1968,kimura_1969,guillermet_1989,grigoriev_1997}, the bulk modulus $B$, the pressure derivate of the bulk modulus $B'$~\cite{katahara}, the Gr\"{u}neisen parameter $\gamma_G^0$~\cite{gschneidner} and the second order Gr\"{u}neisen parameter $q_0$ [Eq.~\ref{eq:q0_final}] at ambient conditions. }
\begin{ruledtabular}
\begin{tabular}{ccccccc}
&$\theta_D$ (K) & $B$ (GPa) & $B'$ & $\gamma_G^0$ & $q_0$  \\
\hline
Nb (calc) & 271 & 163 & 3.52 & 1.55 & 0.14  \\
Nb (expt.) & 270-280 & 169 & 4.02 & 1.59 & 0.16  \\
Ta (calc) & 219 & 194 & 3.787 & 1.427 & 0.067 \\
Ta (expt.) & 229-258 & 194 & 3.80 & 1.64 & 0.48 \\
\end{tabular}
\end{ruledtabular}
\end{table}
	
\begin{figure}[t]
\includegraphics[width=0.45\textwidth]{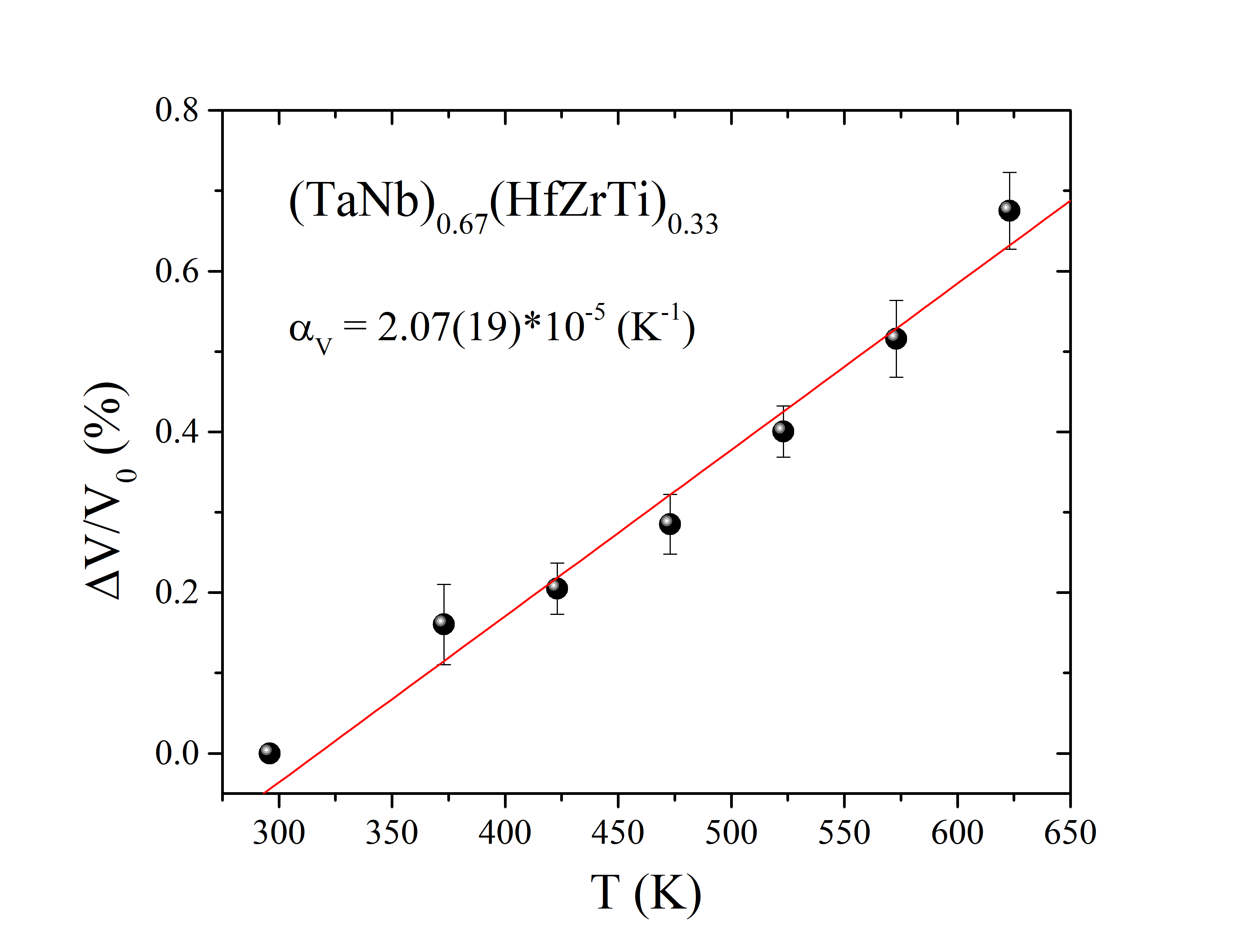}
\caption{\label{fig:thermal}Temperature dependence of a relative change of the unit cell volume. The lattice parameter was obtained by the LeBail method. A cubic $Im$-$3m$ (s.g. no 229) structure was used as a starting model.}
\end{figure}

\begin{figure}[t]
\includegraphics[width=0.45\textwidth]{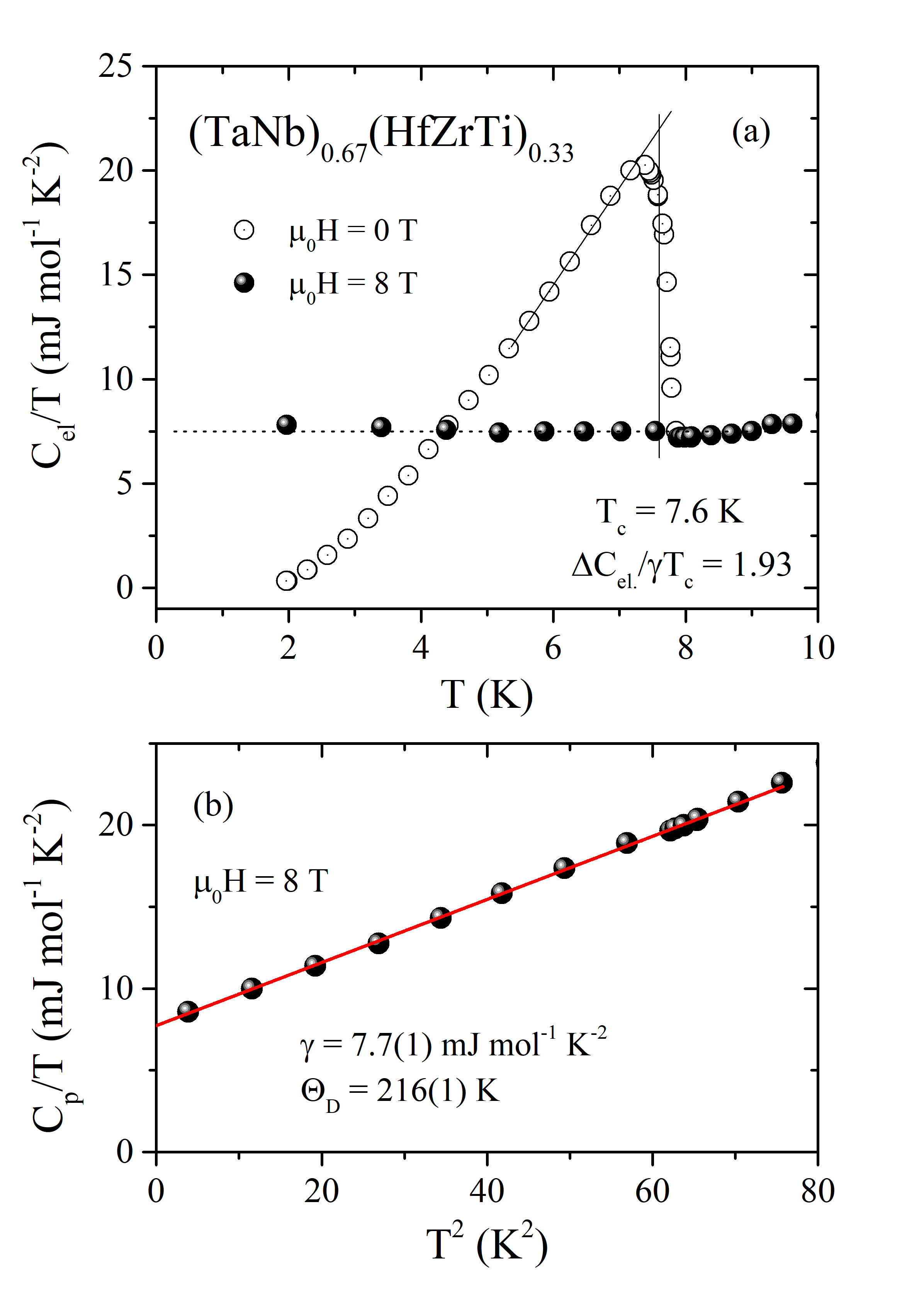}
\caption{\label{fig:heat}Panel (a): temperature dependence of the electronic heat capacity $C_{\rm el}/T$ in zero (open circles) and 8 T (close circles) magnetic field. Panel (b): low temperature experimental data $C_p/T$ vs. $T^2$. The solid red line is a fit by the expression $C_p/T = \gamma + \beta T^2$.}
\end{figure}

Calculations of the phonon densities of states for Nb and Ta were {performed} using a {\sc Quantum ESPRESSO} software~\cite{QE-2009,QE-2017}. We used projected augmented wave (PAW) pseudopotentials~\cite{pseudo2,pseudo},
with the Perdew-Burke-Ernzerhof generalized gradient approximation for the exchange-correlation 
potential~\cite{pbe}. First, phonon densities of states $F(\omega)$ were computed for various pressures and then the Debye temperature was calculated, based on the $m$-th moment of the phonon spectrum:
\begin{eqnarray}
\mu_m = \int_0^{\omega_{\mathsf{max}}} \omega^{m} F(\omega) d\omega \left/ \int_0^{\omega_{\mathsf{max}}} F(\omega) {d\omega}\right.\label{eq:debcalc}\\
\omega_D(m) = \left(\frac{m+3}{3}\mu_m\right)^{1/m}.\label{eq:debcalc2}
\end{eqnarray}
Among many available formulas for the "theoretical" Debye temperature (see, Ref.~\cite{grimvall,grimvall2} for more details) we choose the one, which corresponds to the correct representation of the heat capacity for $T > \theta_D$, i.e. $m = 2$, and $k_B\theta_D = \hbar\omega_D(2)$.
{However, since our materials have a simple acoustic phonon spectrum, $\theta_D$ computed using different values of $m$ do not change in more than 5\%.}
It should be noted that there is no conflict between Eqs.(\ref{eq:debcalc})-(\ref{eq:debcalc2}) and approximation $\langle{\omega^2}\rangle = \frac{1}{2}\theta_D^2$. {To avoid any confusions we explain} the difference between $\langle{\omega^2}\rangle$ (that enters Eq. (\ref{eq:lambda0})) and the second moment of the phonon DOS function in Supplemental Material~\cite{suppl}.

Computed and experimental values of the bulk modulus $B$, its pressure derivative $B'$, the Gr\"{u}neisen parameter $\gamma_G^0$, and the second order Gr\"{u}neisen parameter $q_0$ for Nb and Ta are gathered in Table \ref{tab:grun}. 
The second order Gr\"{u}neisen parameter $q_0$ is calculated {using} Eq. \ref{eq:q0_final}. The $\theta_D^0$ parameter obtained from the phonon calculations {and} at zero pressure {is} 271 K for Nb and 219 K for Ta. The calculated Debye temperature of Nb remains in a very good agreement {to} the experimental values, which span the range of 270-280 K \cite{mcmillan_1968,kimura_1969}. The $\theta_D^0$ of Ta is slightly smaller than the literature values that range from 229 K \cite{guillermet_1989} {\t via} 245 K, \cite{grigoriev_1997} up to 258 K  \cite{mcmillan_1968}. Gr\"{u}neisen model calculations of $\theta(P)$ were performed using the computed $\theta_D^0$ and other parameters, shown in Table~\ref{tab:grun} {and} for several values of $n$, {ranging} from 4 to 16. The values of $\theta(P)$, calculated directly from phonon DOS under pressure (shown in Supplemental Material~\cite{suppl}) and from the Gr\"{u}neisen model for representative values of $n$ are compared in Fig.~\ref{debye_nb} for Nb and Fig.~\ref{debye_ta} for Ta. {In the case of} Nb an almost perfect agreement is found. Larger deviation is seen for Ta, but still the differences between the model and first-principle calculations are smaller than 10\%. It is also worth noting that our model calculations of $\theta_D(P)$ for Ta remain in a very good agreement with the quasi-harmonic approximation calculations of Liu {\it et al.} \cite{liu_2009}. The general observation is that the pressure dependence of $\theta_D$ is quite well captured by the Debye-Gr\"{u}neisen model, which contains only one free parameter, $n$. Moreover, the computed $\theta_D(P)$ are not very sensitive to the particular choice of $n$ due to relatively small $q_0$ values. {In} the case of Nb, where the agreement is better, $n=16$ seems to be the best choice. {Thereofore}, this value will be assumed in the analysis of HEA, where due to the presence of chemical disorder a phonon spectrum was not calculated. 

{To summarize the methodology section, electron-phonon coupling constant $\lambda$ is calculated using Eq.~(\ref{eq:lambda}), McMillan-Hopfield parameters are computed from band-structure results using Eq.~(\ref{eq:eta}), ambient-pressure values of the Debye temperature $\theta_D^0$ and the Gr\"{u}neisen parameters $\gamma_G^0$ and $q_0$ are taken from experiment, and the evolution of $\theta_D$ with pressure is modelled using Eq.~(\ref{eq:gruneisen_v}-\ref{eq:thetap}), where $n = 16$ is assumed.}

\section{Results and discussion}

\begin{table}[t]
\caption{\label{tab3}Volume thermal expansion coefficient $\alpha_V$ (K$^{-1}$), zero-pressure Gr\"{u}neisen parameter (dimensionless) and the bulk modulus $B$ (GPa) of TNHZT, determined in this work, compared to several  refractory HEAs \cite{li_refractory_hea_2018},\cite{ahmad_2017}.}
\begin{ruledtabular}
\begin{tabular}{lccr}
 & $\alpha_V$ &  $\gamma_G^0$  & B \\
\hline
TNHZT (this work) & 2.07 &  1.62 & 177.4 \\
TiZrHfVNb & 3.60  & 1.83 & 79.0 \\
TiZrVNb & 3.34    & 1.65 & 84.2 \\
TiZrVNbMo & 3.32  & 2.19 & 125.0\\
NbTaMoW & 2.67  & 2.40 & 162.5  \\
NbHfZrTi & 2.30	 & - & 88.3 \\
\end{tabular}
\end{ruledtabular}
\end{table}

\subsection{The thermal expansion and heat capacity}
The (TaNb)$_{0.67}$(HfZrTi)$_{0.33}$ as-cast sample was characterized using X-ray diffraction method (XRD). The measurement was first performed at room temperature and then at temperatures from 100$^\circ$C up to 400$^\circ$C with a step of 50 deg. The XRD pattern is shown in the Supplemental Material~\cite{suppl} and contains only sharp Al$_2$O$_3$ (holder) reflections and reflections that were indexed with an I-centered cubic phase. A cubic lattice parameter for TNHZT was refined using the LeBail method {and} HighScore software. A relative change of a unit cell volume ($\Delta V/V_0$) vs. temperature is presented in Figure~\ref{fig:thermal}. The volumetric thermal expansion coefficient was found to be $\alpha = 2.07(19) \times 10^{-5}$~K$^{-1}$ and is comparable {to} those obtained for the constituting metals [for which it changes from 17.1 (Zr) to 27 (Ti), given in  ($10^{-5}$~K$^{-1}$)].

\begin{figure}[t]
\begin{center}
\includegraphics[width=0.48\textwidth]{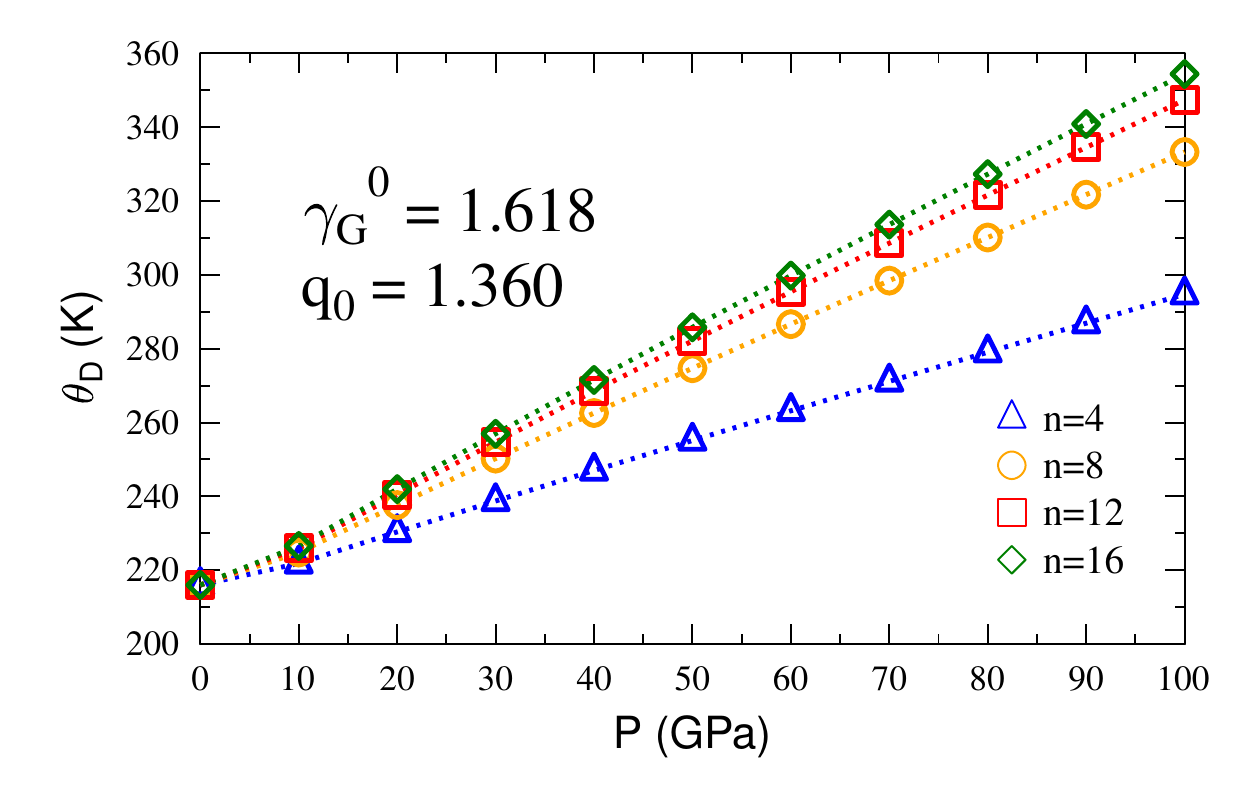}
\end{center}
\caption{\label{debye_tnhzt} Simulated pressure dependence of $\theta_D$ of TNHZT.}
\end{figure}

{The} temperature dependence of the electronic heat capacity, $C_{\rm el}/T$ of (TaNb)$_{0.67}$(HfZrTi)$_{0.33}$ is presented in Fig.~\ref{fig:heat}(a). The experimental data were collected under zero (open circles) and applied magnetic field (close circles). {The} $C_{\rm el}$ was obtained from the relation $C_p = C_{\rm el} + C_{\rm ph}$, where $C_{\rm ph} = \beta T^3$ is the low temperature ($T < \theta_D/50$) phonon contribution. In order to obtain $\beta$, the heat capacity in the normal state was measured and the data are presented in panel (b), plotted as $C_p/T$ vs. $T^2$. 
In the normal state $C_p$ can be {analyzed} by $C_p = \gamma T + \beta T^3$, where $\gamma T$ is the contribution from the conduction electrons. The fit is represented by a solid red line with the fitting parameters: $\gamma = 7.7(1)$ mJ mol$^{-1}$ K$^{-2}$ and $\beta = 0.193(2)$ mJ mol$^{-1}$ K$^{-4}$. {Then}, we can {calculate} the Debye temperature from the relation: $\theta_D = [(12\pi R)/(5\beta)]^{1/3}$, where $R$ is the gas constant. Both {the} Sommerfeld parameter and Debye temperature $\theta_D = 216(1)$~K are in good agreement with those reported previously~\cite{tnhzt2} ($\gamma = 7.97$ mJ mol$^{-1}$ K$^{-2}$, $\theta_D = 225$~K). The sharp anomaly at $T_c = 7.6$~K, seen in the $C_{\rm el}/T$ confirms a bulk nature of superconductivity in the studied sample. A normalized jump {of} the specific heat $\Delta C/\gamma T_c = 1.93$ is comparable to that reported {in Ref. \onlinecite{tnhzt2}}. The estimated value exceeds the {one} expected for weak-coupling BCS superconductors $\Delta C/\gamma T_c = 1.43$, {indicating} that (TaNb)$_{0.67}$(HfZrTi)$_{0.33}$ is a an intermediate- or strong-coupling superconductor. 

\subsection{The Debye temperature under pressure}

Evolution of {the} Debye temperature with pressure, needed to calculate $\lambda$ in our approach, was simulated according to the model described above. {The} ambient-pressure Gr\"{u}neisen parameter $\gamma_G^0$ was calculated {using} Eq. \ref{eq:gruneisen_ambient}. The lattice thermal expansion coefficient and unit cell volume have been measured experimentally {and} heat capacity was approximated {by} the Dulong-Petit law. {The parameters} $B = 177$ GPa and $B' = 2.87$ were determined from the Birch-Murnaghan equation of state \cite{guo-pressure}. Values of those parameters are gathered in Table~\ref{tab3}, along with a data reported for similar alloys~\cite{li_refractory_hea_2018}. The lattice thermal expansion coefficient of TNHZT is relatively low {and it is} accompanied by a large bulk modulus. The obtained value of the Gr\"{u}neisen parameter $\gamma_G^0 = 1.62$ is similar to that found for the other listed alloys. 
Using Eq.(\ref{eq:q0_final}), the second-order Gr\"{u}neisen parameter $q_0 = 1.36$ is obtained. {It is} larger than {$q_0$ of} Nb and Ta, which is a direct consequence of smaller $B'$ {observed} in TNHZT. Unfortunately, there are no other reported values of $q_0$ ({to the best of our knowledge}) among HEAs to compare with. The calculated evolution of $\theta_D$ under pressure are shown in Fig.~\ref{debye_tnhzt} {for different values of $n$}. 
For larger $n$, $\theta_D(P)$ becomes insensitive to choice of $n$ and we assume $n = 16$ in further analysis. It also gave the closest results to the first-principles modeling and quasi-harmonic calculations for Nb and Ta, as described {above}. Finally, $\theta_D$ increases almost linearly with pressure and reaches 360 K at $P = 100$~GPa.

\subsection{Electronic structure}

\begin{table}[t]
\caption{\label{tab:el}Electronic properties of (TaNb)$_{67}$(HfZrTi)$_{33}$. M$_i$ is given in u, c$_i$ in \%, N(E$_F$) in Ry$^{-1}$, $\eta$ in mRy/a${_B^2}$.}
\begin{ruledtabular}
\begin{tabular}{ccccccccc}
 & c$_i$ & M$_i$ & N($E_F$)& $\eta_i$  &$\eta_{sp}$ &$\eta_{pd}$ &$\eta_{df}$ \\
\hline
Ta & 33.5 & 181 & 16.54  & 151.79 & 0.83 & 49.08 & 101.90 \\
Nb & 33.5 & 93 & 18.10  & 157.09 & 4.46 & 53.19 & 99.45 \\
Hf & 11 & 179 & 15.20  & 156.30 & 1.58 & 66.16 & 88.60\\
Zr & 11 & 91 & 16.21  & 165.01 & 6.55 & 73.16 & 85.28 \\
Ti & 11 & 41 & 24.78 & 119.17 & 4.80 & 43.56 & 70.80
\end{tabular}
\end{ruledtabular}
\end{table}

\begin{figure*}[htb]
\includegraphics[width=1.0\textwidth]{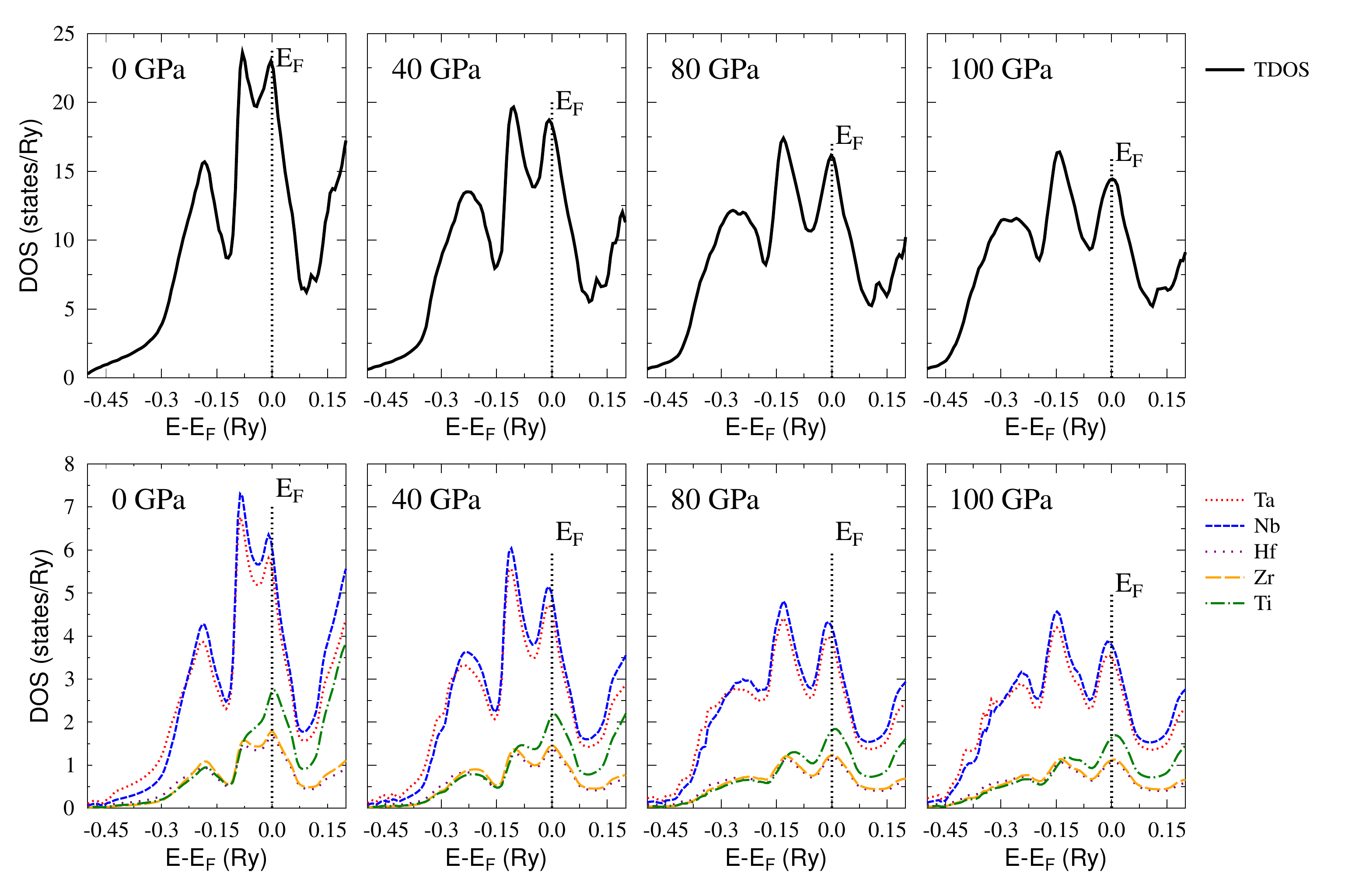}
\caption{\label{tdos-pdos}Total and atomic densities of states of TNHZT alloy, calculated under various pressures in the range of 0 to 100 GPa. Solid black line represents the total DOS. Atomic densities of states are plotted with colors and weighted over its concentrations. }
\end{figure*}	

\begin{figure}[b]
\begin{center}
\includegraphics[width=0.48\textwidth]{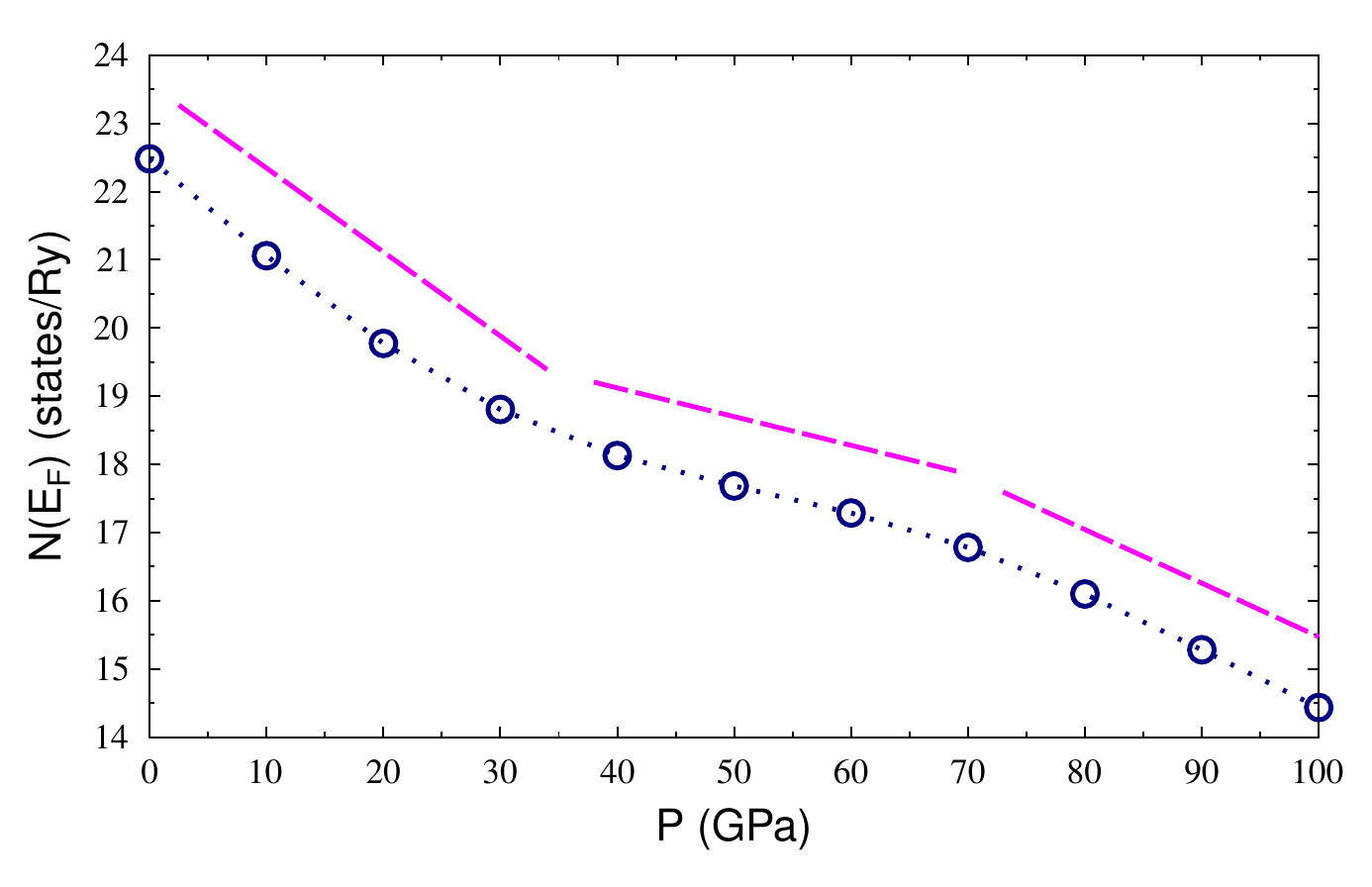}
\end{center}	
\caption{Variation of the density of states at the Fermi level under hydrostatic pressure from 0 to 100 GPa. Dashed lines are the linear trend lines, described in the text.}
\label{nef_tdos}
\end{figure}

{Figure~\ref{tdos-pdos} presents total and atomic densities of states of TNHZT, calculated under various pressures. In (TaNb)$_{67}$(HfZrTi)$_{33}$ and at ambient conditions, the Fermi level is located in the DOS peak, similarly to the first superconducting HEA, Ta$_{34}$Nb$_{33}$Hf$_{8}$Zr$_{14}$Ti$_{11}$~\cite{jasiewicz_2016}.}
Main contributions to the total DOS {originate} from the $d$-shells of all consistuent atoms (3$d$ for Ti, 4$d$ for Zr and Nb, and 5$d$ for Hf and Ta). {As the pressure increases, DOS strongly decreases. It is mainly due to enhanced hybridization and decrease of the unit cell volume.} {Furthermore}, applied pressure increases separation of the two highest DOS peaks (one located at the Fermi level and the second one below $E_F$) and shifts electronic states to a lower energy range (i.e. increases the bandwidth). {In addition,} a shift of electronic states causes a gradual decrease of the third DOS maximum, lying in the lowest energy range. Atom with the largest contribution to the total DOS at $E_F$ {is Ti (see also Table~\ref{tab:el}) for both ambient and elevated pressure conditions}. 
Fig.~\ref{nef_tdos} shows the gradual decrease of the $N(E_F)$ value with pressure, from about 22.5~Ry$^{-1}$ to 14.5~Ry$^{-1}$ at 100 GPa.  
The ambient-pressure value corresponds to the non-interacting Sommerfeld parameter $\gamma_0 = 3.9$ mJ mol$^{-1}$ K$^{-2}$. Comparison {to} the experimental value of $\gamma = 7.7(1)$ mJ mol$^{-1}$ K$^{-2}$ gives the electron-phonon enhancement parameter $\lambda = \gamma/\gamma_0 - 1 = 0.97$, {which is} very close to the value obtained for Ta$_{34}$Nb$_{33}$Hf$_{8}$Zr$_{14}$Ti$_{11}$~\cite{jasiewicz_2016}.

\begin{figure}[t]
\begin{center}
\includegraphics[width=0.48\textwidth]{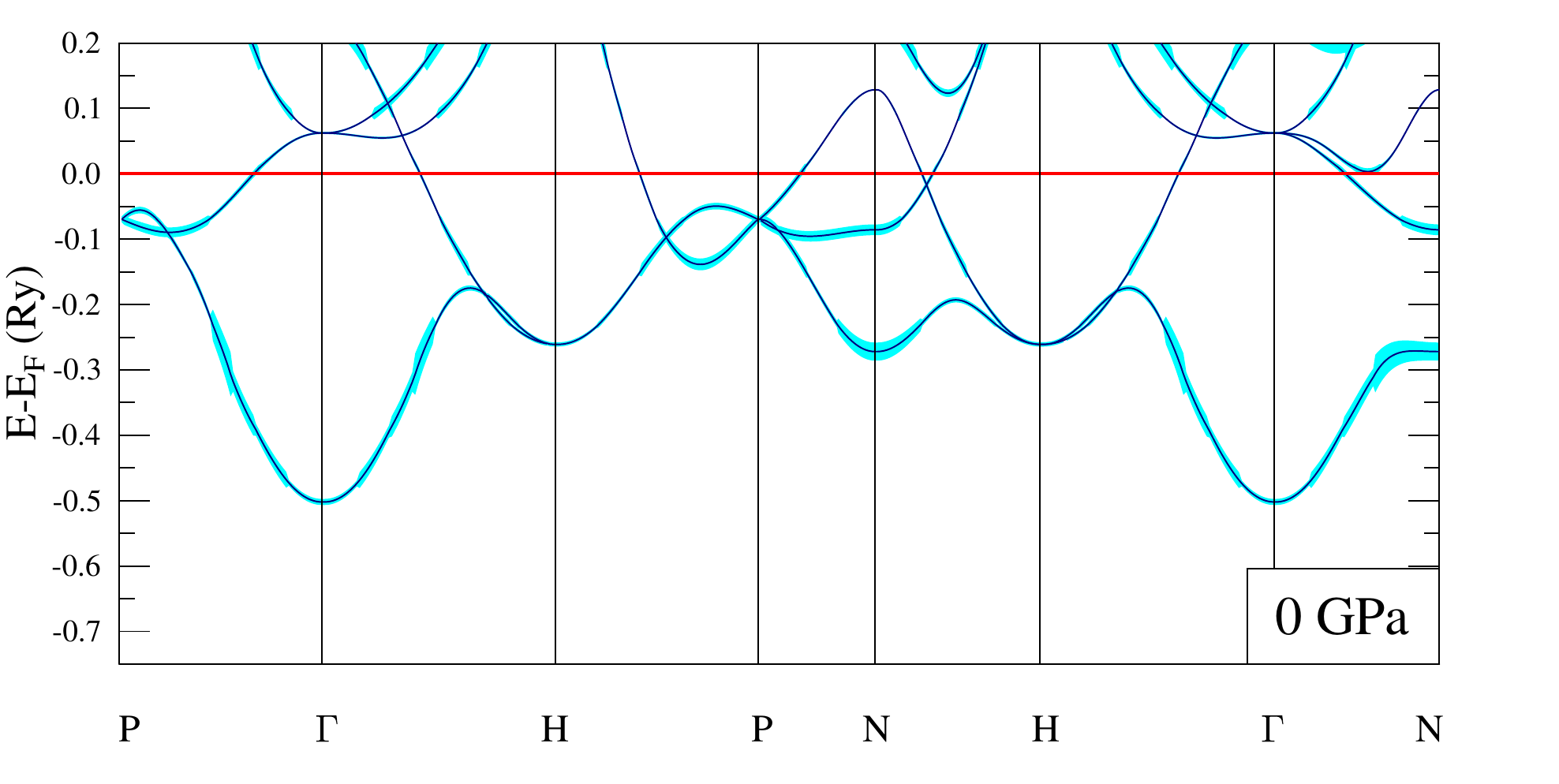}\\
\includegraphics[width=0.48\textwidth]{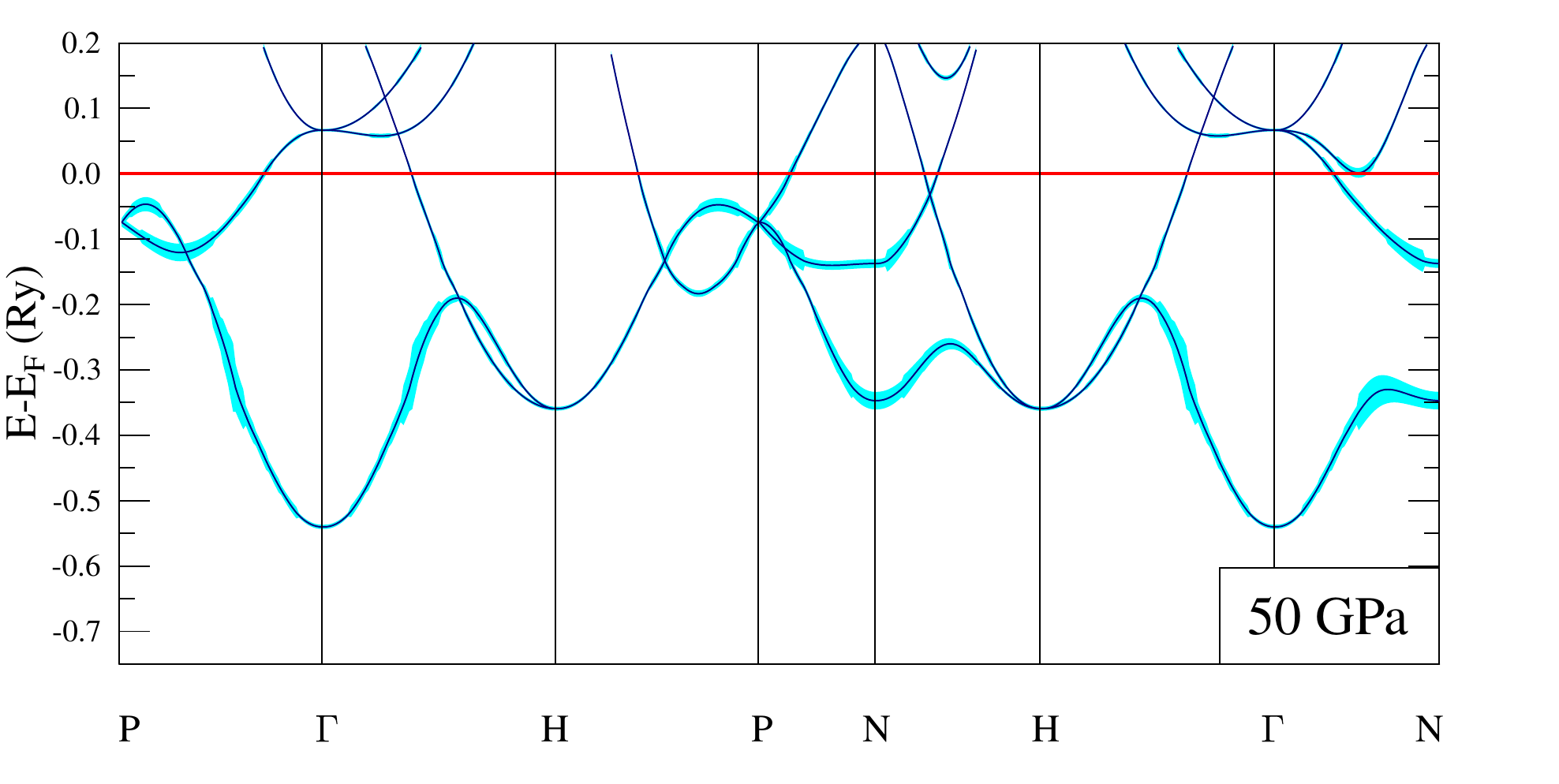}\\
\includegraphics[width=0.48\textwidth]{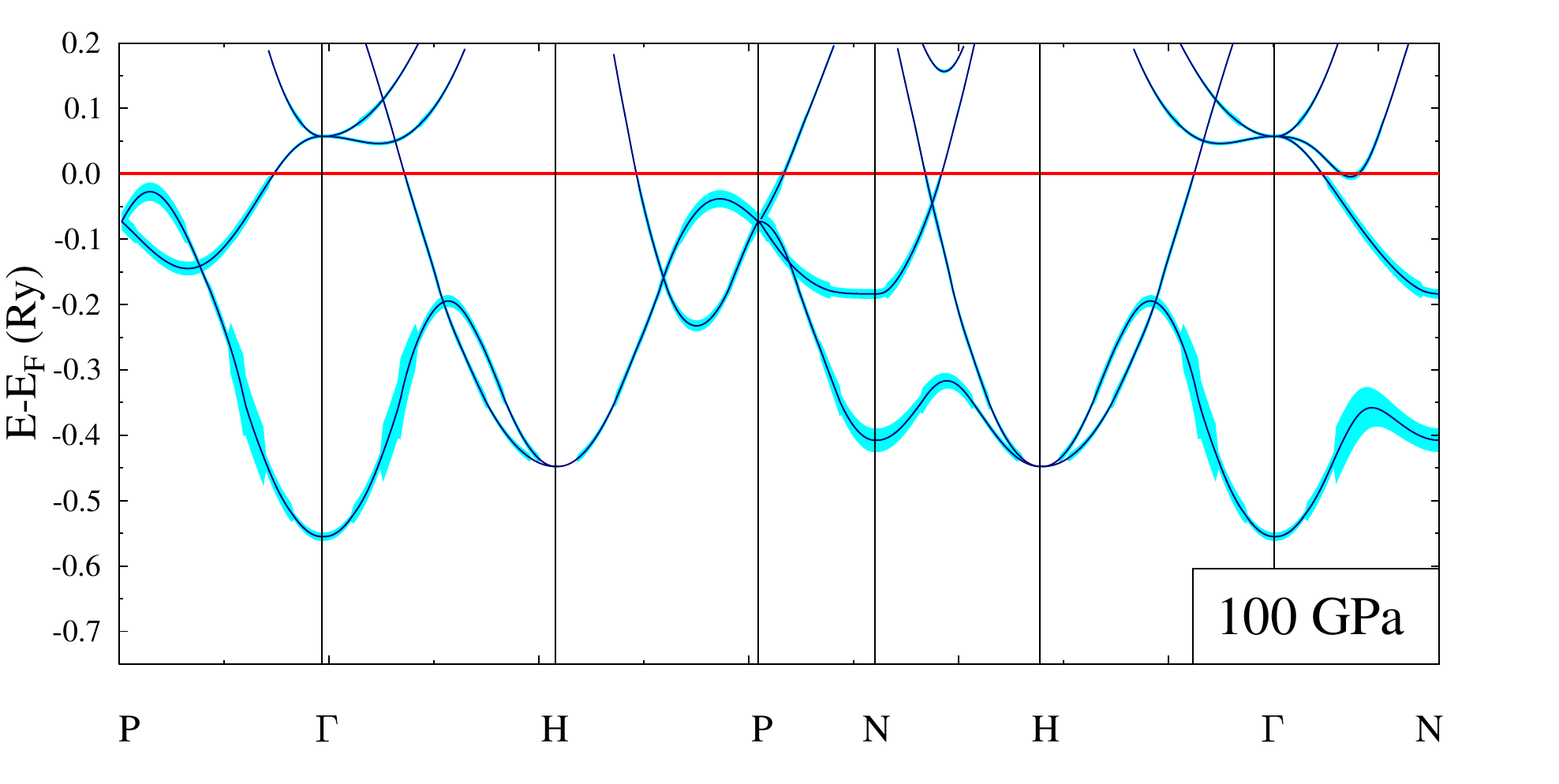}
\end{center}	
\caption{Electronic dispersion relations for $P = 0, 50$ and $100$ GPa. Shading describes the band smearing and corresponds to the imaginary part of the complex energy.}
\label{bands_press}
\end{figure}

In the $N(E_F)$ {\it versus} $P$ relation (Fig.~\ref{nef_tdos}) we can distinguish three regions. At first, $N(E_F)$ quickly decreases with a slope of -0.123 Ry$^{-1}$/GPa {up to 40 GPa}. Above 40 GPa, the decrease becomes slower (-0.042 Ry$^{-1}$/GPa), and {then}, above 70 GPa, the slope becomes more negative, reaching -0.079 Ry$^{-1}$/GPa. Interestingly, this evolution is correlated with the observed modifications of $T_c$ under pressure, where $T_c$ increases monotonically up to 10 K at around 50 GPa. {Above that pressure, the transition temperature}, remains practically constant. To analyze this trend {of} $N(E_F)$, electronic dispersion relations were computed using the complex energy band technique, attainable in the KKR-CPA formalism~\cite{bansil-complex,butler_1985,bw_scripta}. In this method, the real part of electron energy shows the band center, whereas the imaginary part describes the band smearing effects caused by a chemical disorder. A bandwidth is related to the electronic life time {that} is finite due to the presence of the disorder-induced electron scattering, $\tau = \hbar/2Im(E)$. 
As we have shown in Ref.~\cite{jasiewicz_2016} in the case of Ta$_{34}$Nb$_{33}$Hf$_{8}$Zr$_{14}$Ti$_{11}$, electronic bands were rather sharp with small smearing effect, {in spite of the high level of disorder.}
As seen in Fig.~\ref{bands_press} the same situation is found here, especially near $E_F$, where the small imaginary part of energy gives $\tau \sim 0.5 - 1\times 10^{-14}$~s. {Also, the smearing near $E_F$ does not change much under pressure, although it increases for the lower-lying states.}

\begin{figure}[htb]
\begin{center}
\includegraphics[width=0.48\textwidth]{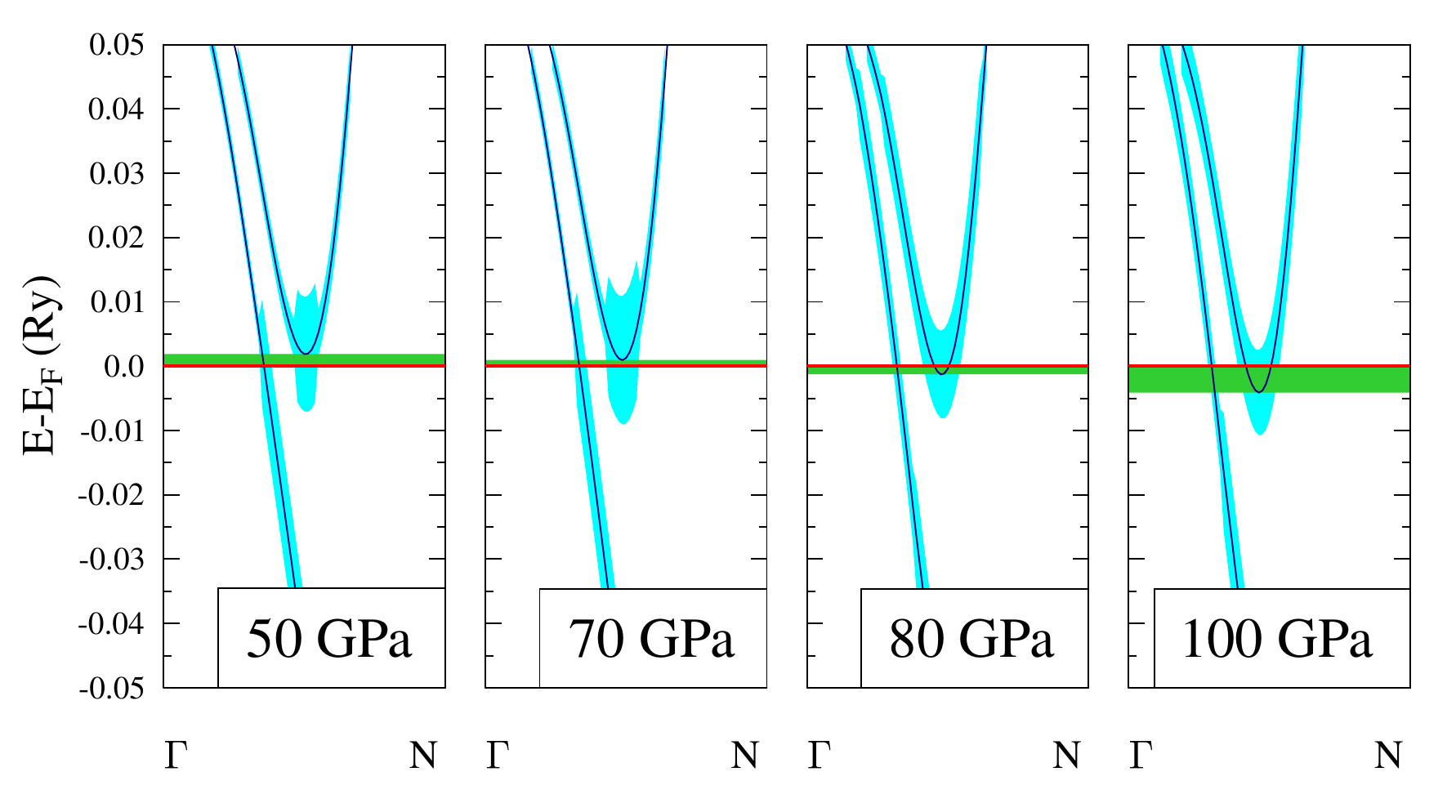}		
\end{center}	
\caption{Electronic dispersion relations near $E_F$ in $\Gamma-N$ direction, showing the Lifshitz transition. The distance between the center of the band and the Fermi level is marked in green. Shading describes the band smearing and corresponds to the imaginary part of the complex band energy.}
\label{band_zoom_ng}
\end{figure}

\begin{figure*}[htb]
			\begin{center}
				\includegraphics[width=0.24\textwidth]{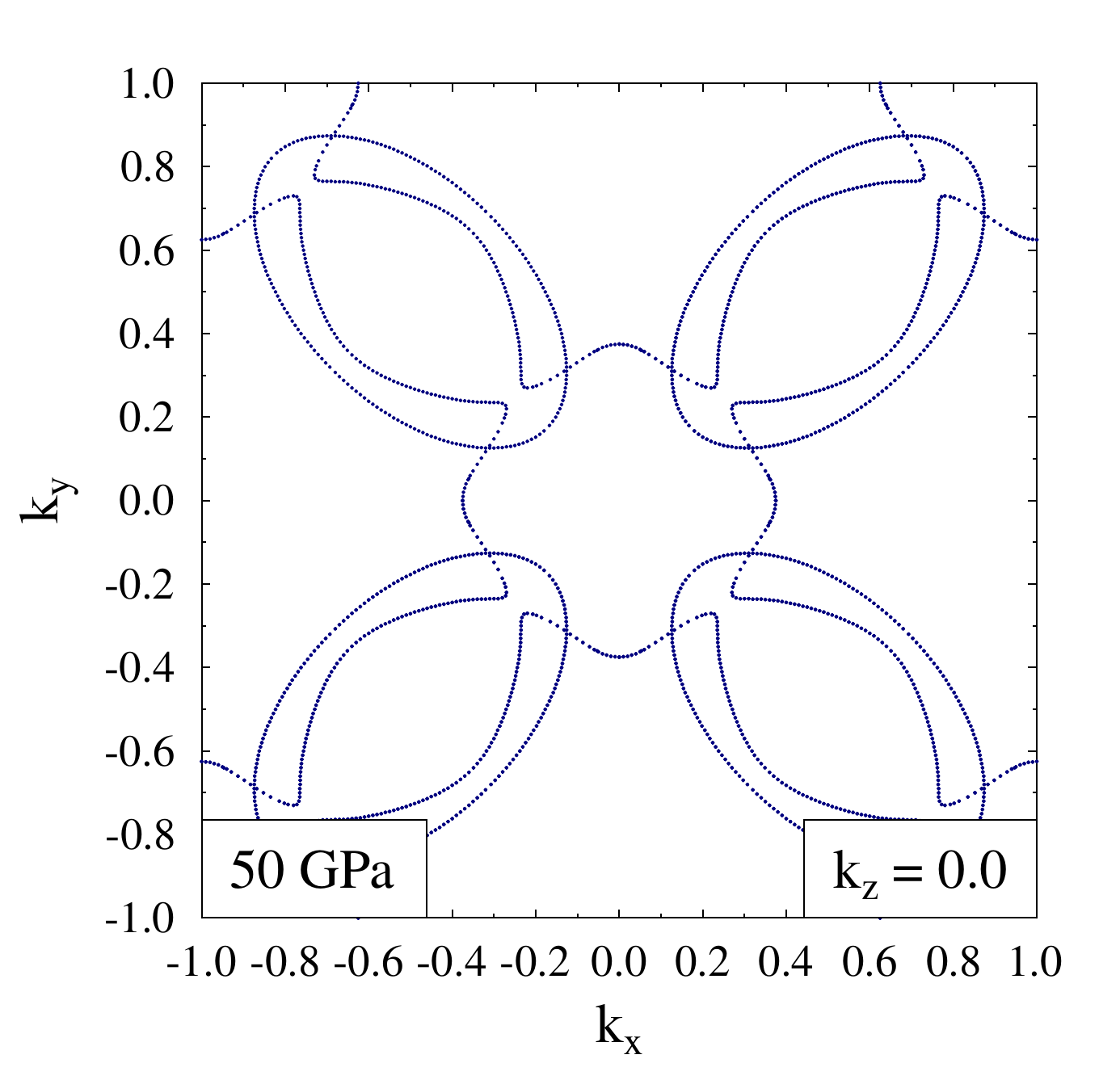}
				\includegraphics[width=0.24\textwidth]{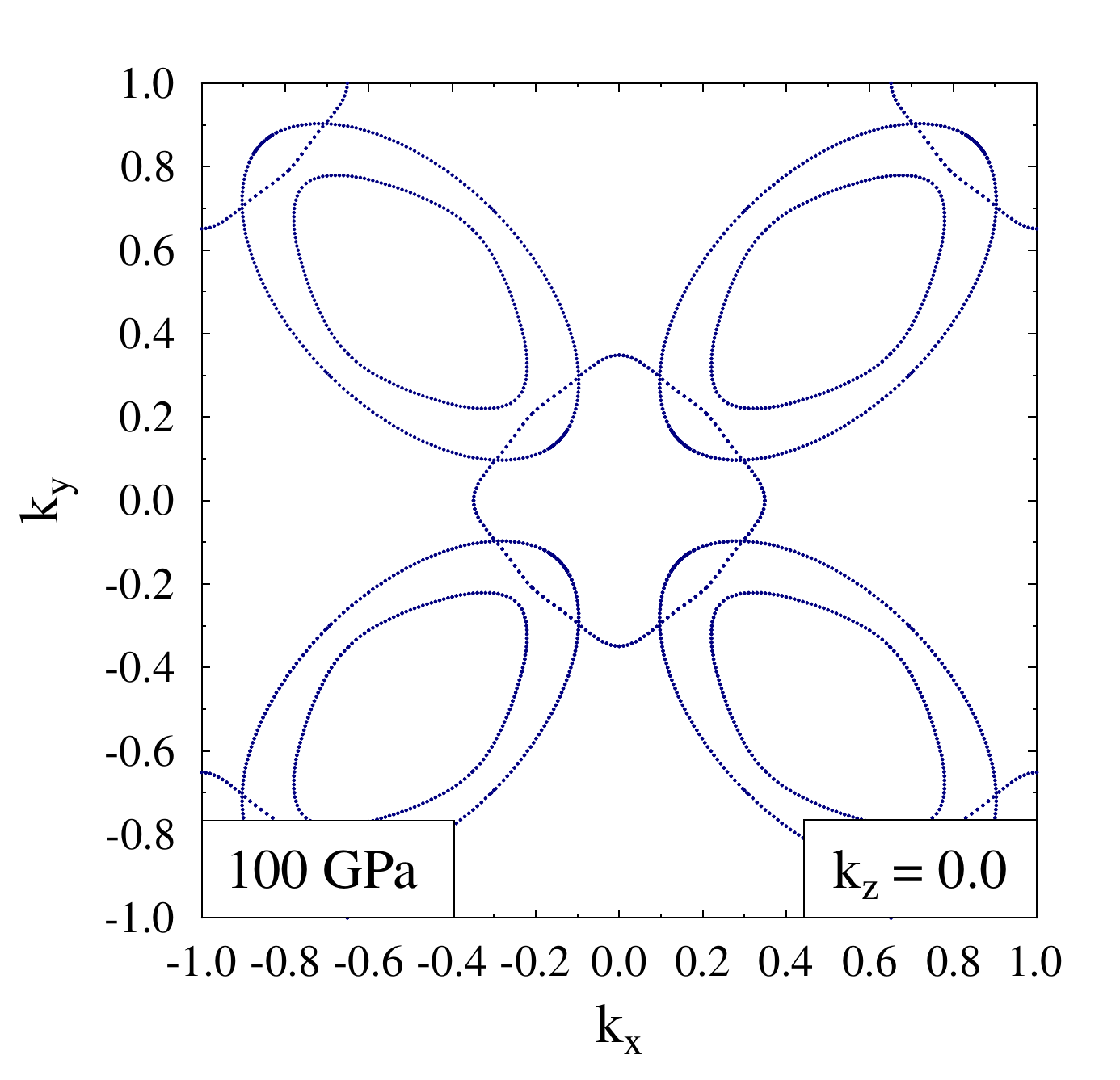}
				\includegraphics[width=0.24\textwidth]{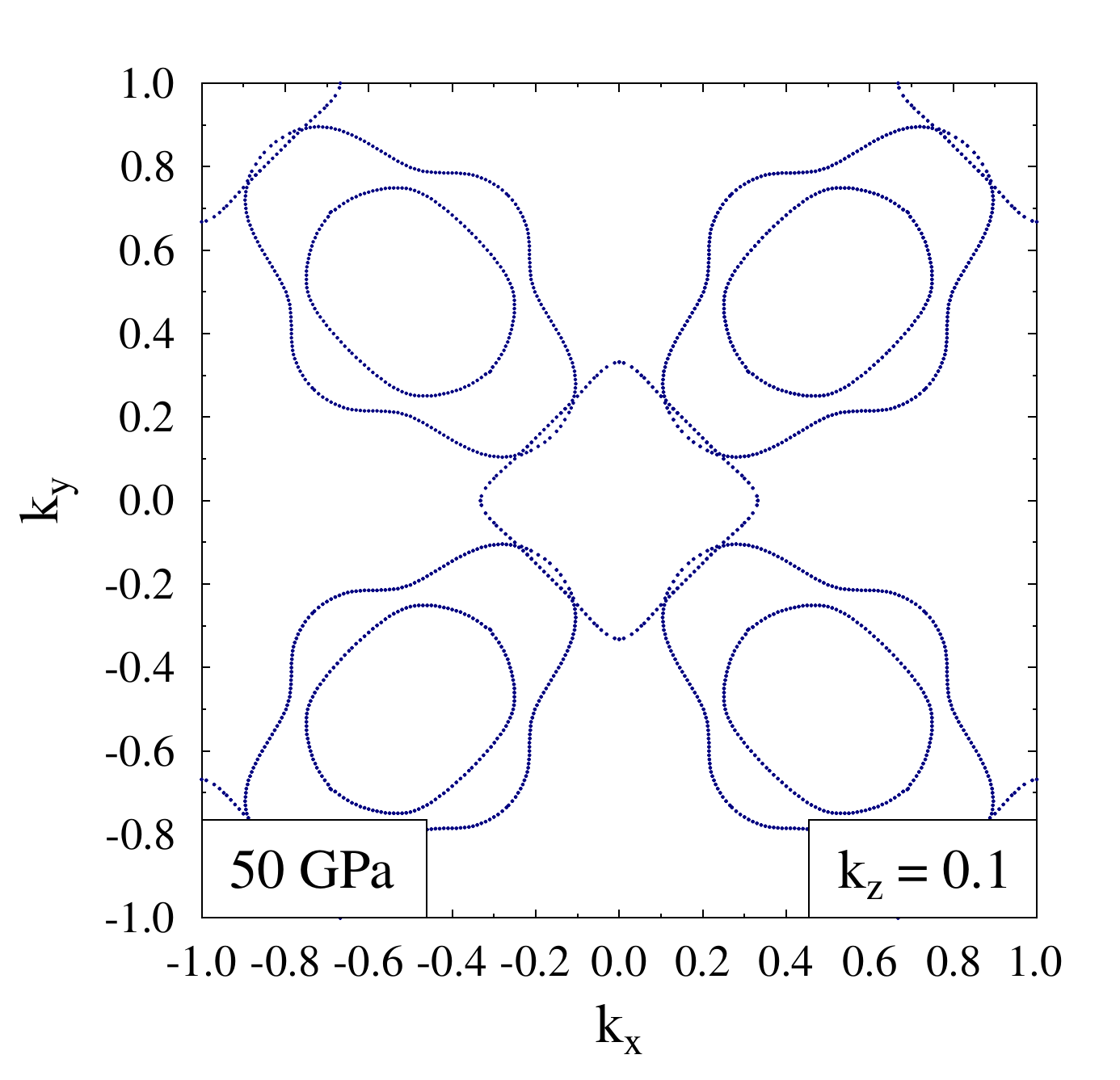}
				\includegraphics[width=0.24\textwidth]{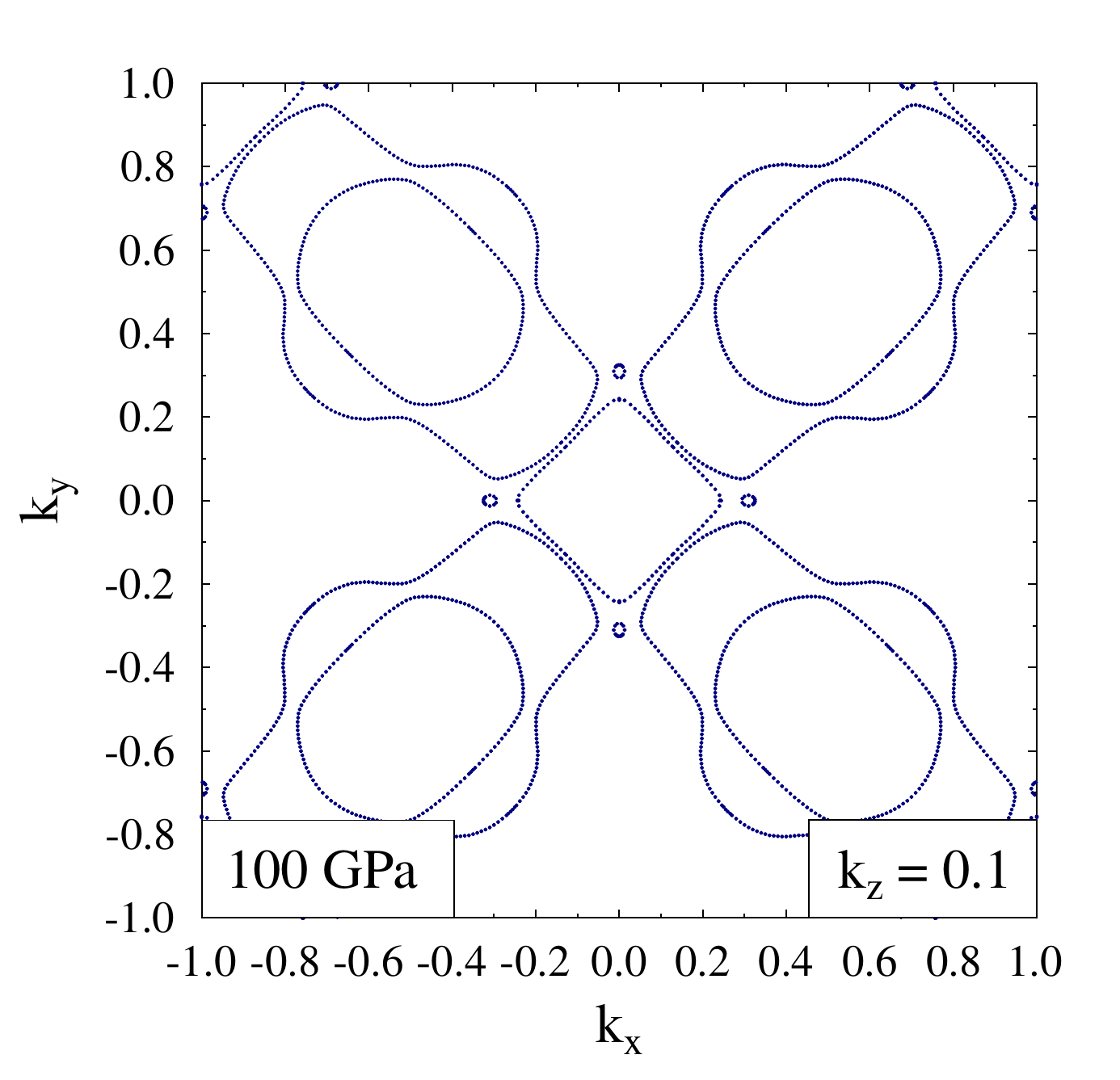}\\
				\includegraphics[width=0.24\textwidth]{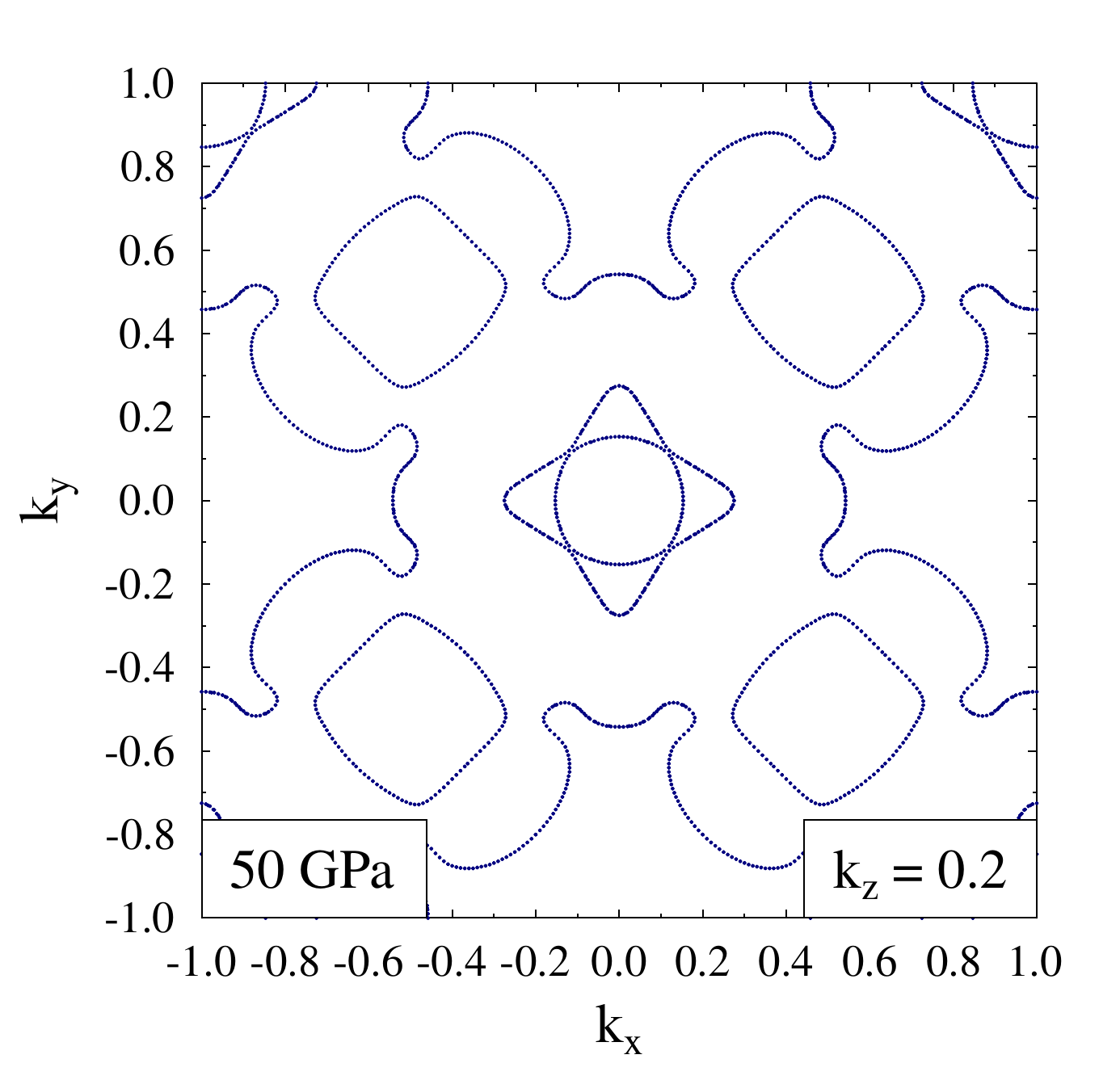}
				\includegraphics[width=0.24\textwidth]{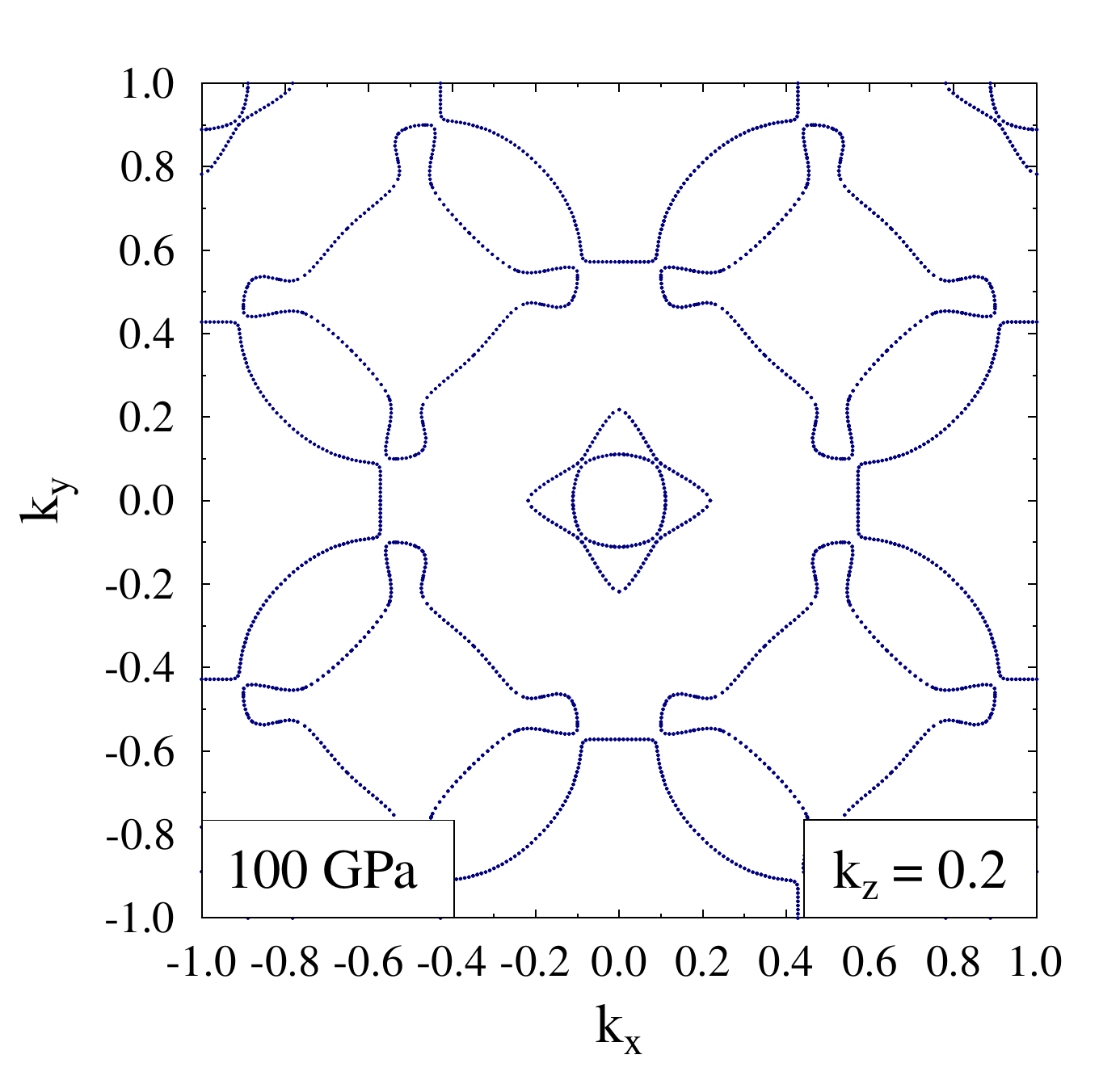}
				\includegraphics[width=0.24\textwidth]{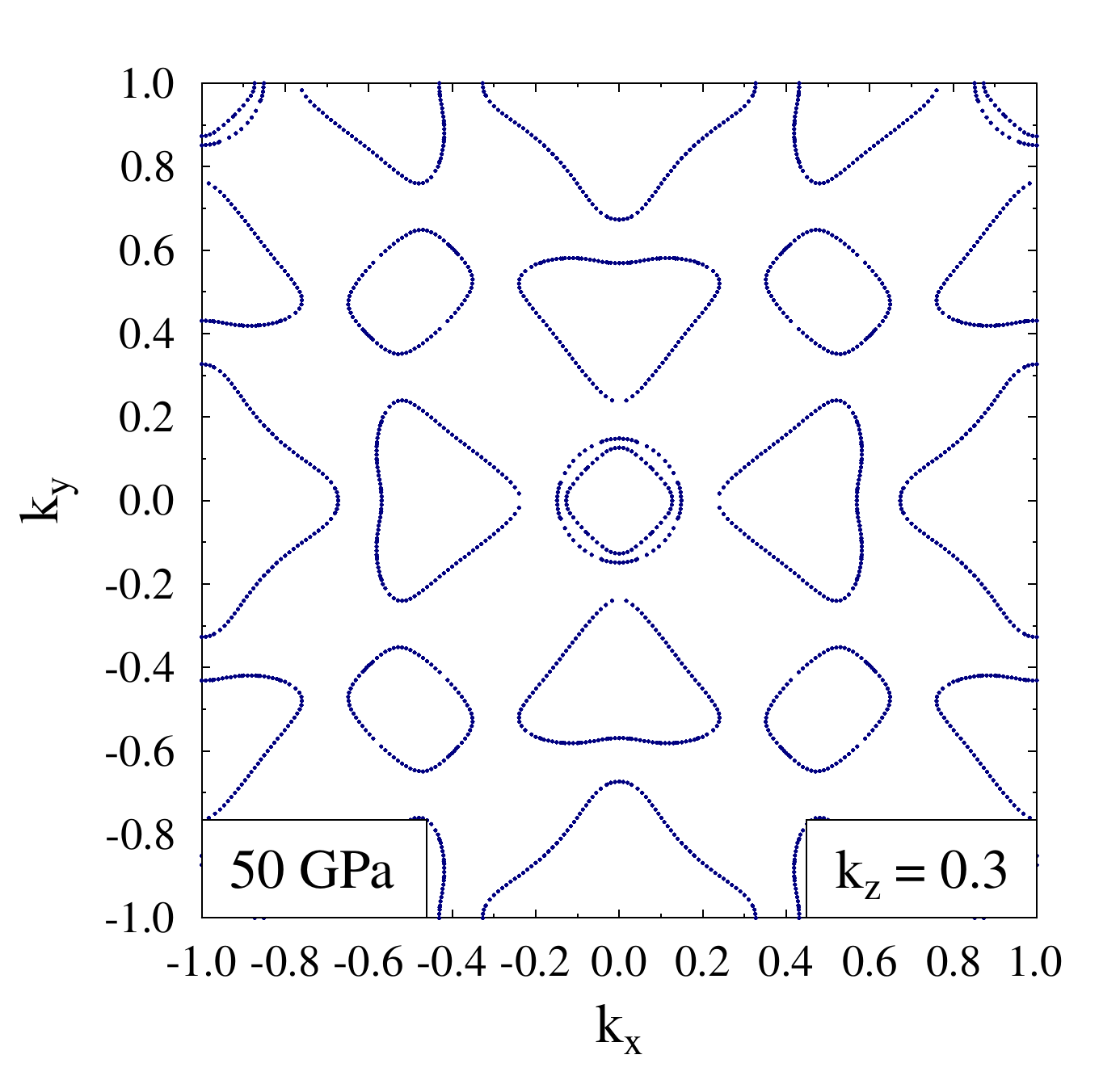}
				\includegraphics[width=0.24\textwidth]{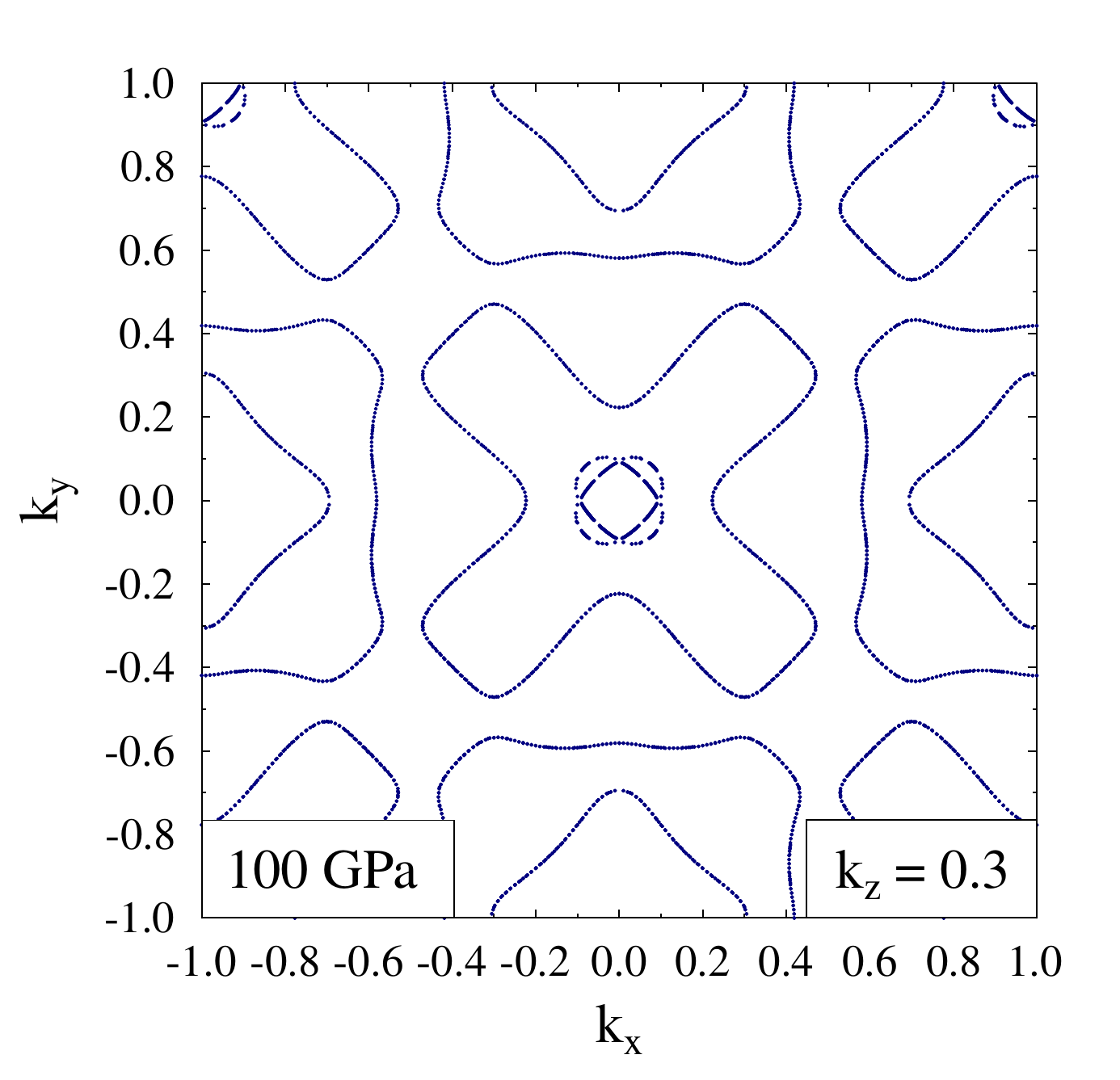}\\
				\includegraphics[width=0.24\textwidth]{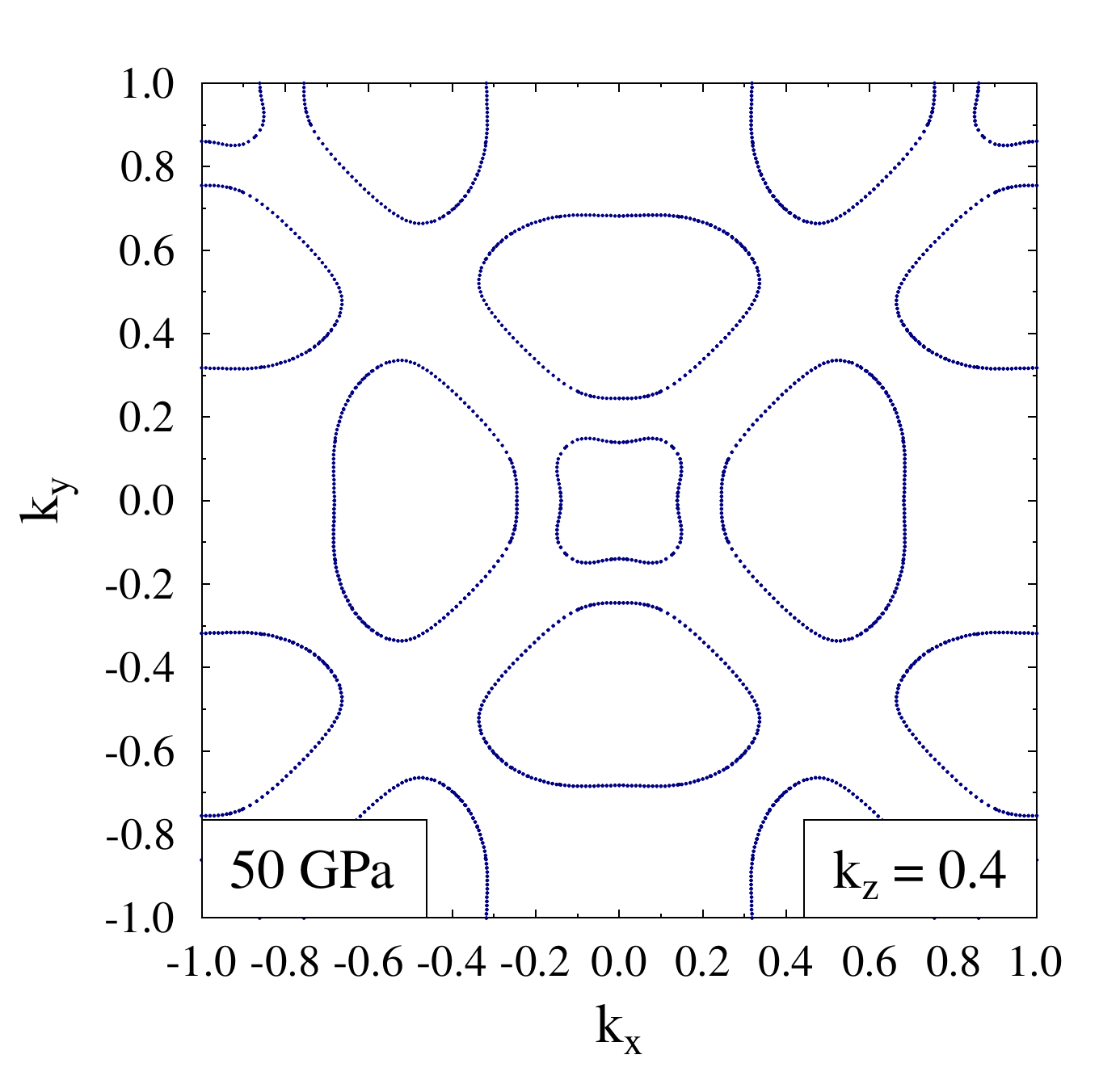}
				\includegraphics[width=0.24\textwidth]{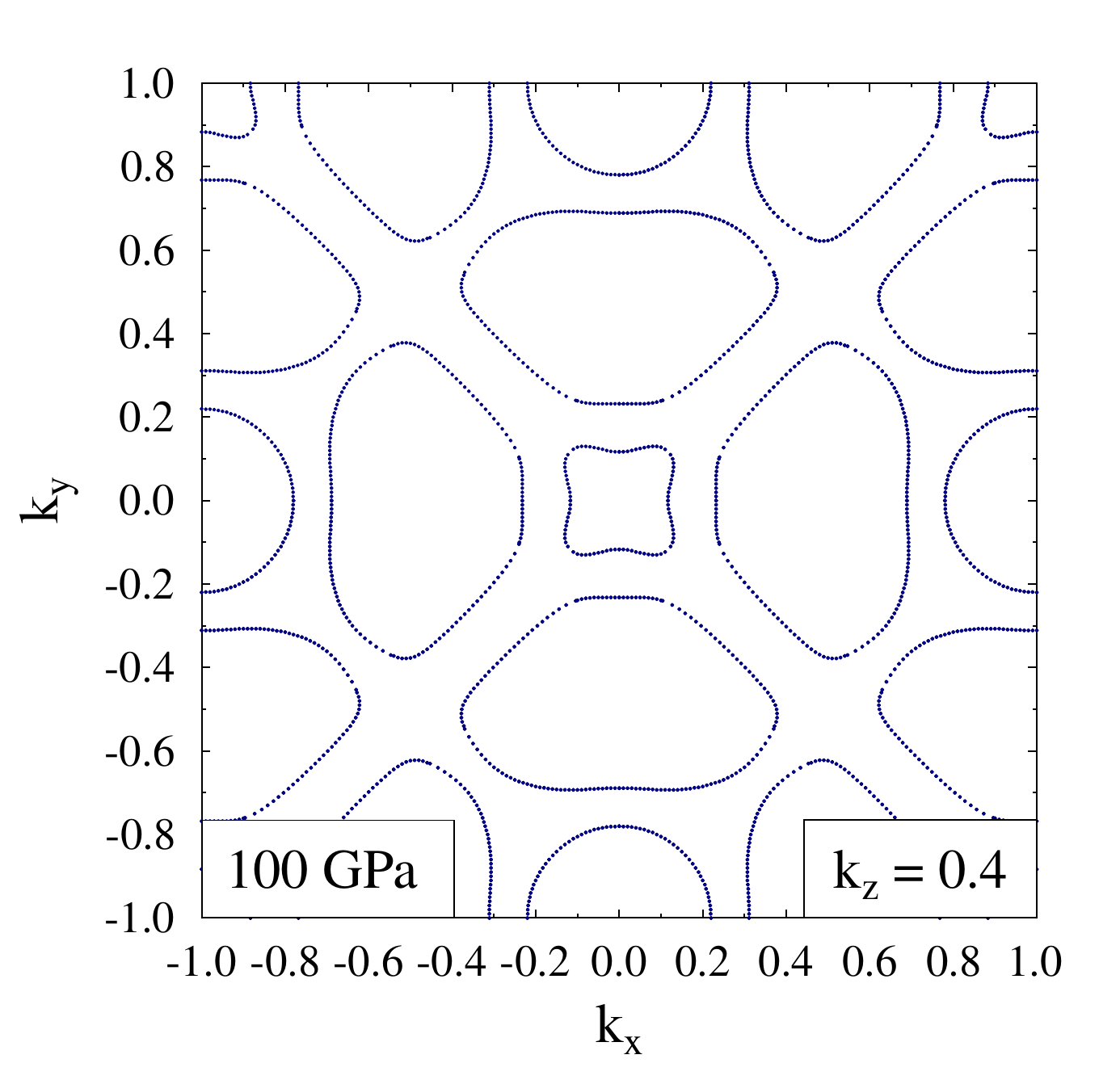}
				\includegraphics[width=0.24\textwidth]{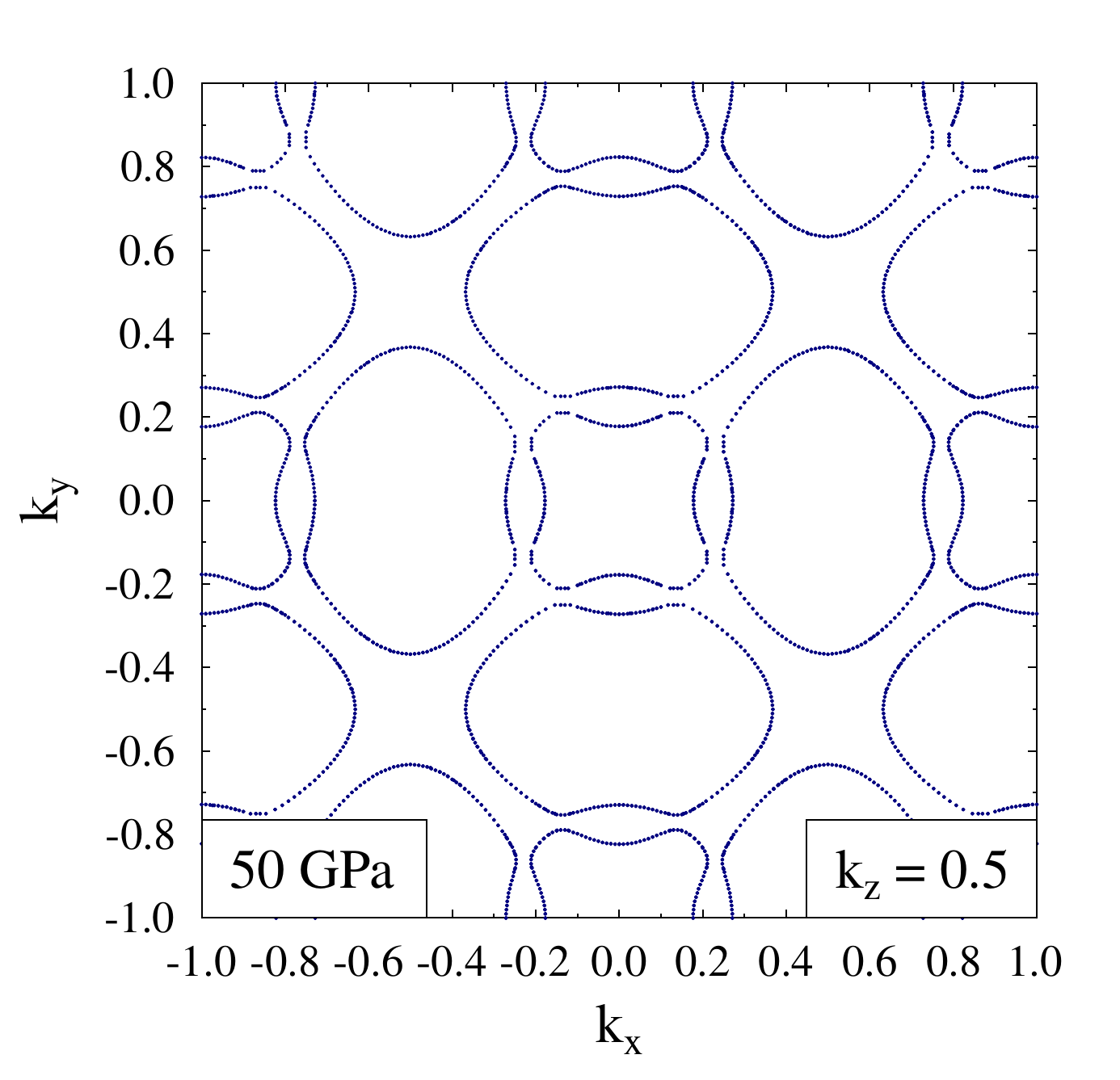}
				\includegraphics[width=0.24\textwidth]{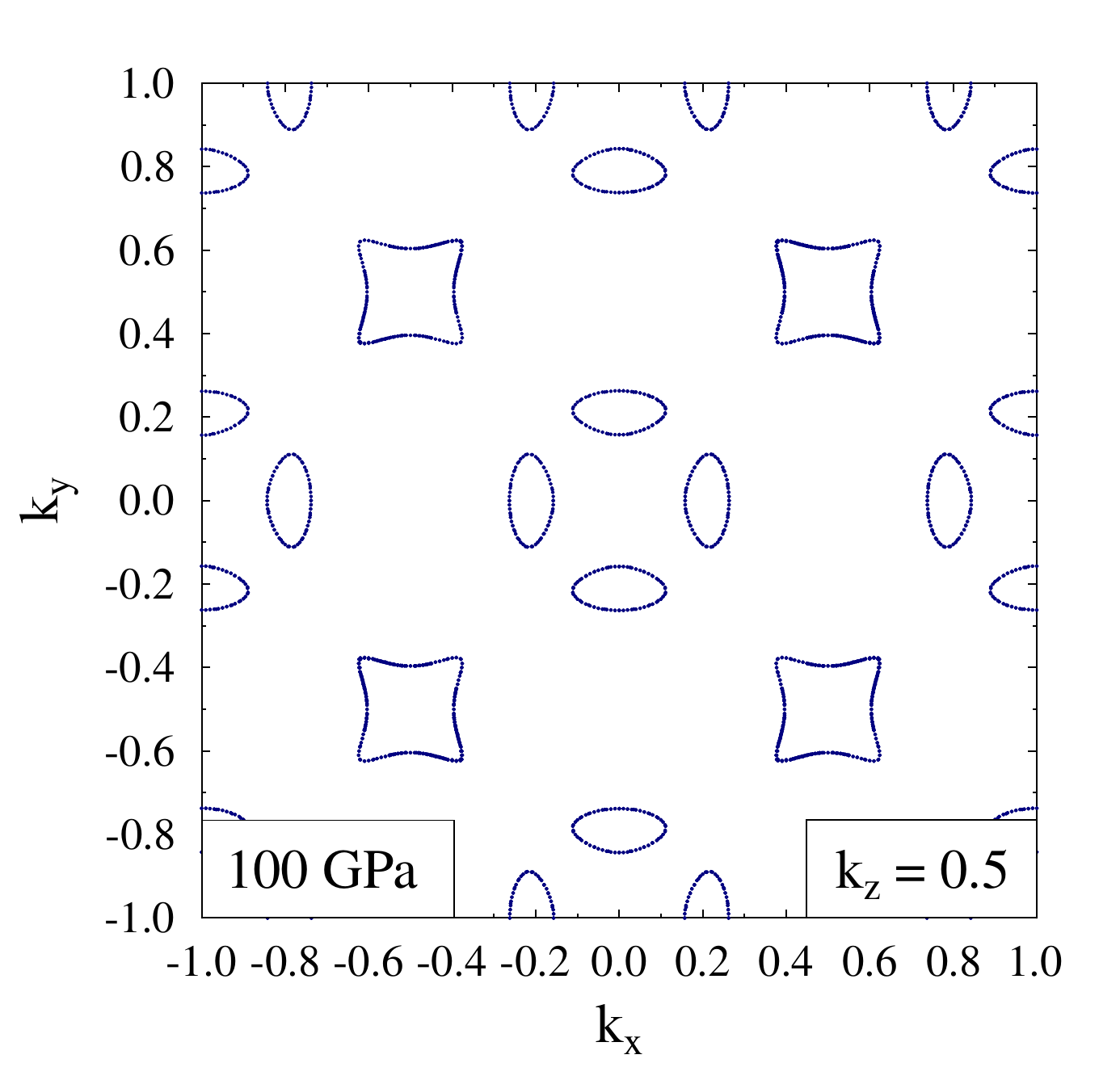}
			\end{center}	
			\caption{Cross sections of Fermi surface, calculated under pressures of 50 and 100 GPa. $k$ is given in $2\pi/a$.}
			\label{fermi_surf_50-100}
		\end{figure*}

On the whole, upon external pressure both empty and occupied electronic bands move towards $E_F$. Interesting evolution is found in the $N-\Gamma$ direction, where the local minimum of one of the bands is near $E_F$ (peak in DOS is associated with this band). As shown in Fig.~\ref{band_zoom_ng}, above 50 GPa this band comes very close to $E_F$ and its center actually crosses $E_F$ at around $\sim$70 GPa, leading to a Lifshitz transition~\cite{lifshitz_1960} (change of the Fermi surface topology). This topological transition is also visualized in Fig.~\ref{fermi_surf_50-100}, where $k_x-k_y$ cross-sections of the Fermi surface are plotted for 0, 50 and 100 GPa. The appearance of an additional band at the Fermi level is correlated with the slowing down of the decrease in $N(E_F)$ around 40 GPa, as discussed above. The fact that the band is actually blurred by the disorder leads to smearing of this transition and the band starts to contribute to DOS at lower pressures. 

What is worth noting, two transitions in topology of electronic states under pressure were reported theoretically \cite{tse_2004} for pure Nb; one slight change in the Fermi surface shape at 5-6 GPa and more prominent one around 60 GPa, connected to similar shift in electronic band in $N-\Gamma$. In an earlier experiment~\cite{struzhkin_1997}, $T_c$ of Nb was reported to show anomalies around these pressures (increase by 0.7~K and decrease by 1~K, respectively) and changes in the topology of the Fermi surfaces {were} given as an explanation for these anomalies in Ref.~\cite{tse_2004}.
However, in another theoretical work~\cite{ostanin_2000}, where relativistic full potential LMTO calculations were presented, only the second transition was observed, at around 60 GPa of hydrostatic pressure. The first anomaly in $T_c$, reported in Ref.~\cite{struzhkin_1997} was ascribed to {the presence of} non-uniform pressure conditions or polycrystalline sample effects. In our case the $T_c$ increases monotonically up to about 50 GPa and remains practically constant above that pressure. {This} trend may be correlated to the observed Lifshitz transition, which is additionally smeared by the disorder effects.

\subsection{Electron-phonon coupling}

Values of the {ambient-pressure} McMillan-Hopfield parameters are gathered in Table \ref{tab:el}. Titanium, despite highest $N(E_F)$ has the lowest contribution to the electronic part of the EPC, while the highest belongs to Zirconium. Interestingly, Zr atoms also present almost equal contribution to $\eta$ from the $p-d$ and $d-f$ scattering channels. For other constituent atoms, $d$-$f$ scattering channel gives the largest contribution to $\eta$ and it is typical for transition metal elements. To have a reference point, $\eta$ for pure Nb is about 165 mRy/a$_B^2$~\cite{kaprzyk_1996}, and calculated $\eta_i$ are slightly smaller than for Ta$_{34}$Nb$_{33}$Hf$_{8}$Zr$_{14}$Ti$_{11}$~\cite{jasiewicz_2016}. Using Eq.~\ref{eq:lambda}, the calculated $\eta_i$ and the experimental Debye temperature $\theta_D = 216$~K, we get the ambient pressure electron-phonon coupling constant 
$\lambda = 1.1$. {This value is} in close agreement with $0.97$ determined above from the renormalization of the Sommerfeld coefficient $\gamma$.

\begin{figure}[t]
\begin{center}
\includegraphics[width=0.48\textwidth]{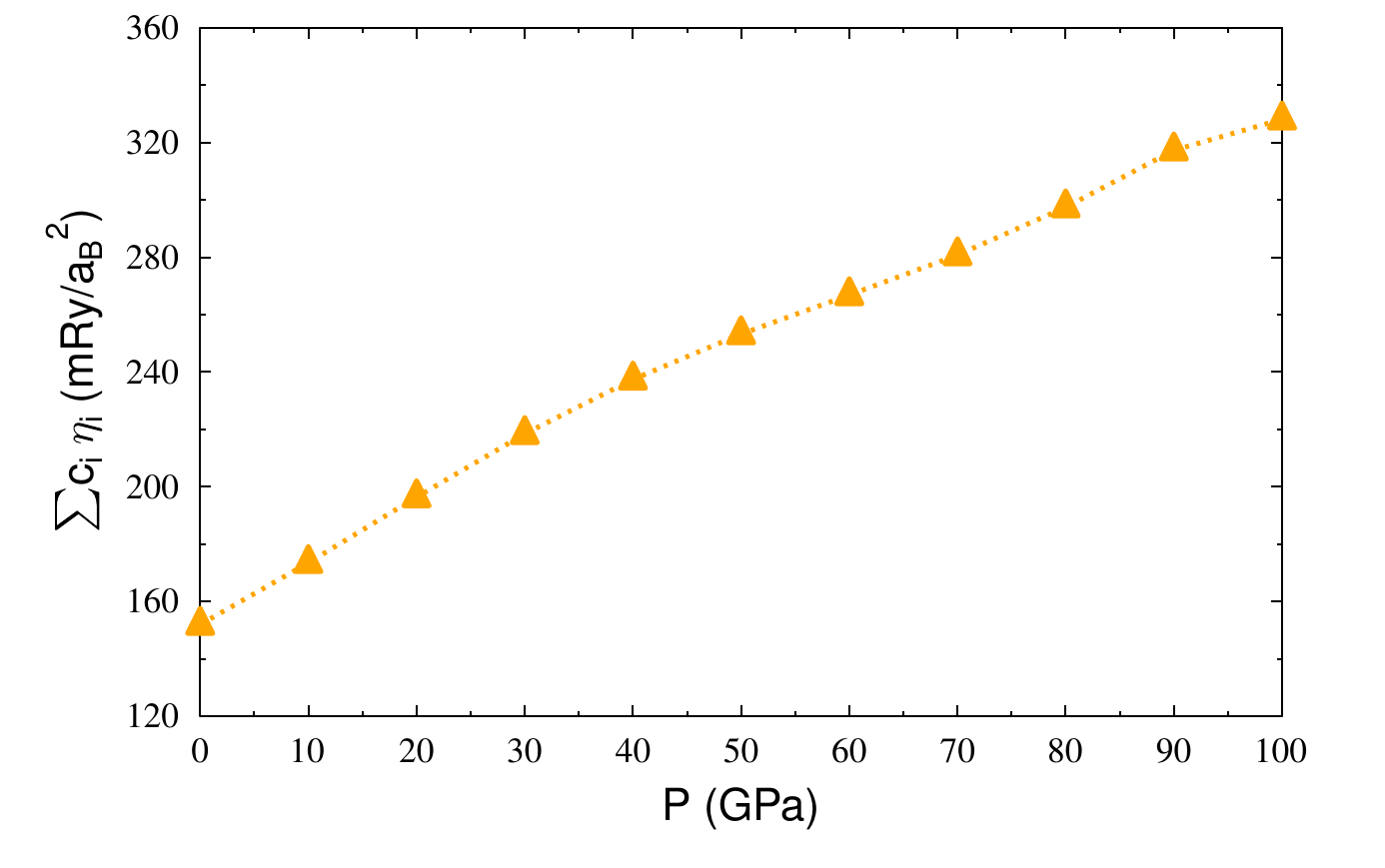}	
\includegraphics[width=0.48\textwidth]{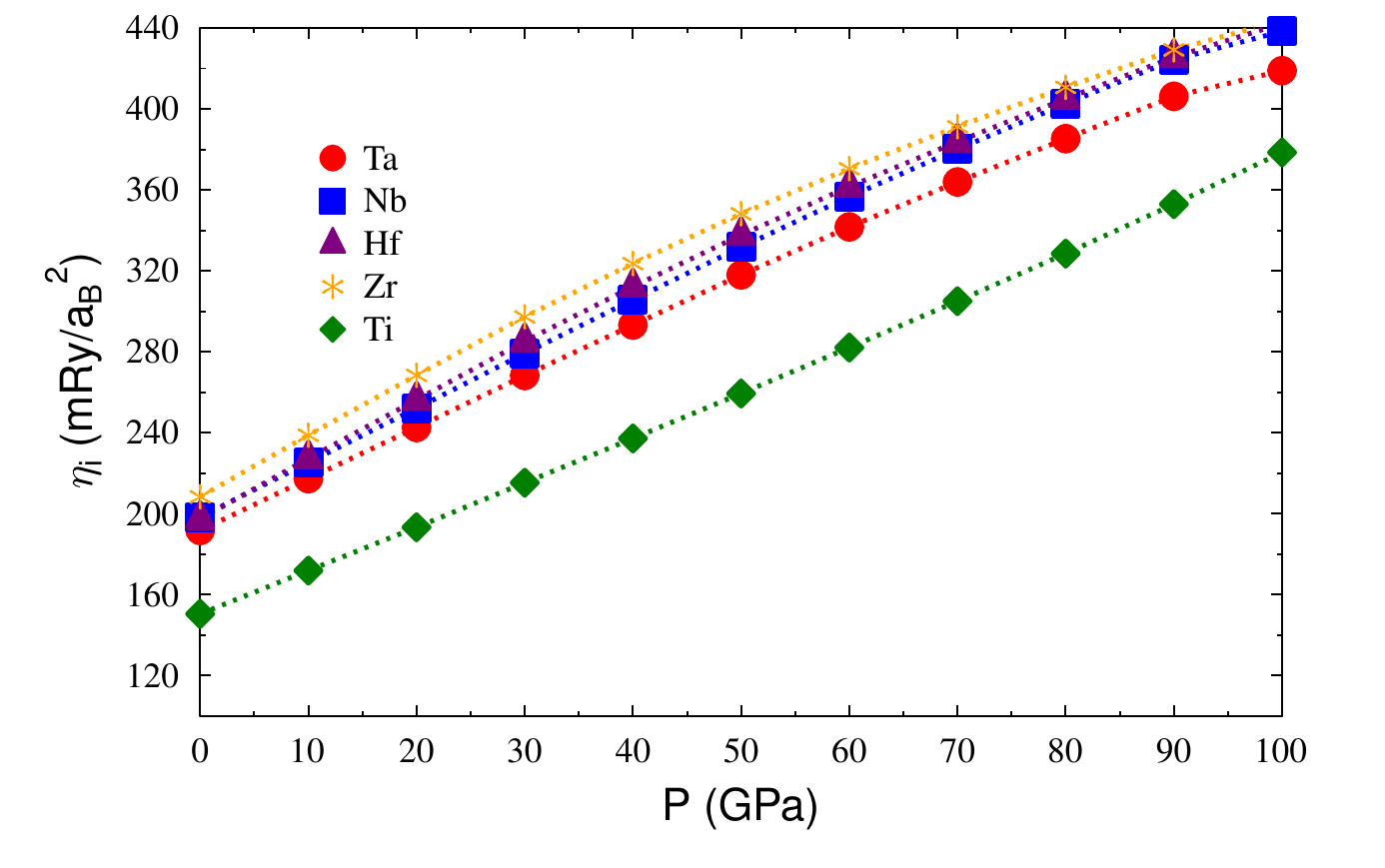}
\end{center}	
\caption{Pressure evolution of the McMillan-Hopfield parameters: concentration-weighted sum (top) and $\eta_i$ per atom (bottom).}
\label{mcmh_sc}
\end{figure}

\begin{figure*}[htb]
\begin{center}
\includegraphics[width=0.96\textwidth]{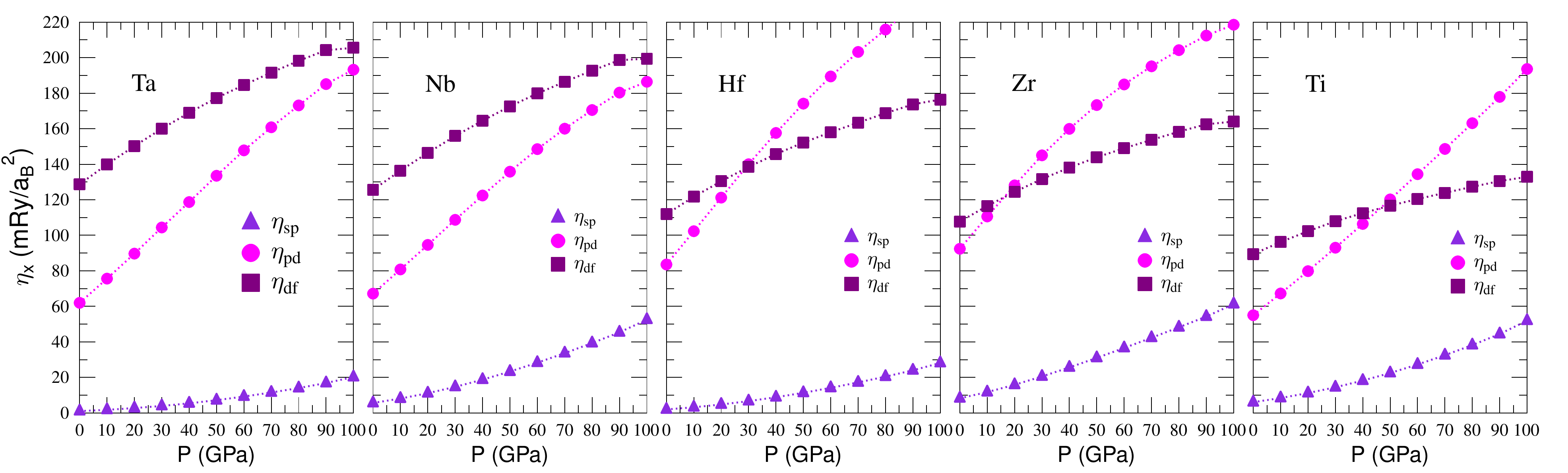}
\end{center}	
\caption{Evolution of the McMillan-Hopfield parameters of each of the atom, decomposed over the $l\rightarrow l+1$ scattering channels.}
\label{mcmh_atom_channels}
\end{figure*}

Evolution of the McMillan-Hopfield parameters with pressure is shown in Fig.~\ref{mcmh_sc}; a concentration-weighted sum in the top panel and $\eta_i$ per atom in the bottom panel. In both cases the evolution is very smooth, with gradual increase in $\eta$. What can be noticed in Fig.~\ref{mcmh_sc}(a) is the slight change {of} slope of the curve, above 40-50 GPa, which resembles the one seen in $N(E_F)$ variation in Fig.~\ref{nef_tdos}. Nevertheless, the evolution {of} $\eta$ is rather typical, as $\eta$ generally increases with pressure~\cite{wiendlocha_2006,ratti_1974,bw-heusler}. 
Less obvious is the change {of} the distribution of $\eta$ among the $s-p-d-f$ scattering channels, which is plotted in~Fig.~\ref{mcmh_atom_channels}. For the group 4 elements, i.e. Hf, Zr, Ti, a change of the dominating scattering channel to $p-d$ at high pressures is observed. Such a {behavior} is not observed for Ta and Nb atoms, {although} values of $\eta_{pd}$ and $\eta_{df}$ become close to each other. 
{The increase in $\eta_i$ is related to the increase in the matrix elements in Eq.(\ref{eq:eta}), which we additionally plotted in Supplemental Material~\cite{suppl}.}

\begin{figure}[htb]
\begin{center}
\includegraphics[width=0.48\textwidth]{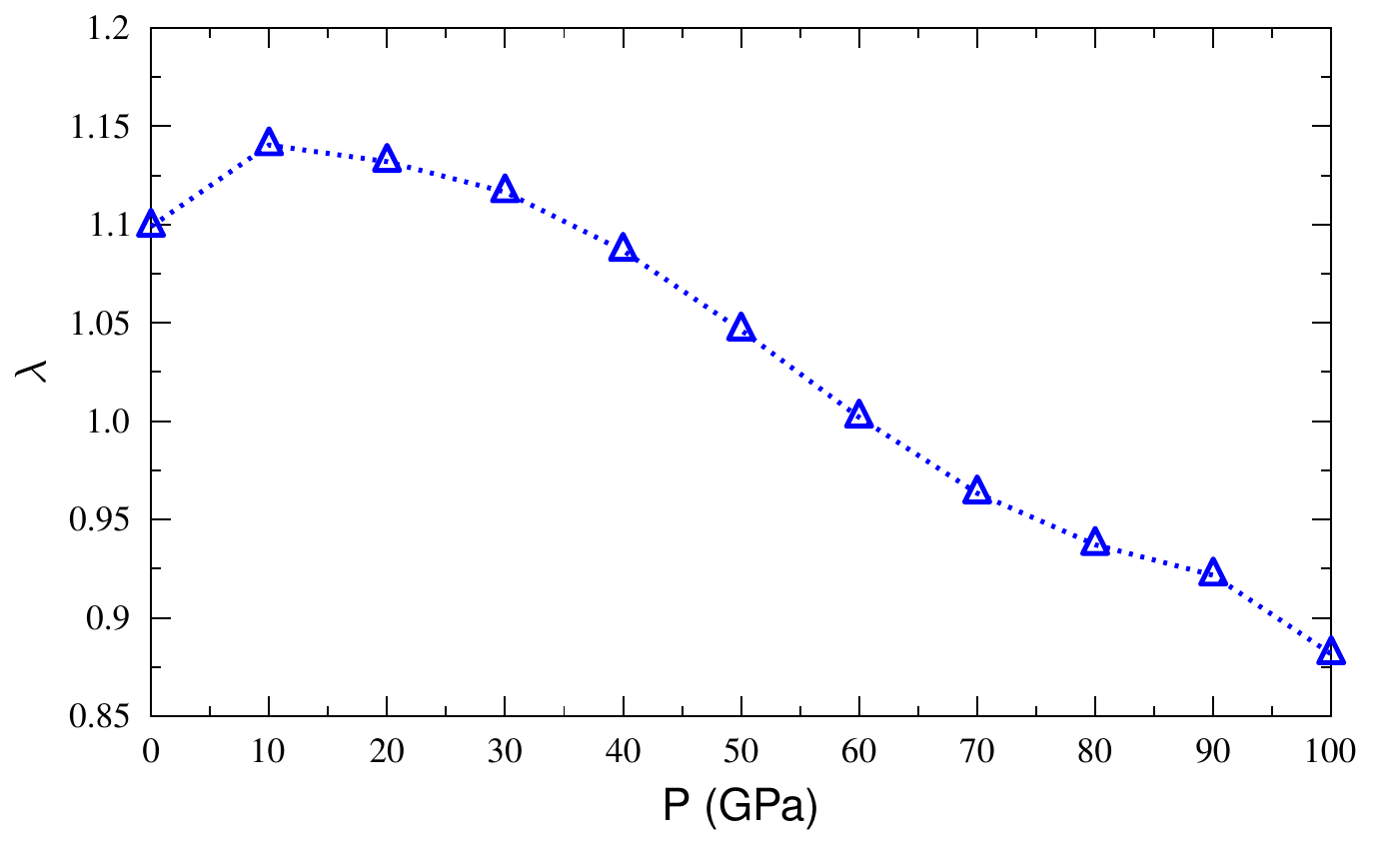}
\end{center}
\caption{\label{epc_tnhzt} Calculated evolution of the electron-phonon coupling parameter $\lambda$ with pressure. }
\end{figure}

The pressure evolution of $\lambda$, obtained {based} on computed $\eta_i(P)$ parameters (Fig.~\ref{mcmh_sc}), simulated evolution of Debye temperature $\theta_D(P)$ (Fig.~\ref{debye_tnhzt}) and Eq.\ref{eq:lambda} is shown in Fig.~\ref{epc_tnhzt}. After an initial increase, we observe a general decrease in $\lambda$. 
This is due to the fact that the evolution of the electron-phonon coupling constant $\lambda$ with pressure is the {result} of two competitive effects; an increase of the McMillan-Hopfield $\eta$ and increase of the phonon frequencies $\omega$, here represented by the Debye temperature $\theta_D$. 
Taking the derivative of $\ln \lambda$ from Eq.(\ref{eq:lambda}) we get:
\begin{equation}
\frac{d\ln\lambda}{dP} = -\frac{1}{\tilde{B}}\left(\frac{d\ln\eta}{d\ln V} + 2\gamma_G\right),
\end{equation}
where $\eta = \sum_ic_i\eta_i$ {and} $\gamma_G = -\frac{d\ln\theta_D}{d\ln V}$. {The} simplified pressure-volume equation of state $V = V_0\exp(-P/\tilde{B})$ was used to convert the pressure derivative into the volume one. The value of such-defined $\tilde{B}$ "bulk modulus" is of no importance here for the qualitative discussion. The McMillan-Hopfield parameters increse when the unit cell volume decreases, thus $\frac{d\ln\eta}{d\ln V}$ is negative
\cite{wiendlocha_2006,ratti_1974,bw-heusler} and its value is usually between -1.0 and -3.0. From the equation above, we {can} see that $\lambda(P)$ would be an increasing function of pressure for the case where $-\frac{d\ln\eta}{d\ln V} > 2\gamma_G$. In our case $2\gamma_G \simeq 3.0$ and $-\frac{d\ln\eta}{d\ln V} < 3.0$ for all pressures above 20 GPa, {and, therefore,} a decreasing $\lambda(P)$ function is expected. {This is exactly} what we can see in Fig.~\ref{epc_tnhzt}, where $\lambda$ {decreases} with pressure above 10 GPa. Only at 10 GPa, due to the strong increase of $\eta$, an increase of calculated $\lambda$ is observed, since the condition $-\frac{d\ln\eta}{d\ln V} > 2\gamma_G$ is fulfilled. At ambient conditions we have $\lambda = 1.10$. {It raises to $\lambda \simeq 1.15$ at 10 GPa and then gradually decreases for larger pressures, reaching $0.88$ at 100 GPa. }

\begin{figure}[htb]
\begin{center}
\includegraphics[width=0.48\textwidth]{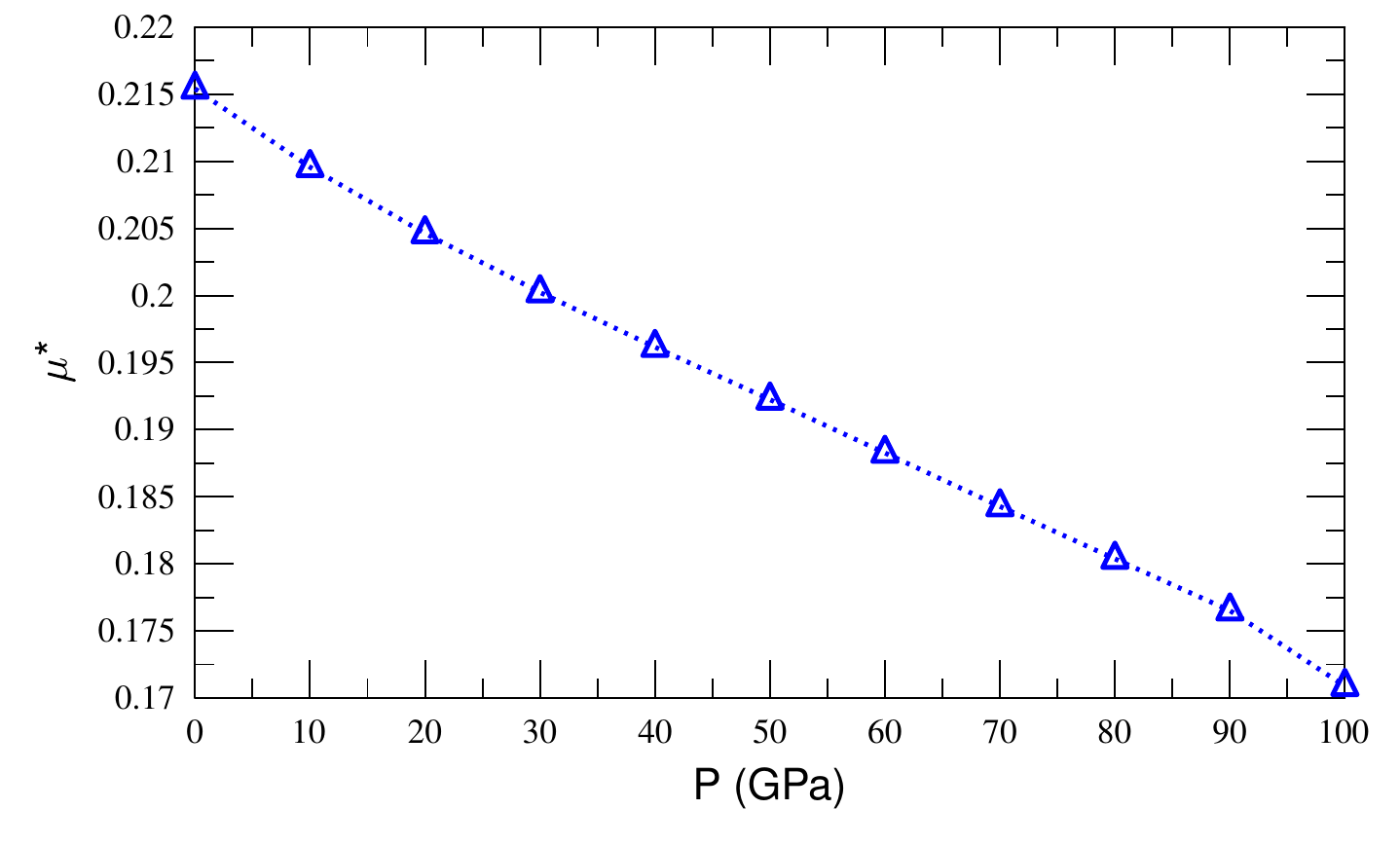}
\end{center}
\caption{\label{mu_press} Variation of the Coulomb pseudopotential parameter $\mu^*$ with pressure, calculated using Eq.(\ref{eq:garland}) and $N(E_F$) as in Fig.~\ref{nef_tdos}.}
\end{figure}

Finally, the superconducting critical temperature may be calculated using McMillan formula~\cite{mcmillan_1968}: 
\begin{equation}
T_c = \frac{\theta_D}{1.45}\exp\left[-\frac{1.04(1+\lambda)}{\lambda-\mu^*(1+0.62\lambda)}\right].
\label{eq:tc}
\end{equation}

\begin{figure}[t]
\begin{center}
\includegraphics[width=0.48\textwidth]{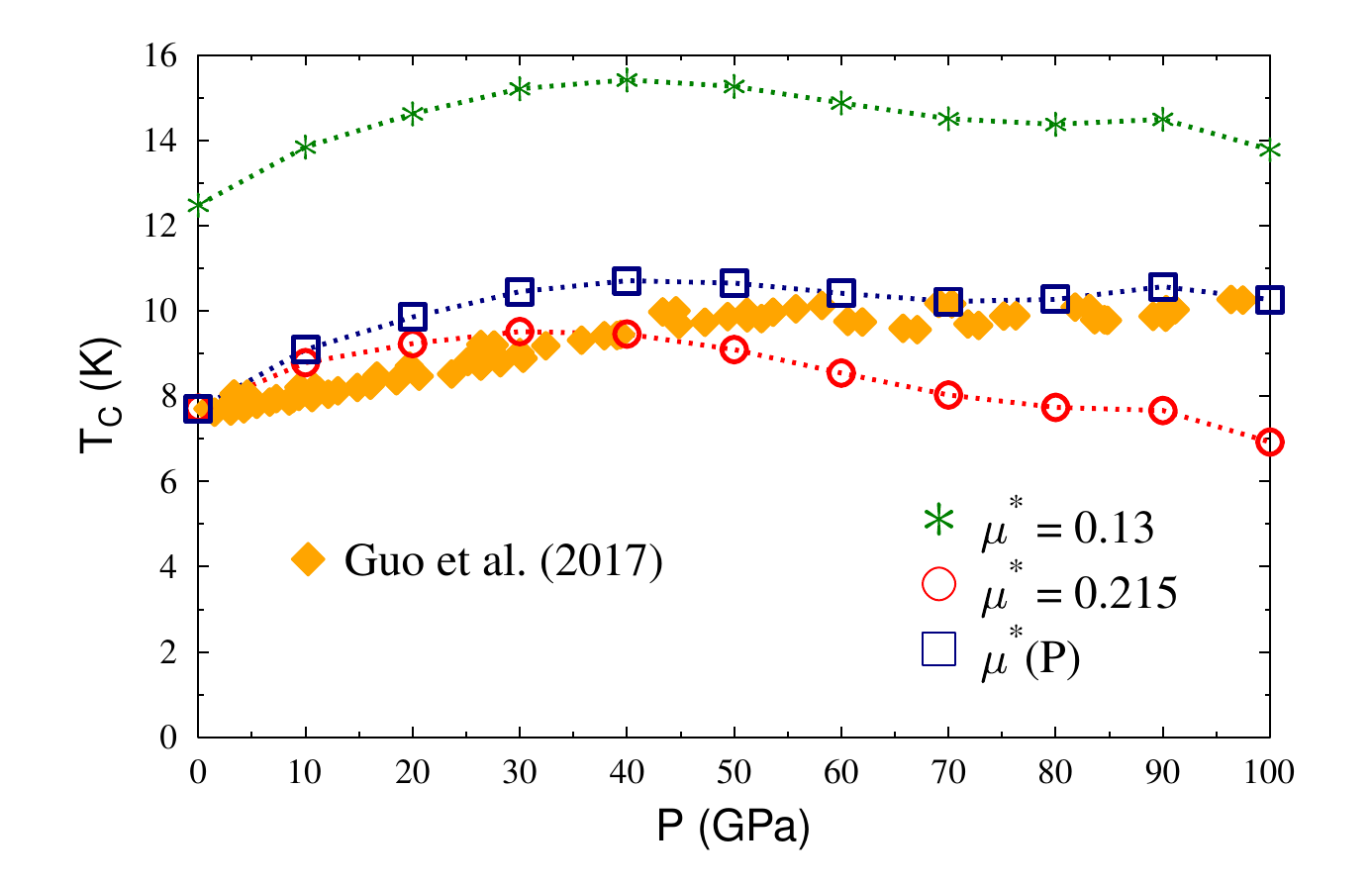}
\end{center}
\caption{\label{tc_tnhzt}Calculated pressure dependence of $T_c$ of (TaNb)$_{0.67}$(HfZrTi)$_{0.33}$ for two constant {values of $\mu^* = 0.13$ and $\mu^* = 0.215$}, 
variable $\mu^*(P)$ (see, Fig~\ref{mu_press}) and compared to experiment~\cite{guo-pressure}.}
\end{figure}

The last parameter, which has to be determined is the Coulomb pseudopotential parameter, $\mu^*$. In zero pressure, the most-commonly used value of 0.13 would lead to an overestimated $T_c = 12.5$~K. To get the experimental zero pressure $T_c = 7.7$~K one has to {use} $\mu^* = 0.215$. Similar value was used for pure Nb to reproduce the experimental critical temperature {based on the calculated} Eliashberg function~\cite{savrasov_1996, ostanin_2000}. Even larger values of $\mu^*$ were postulated for other materials {such as} Nb$_3$Ge ($\mu^* = 0.24$) \cite{carbotte_1990}, V ($\mu^* = 0.3$),\cite{savrasov_1996} or MgCNi$_3$ ($\mu^* = 0.29$)\cite{szczesniak_2015}. 
Thus, to explore the variation of $T_c$ with pressure we assume $\mu^*(0) = 0.215$. For $P > 0$, $T_c(P)$ was calculated in two ways. First, $\mu^* = 0.215$ was kept constant in the whole pressure range. Next, $\mu^*(P)$ dependence was assumed to originate from the pressure dependence of $N(E_F)$ and calculated using the Benneman and Garland formula~\cite{garland_1972}:
\begin{equation}\label{eq:garland}
\mu^*=\frac{AN(E_F)}{1+N(E_F)},
\end{equation}
where $N(E_F)$ is in eV$^{-1}$ per atom.
Originally,  Benneman and Garland set $A = 0.26$ to get $\mu^* = 0.13$ for the typical case of a metal with $N(E_F) = 1$ eV$^{-1}$ per atom. {Therefore}, in our case where $N(E_F) = 1.65$ eV$^{-1}$ for $P = 0$ and postulated $\mu^*(0) = 0.215$ we use $A = 0.345$, and simulate $\mu^*(P)$ dependence according to $N(E_F)$ variation with pressure (Fig.~\ref{nef_tdos}) by using Eq.(\ref{eq:garland}). Fig.~\ref{mu_press} shows the $\mu^*(P)$ dependence {that} decreases smoothly with pressure and drops to 0.17 at 100 GPa. Finally, Fig. \ref{tc_tnhzt} shows the computed critical temperature $T_c(P)$, where {$T_c$ for the "standard" $\mu^* = 0.13$ as also included}.   

In general, under the assumption of relatively large $\mu^*(0) = 0.215$, our calculations quite well predict the variation of $T_c$ with pressure (but only when variable $\mu^*(P)$ is used). In spite of the decrease in the computed $\lambda$ above 10 GPa, $T_c$ increases up to 40 - 50 GPa and then remains almost constant up to 100 GPa, just like {it is} observed in the experiment. This counter-intuitive observation shows the delicate balance between $T_c$, $\theta_D$, and $\lambda$, since an increase in $\theta_D$ leads to a quadratic increase in the denominator of Eq.(\ref{eq:lambda}) (tendency to decrease $\lambda$) and linear increase {of} $T_c$ via the multiplicator in McMillan's Eq.(\ref{eq:tc}). In stabilization of $T_c$ above 40 GPa, the decrease of $\mu^*$, which results from the decrease in $N(E_F)$ occurs to be equally important, since for the constant $\mu^*$ decrease in $T_c$ is predicted {by the theoretical} calculations. This shows that up to a studied pressure of 100 GPa, the evolution of $T_c$ with pressure in (TaNb)$_{0.67}$(HfZrTi)$_{0.33}$ can be explained by the classical electron-phonon mechanism. {This is} surprisingly well captured by a combination of coherent potential approximation, rigid muffin tin approximation, and ''averaged'' phonon spectrum. Thus, structural local short-range ordering effects or local distortions of the crystal structure {that} are likely present in the studied samples, seem not to have a large impact on superconductivity. This may be understood as the superconducting coherence length {being} typically much larger than the structural anomalies' length scale. {Based} on the upper critical field data from Ref.~\cite{tnhzt2} ($\mu_0H_{c2}$ = 7.75 T) the superconducting coherence length may be estimated as 65~\AA. {This is} is roughly 20 times the lattice parameter of the system. On this length scale the possible local crystal structure distortions or chemical inhomogeneities are averaged out, and {therefore} an effective medium theory {that} we apply here, works well.

\section{Summary}

In summary, we have studied pressure effects of the electronic structure, electron-phonon interaction, and superconductivity of the high entropy alloy (TaNb)$_{0.67}$(HfZrTi)$_{0.33}$ in a pressure range from 0 to 100 GPa. With increasing pressure the total density of states at the Fermi level $N(E_F)$ gradually decreases. Lifshitz transition is observed around 70 GPa when one of the bands starts crossing the Fermi level. Due to disorder-induced band smearing effects, however, the transition is not sharp, since {these} bands contribute to $N(E_F)$ also at lower pressures {(even below 50 GPa).} 
As in the experimental studies, $T_c(P)$ changes the slope above 50 GPa {and} this effect may be correlated with the calculated band structure evolution and the Lifshitz transition. The effects of pressure on the lattice dynamics were simulated using the Debye-Gr\"{u}neisen model, where $\gamma_G$ parameter was additionally determined. The calculated McMillan-Hopfield parameters increase with pressure but due to concurrent effect of the lattice stiffening and increase of the Debye temperature, the electron-phonon coupling parameter $\lambda$ decreases above 10 GPa. In spite of this, the calculated superconducting $T_c$ increases up to 40 - 50 GPa and later is stabilized at the larger value {of $\lambda$} than {observed} at the ambient conditions. This non-decreasing $T_c$ results from the increase of the Debye temperature and decrease of $N(E_F)$, which {is caused by} the monotonic decrease of the Coulomb pseudopotential parameter $\mu^*$. Our results are in good agreement with the experimental trend {and} shows that up to a studied pressure of 100 GPa the evolution of $T_c$ with pressure in (TaNb)$_{0.67}$(HfZrTi)$_{0.33}$ can be well explained by the classical electron-phonon mechanism. {This implies that} the electronic structure of the system is well described by the coherent potential approximation. An excellent additional test of our theoretical results would be a measurement of the electronic heat capacity under pressure, which would allow {verification of} the observed decrease in both $N(E_F)$ and $\lambda$.

\section{Acknowledgments}
KJ and BW were supported by the National Science Center (Poland), grant No. 2017/26/E/ST3/00119. K.~Gofryk acknowledges support from DOE's Early Career Research Program. JT was supported by the AGH-UST statutory tasks No. 11.11.220.01/5 within subsidy of the Ministry of Science and Higher Education. 
\bibliography{hea-biblio}
\begin{widetext}
\section{Supplemental Material for\\ Pressure effects on the electronic structure and superconductivity of (TaNb)$_{0.67}$(HfZrTi)$_{0.33}$ high entropy alloy}

\section{Phonon frequency moments}	

The electron-phonon coupling constant $\lambda$ is defined as:
\begin{equation}\label{eq:lam2}
\lambda=2\int_0^{\omega_{\rm max}} \frac{\alpha^2F(\omega)}{\omega} \text{d}\omega,
\end{equation}
where $\alpha^2F(\omega)$ is the Eliashberg electron-phonon interaction function. 
Defining the McMillan-Hopfield parameter as
\begin{equation}
\eta = 2M\int_0^{\omega_{\rm max}} \omega \alpha^2F(\omega) \text{d}\omega,
\end{equation}
and the "second moment" of Eliashberg function as:
\begin{equation}\label{mom2}
\langle \omega^2 \rangle = \int_0^{\omega_{\mathsf{max}}} \omega \alpha^2F(\omega) d\omega \left/ \int_0^{\omega_{\mathsf{max}}} \alpha^2F(\omega) \frac{d\omega}{{\omega}} \right.,
\end{equation}
we arrive at the identity:
\begin{equation}\label{eq:lam3}
\lambda=\frac{\eta}{M\langle \omega^2 \rangle}.
\end{equation}
Note, that $\langle \omega^2 \rangle$ is the second moment of $\frac{\alpha^2F(\omega)}{\omega}$ function, not the usual average square of $\alpha^2F(\omega)$.
The advantage of writing $\lambda$ as in Eq. (\ref{eq:lam3}) is twofold. First, (see, Ref. 58 in the main text), as 
$\alpha^2F(\omega) \propto 1/\omega$, the McMillan-Hopfield parameter $\eta$ is independent of the phonon frequencies, and in the rigid muffin tin approximation may be computed directly from the electronic band structure of the system (Eq. (2) in the main text).
Second, as the $\langle \omega^2 \rangle$ parameter is the ratio of two integrals, for the system, where the electron-phonon interaction weakly depends on the phonon frequency (e.g. in niobium), it may be estimated using the "bare" phonon density of states $F(\omega)$ function. In such a case we may write $\alpha^2F(\omega) = \alpha^2\times F(\omega)$, and 
\begin{equation}\label{mom3}
\langle \omega^2 \rangle = \int_0^{\omega_{\mathsf{max}}} \omega \alpha^2F(\omega) d\omega \left/ \int_0^{\omega_{\mathsf{max}}} \alpha^2F(\omega) \frac{d\omega}{{\omega}} \right. = \int_0^{\omega_{\mathsf{max}}} \omega F(\omega) d\omega \left/ \int_0^{\omega_{\mathsf{max}}} F(\omega) \frac{d\omega}{{\omega}} \right.,
\end{equation}
since $\alpha^2$ term cancels out. 
Thus, in RMTA one may effectively decouple calculations of $\lambda$ into purely electronic and phonon parts. Moreover, if $F(\omega)$ is difficult to compute directly, as in our case of the high entropy alloy, and one assumes, that it follows the Debye model $F(\omega) = \frac{9\omega^2}{\omega_D^3}$, one has
\begin{equation}
\langle \omega^2 \rangle = \int_0^{\omega_D} \omega \frac{9\omega^2}{\omega_D^3} d\omega \left/ \int_0^{\omega_D} \frac{9\omega^2}{\omega_D^3} \frac{d\omega}{{\omega}} \right. = \frac{\omega_D^2}{2}.
\end{equation}
Thus, the experimental Debye temperature $\theta_D$ ($k_B\theta_D = \hbar \omega_D$) may be used to estimate $\langle \omega^2 \rangle$.
On the other hand, in contrast to Eq.(\ref{mom2}), the Debye temperature may be calculated from the phonon DOS function $F(\omega)$ basing on the usual definition of the $m-$th moment of the phonon DOS (see, Ref. 58 in the main text)
\begin{eqnarray}
\mu_m = \int_0^{\omega_{\mathsf{max}}} \omega^{m} F(\omega) d\omega \left/ \int_0^{\omega_{\mathsf{max}}} F(\omega) {d\omega}\right.\label{eq:debcalcs}\\
\omega_D(m) = \left(\frac{m+3}{3}\mu_m\right)^{1/m}.\label{debcalc2}
\end{eqnarray}
For the Debye spectrum, Eq.(\ref{debcalc2}) always gives $\omega_D$. In our benchmark calculations for Nb and Ta, we took $m=2$, as it corresponds to the ''heat capacity Debye temperature'' (see, Ref. 58 in the main text). 

\newpage
\section{Phonon DOS for Nb and Ta under pressure}	
In the figures below we show the computed phonon DOS functions $F(\omega)$ for Nb and Ta under pressure, used for testing of the Debye temperature $\theta_D(p)$ calculations within the Gr\"uneisen model. In each figure, lattice parameter (in atomic $a_B = 0.529$~\AA~units), Debye temperature and pressure (in kbar) are shown.
\begin{figure}[h!]
\begin{center}
\includegraphics[width=0.31\textwidth]{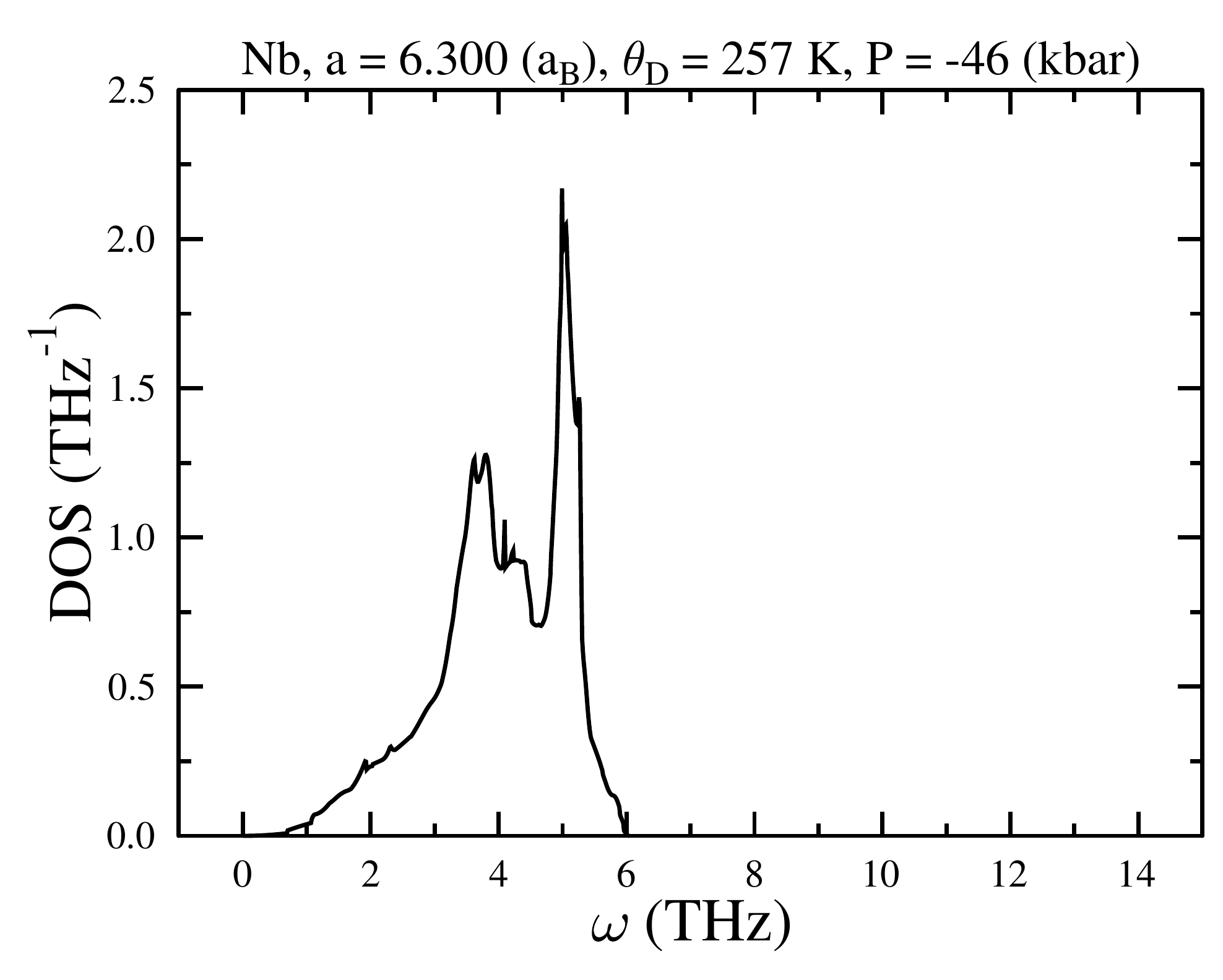}
\includegraphics[width=0.31\textwidth]{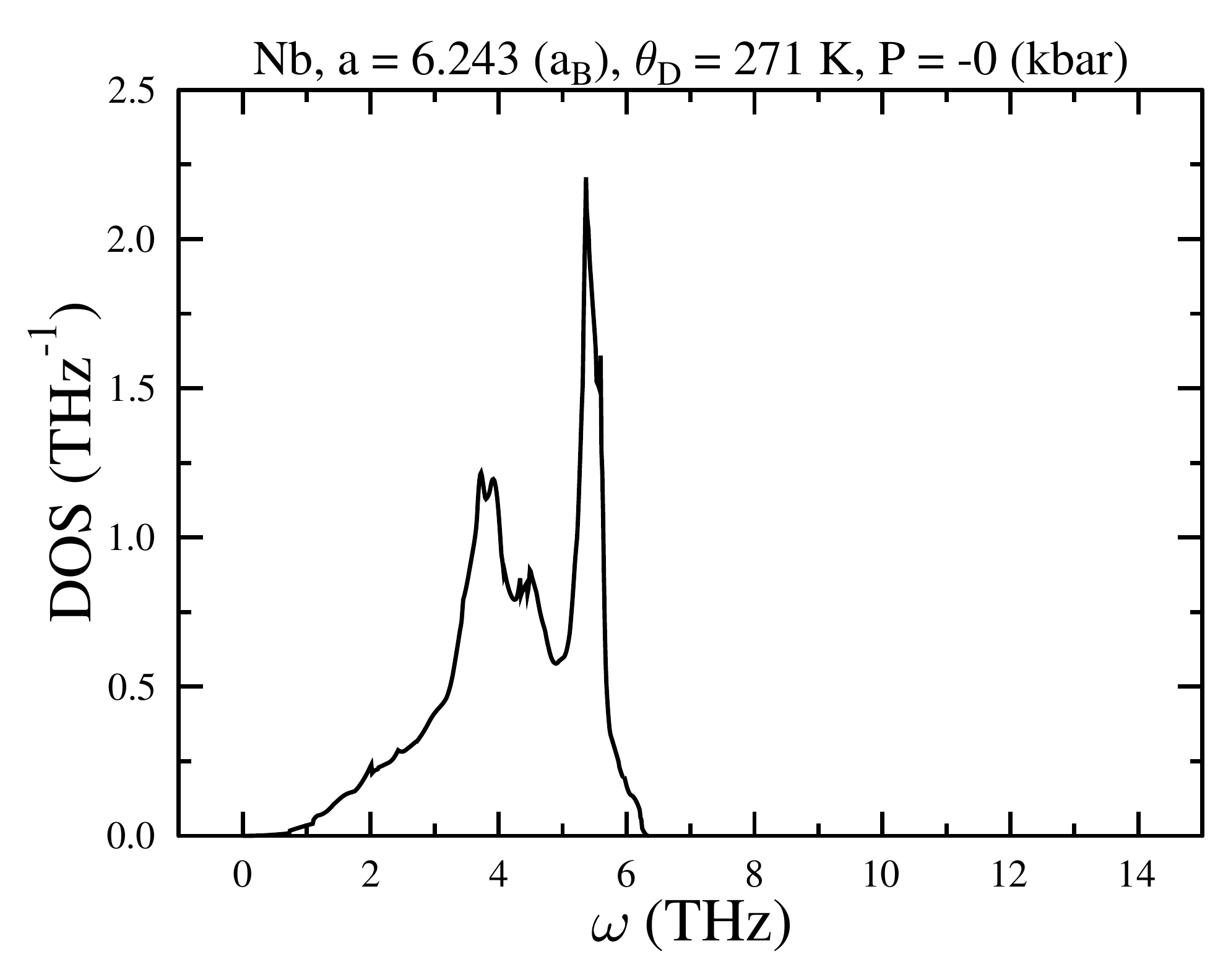}
\includegraphics[width=0.31\textwidth]{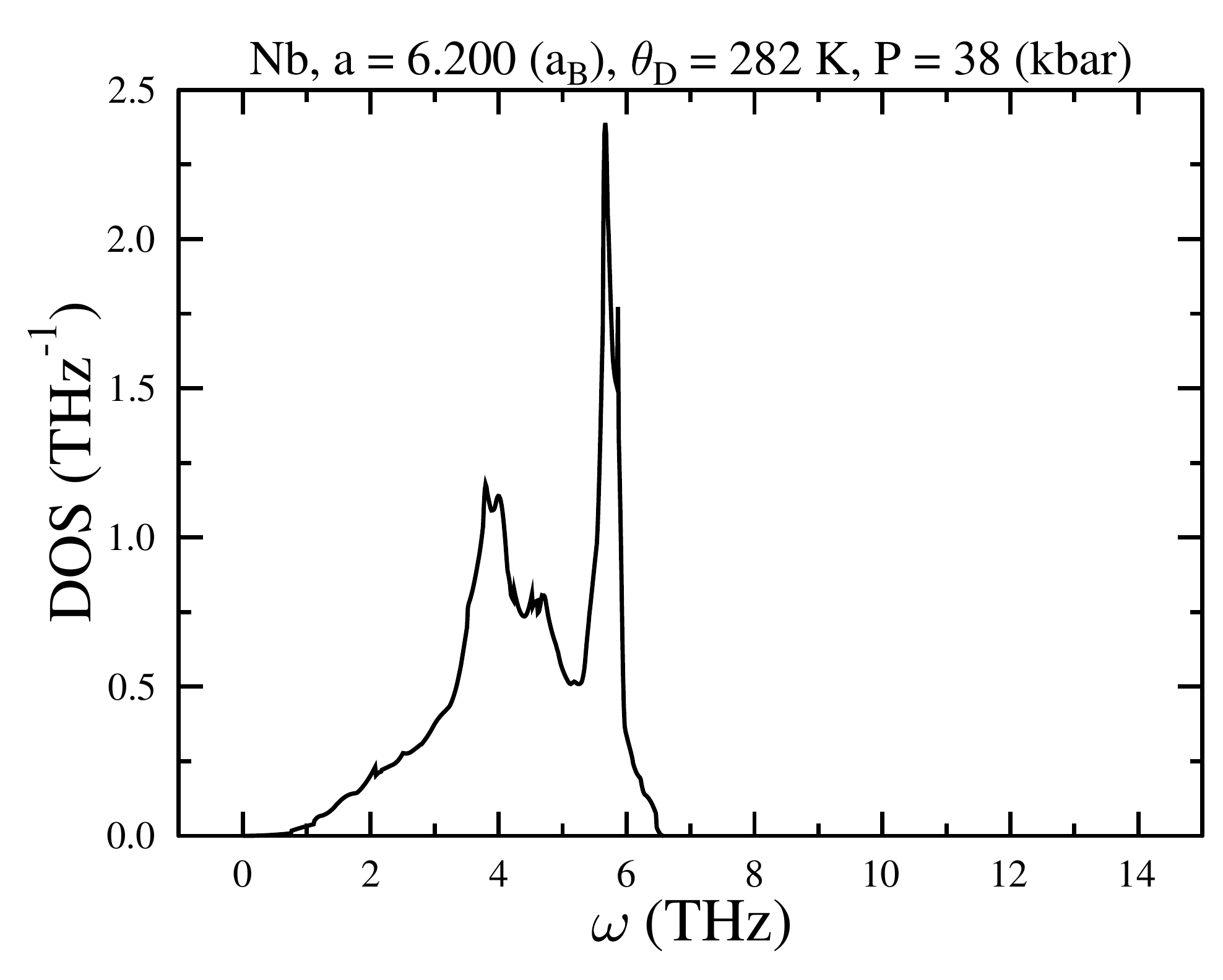}\\
\includegraphics[width=0.31\textwidth]{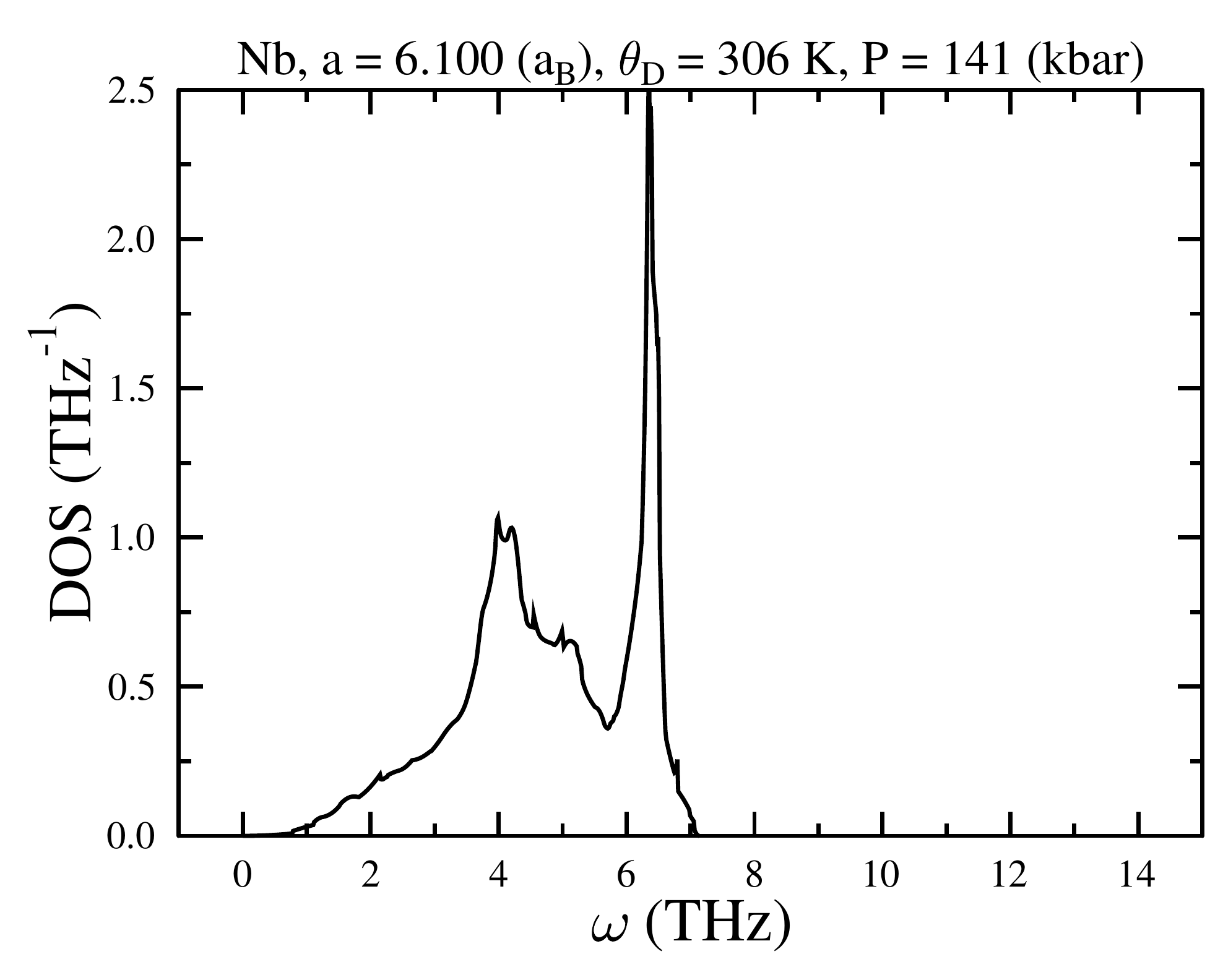}
\includegraphics[width=0.31\textwidth]{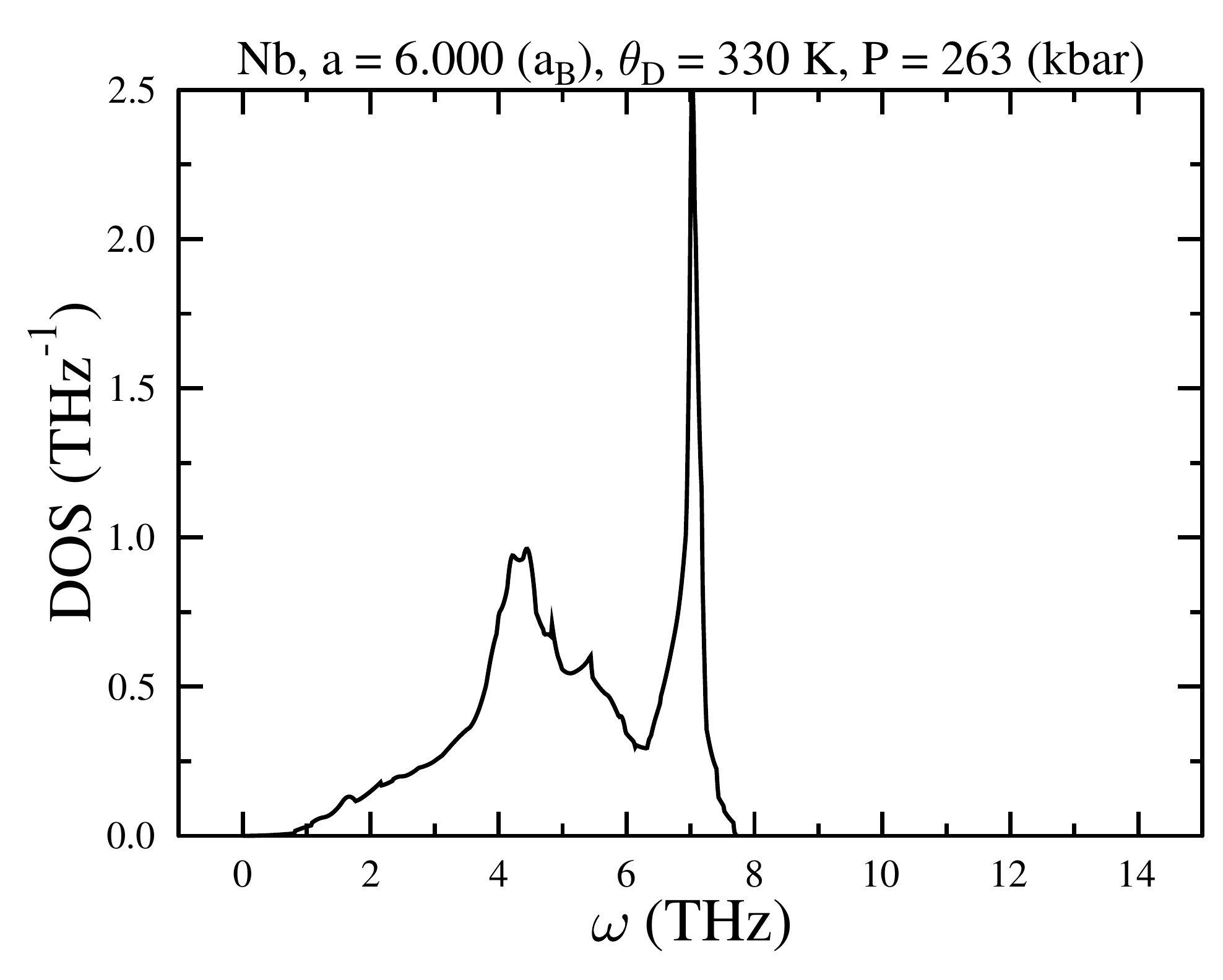}
\includegraphics[width=0.31\textwidth]{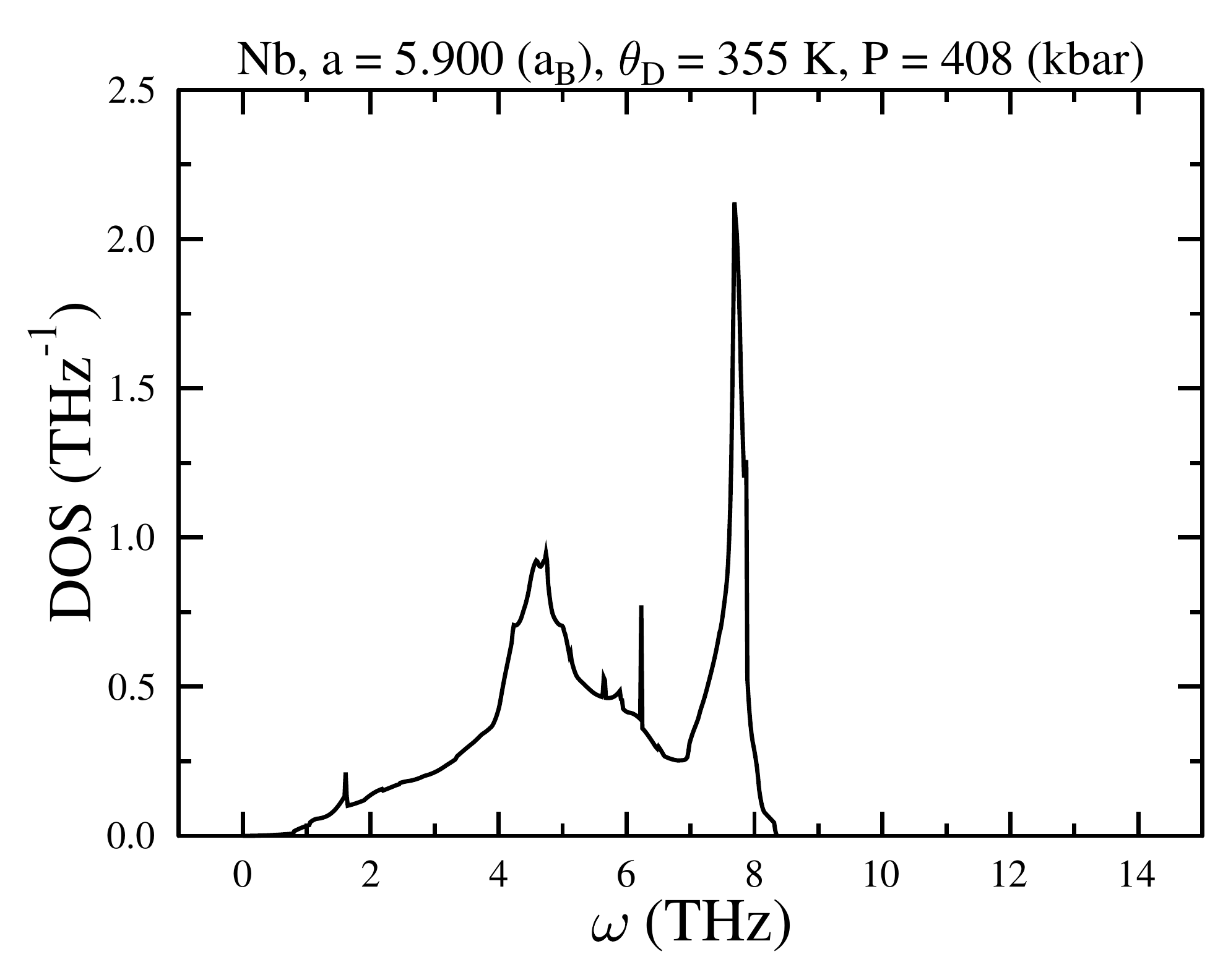}\\
\includegraphics[width=0.31\textwidth]{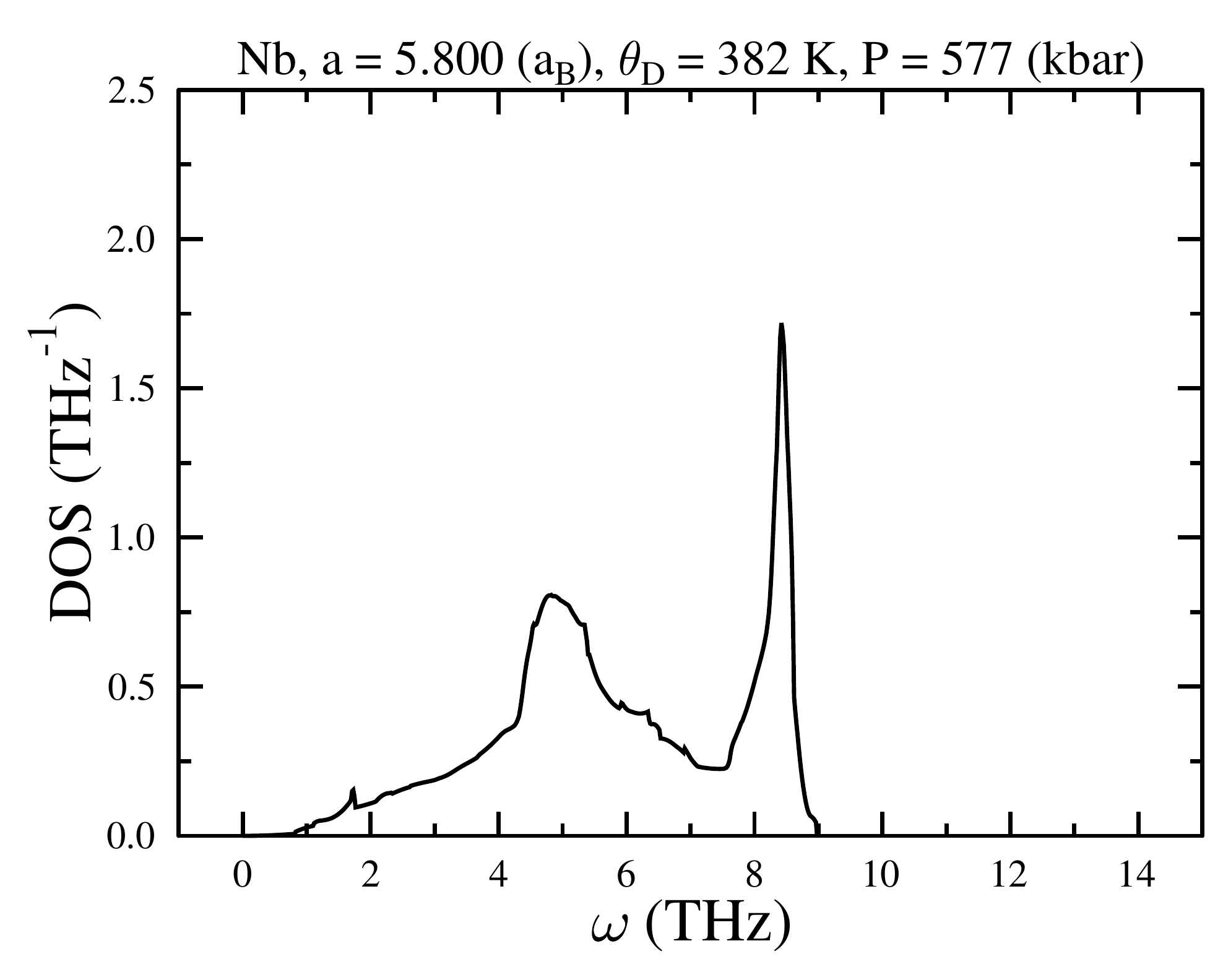}
\includegraphics[width=0.31\textwidth]{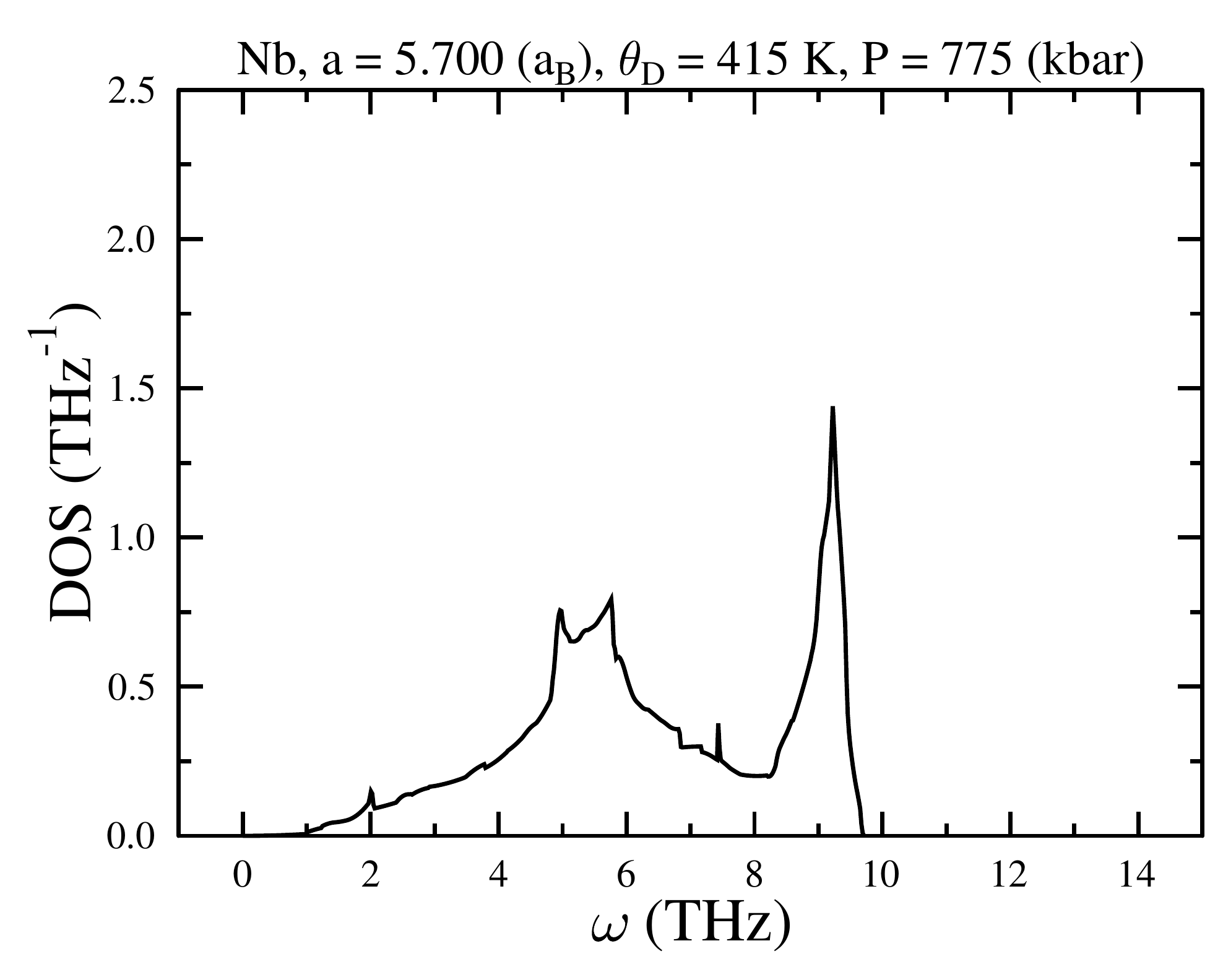}
\includegraphics[width=0.31\textwidth]{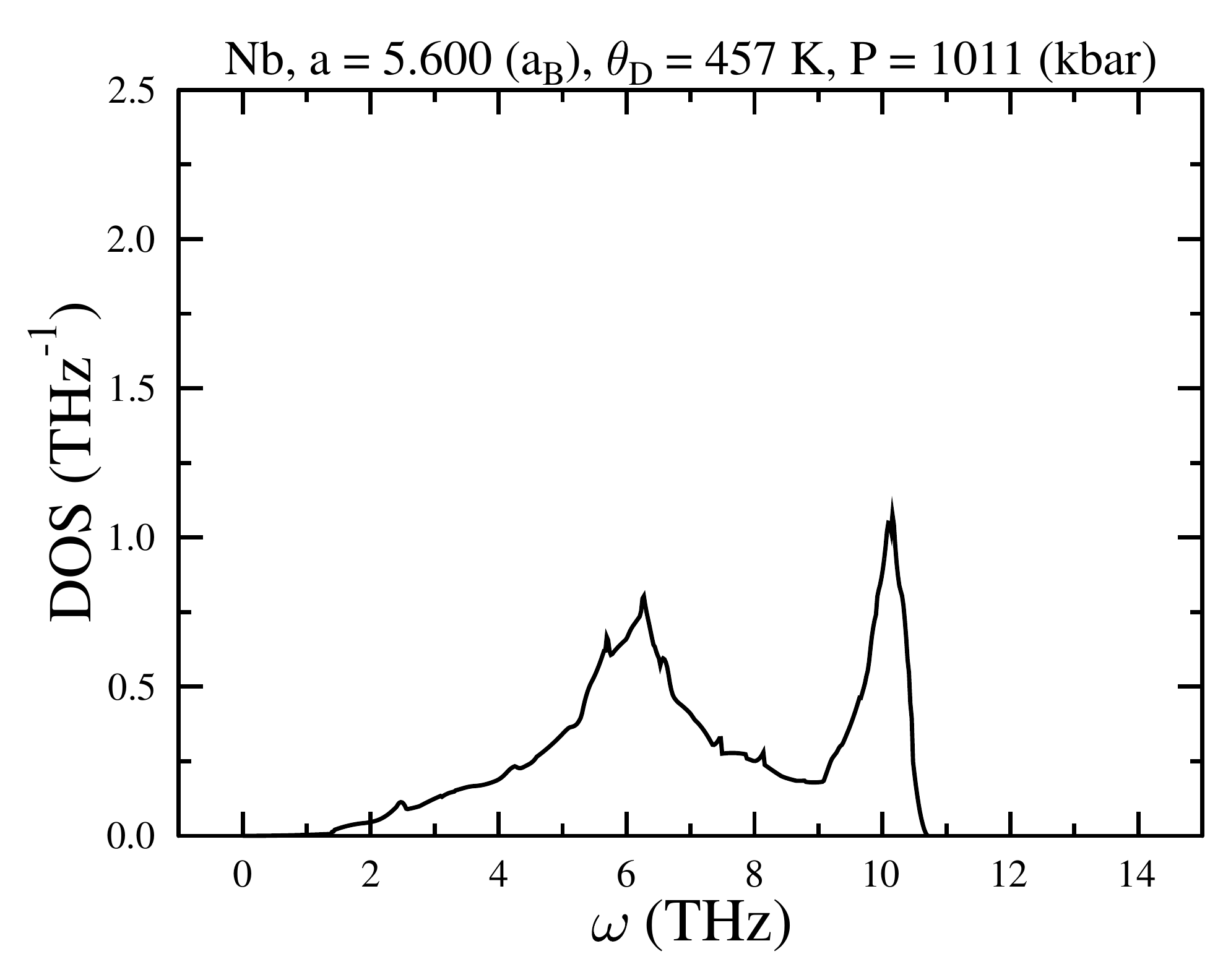}
\end{center}
\caption{Phonon DOS for Nb}
\label{nb}
\end{figure}

\begin{figure}[h!]
\begin{center}
\includegraphics[width=0.31\textwidth]{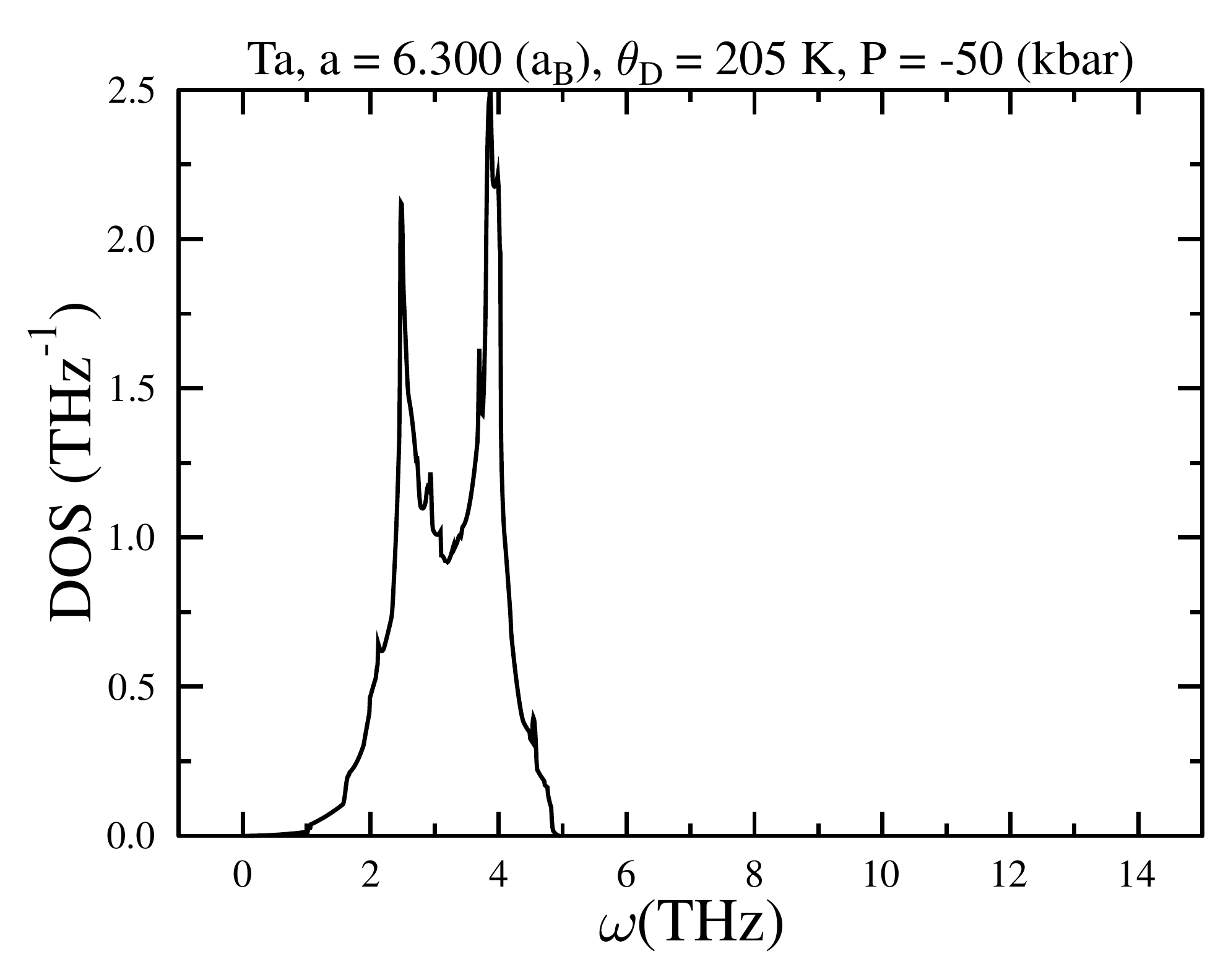}
\includegraphics[width=0.31\textwidth]{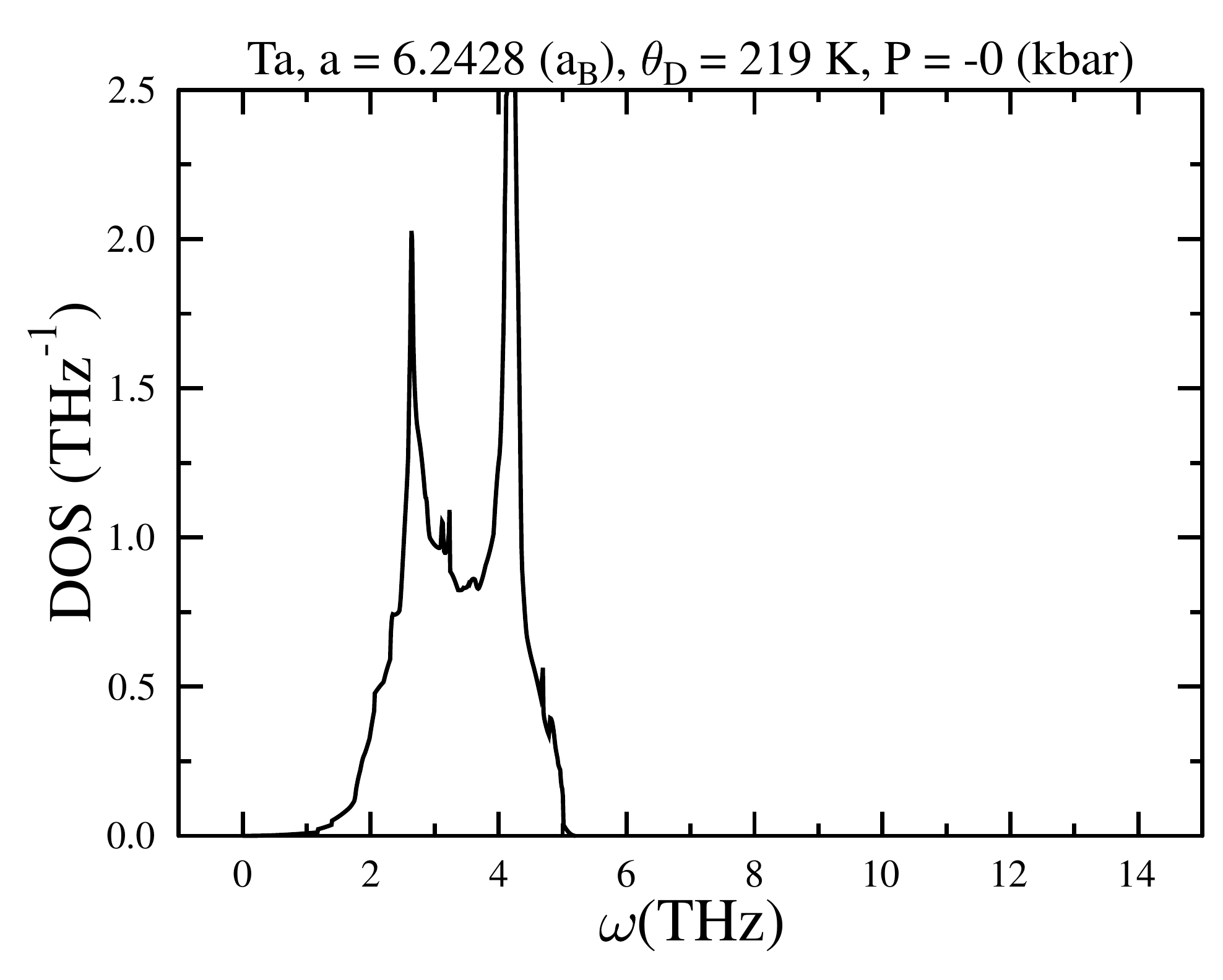}
\includegraphics[width=0.31\textwidth]{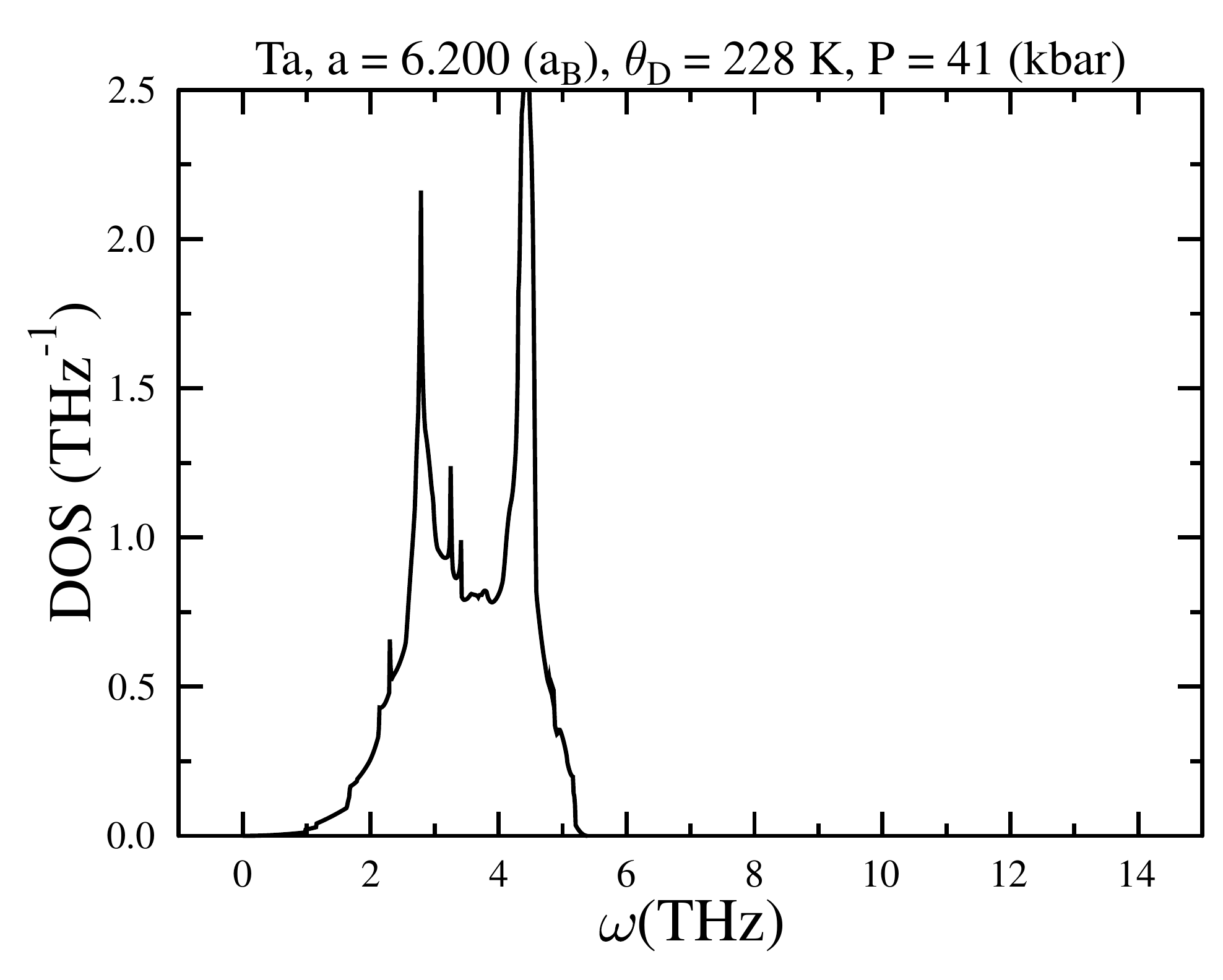}\\
\includegraphics[width=0.31\textwidth]{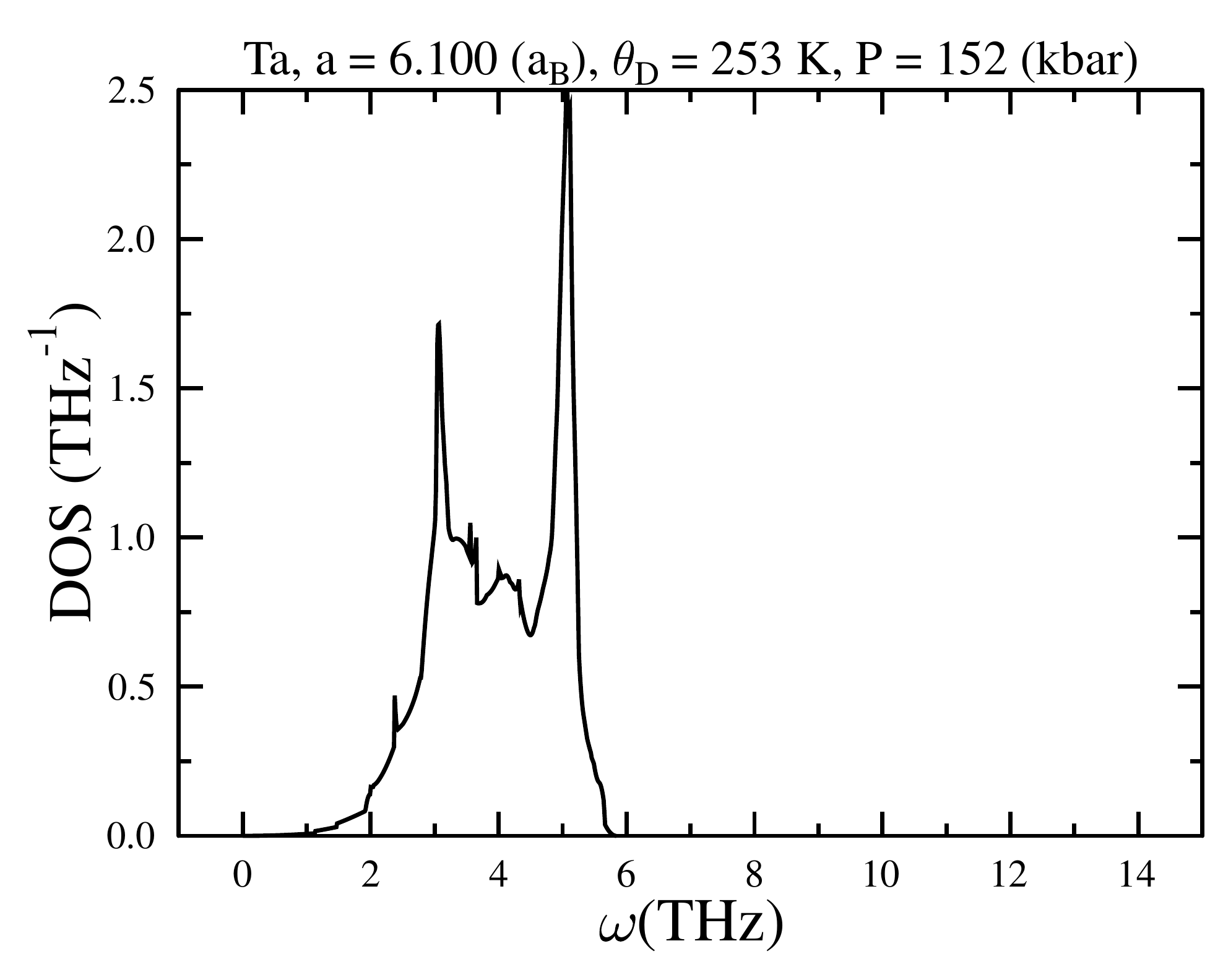}
\includegraphics[width=0.31\textwidth]{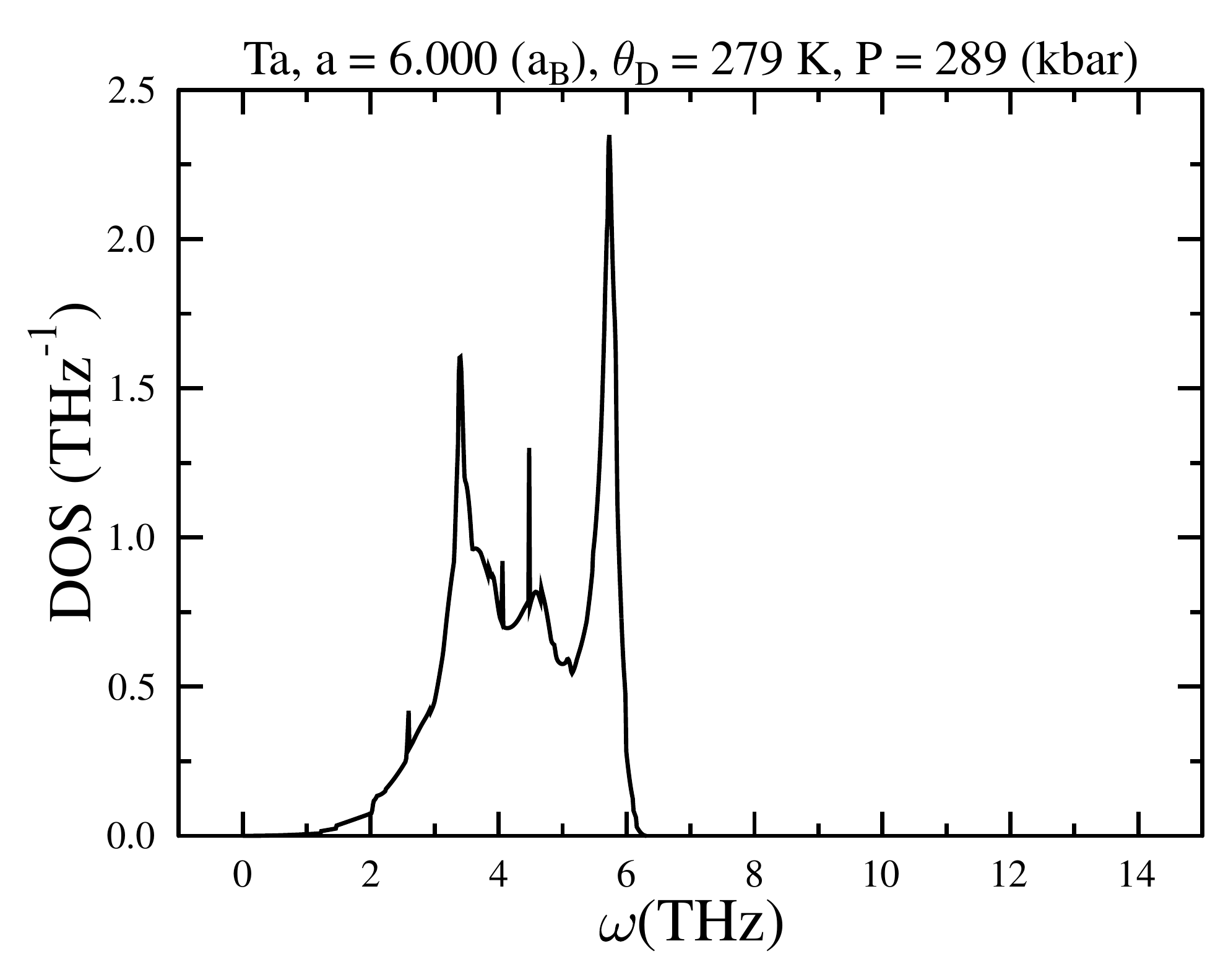}
\includegraphics[width=0.31\textwidth]{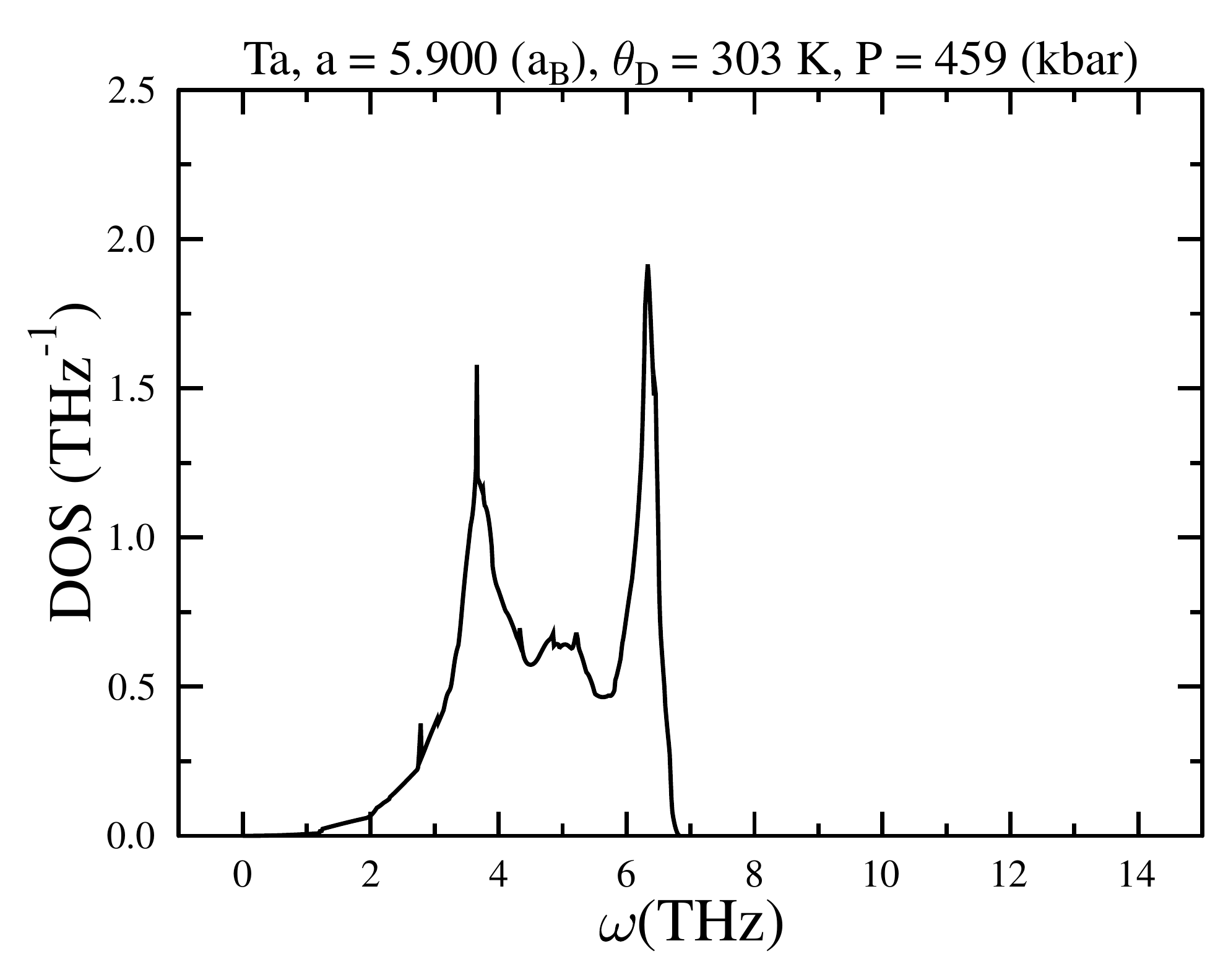}\\
\includegraphics[width=0.31\textwidth]{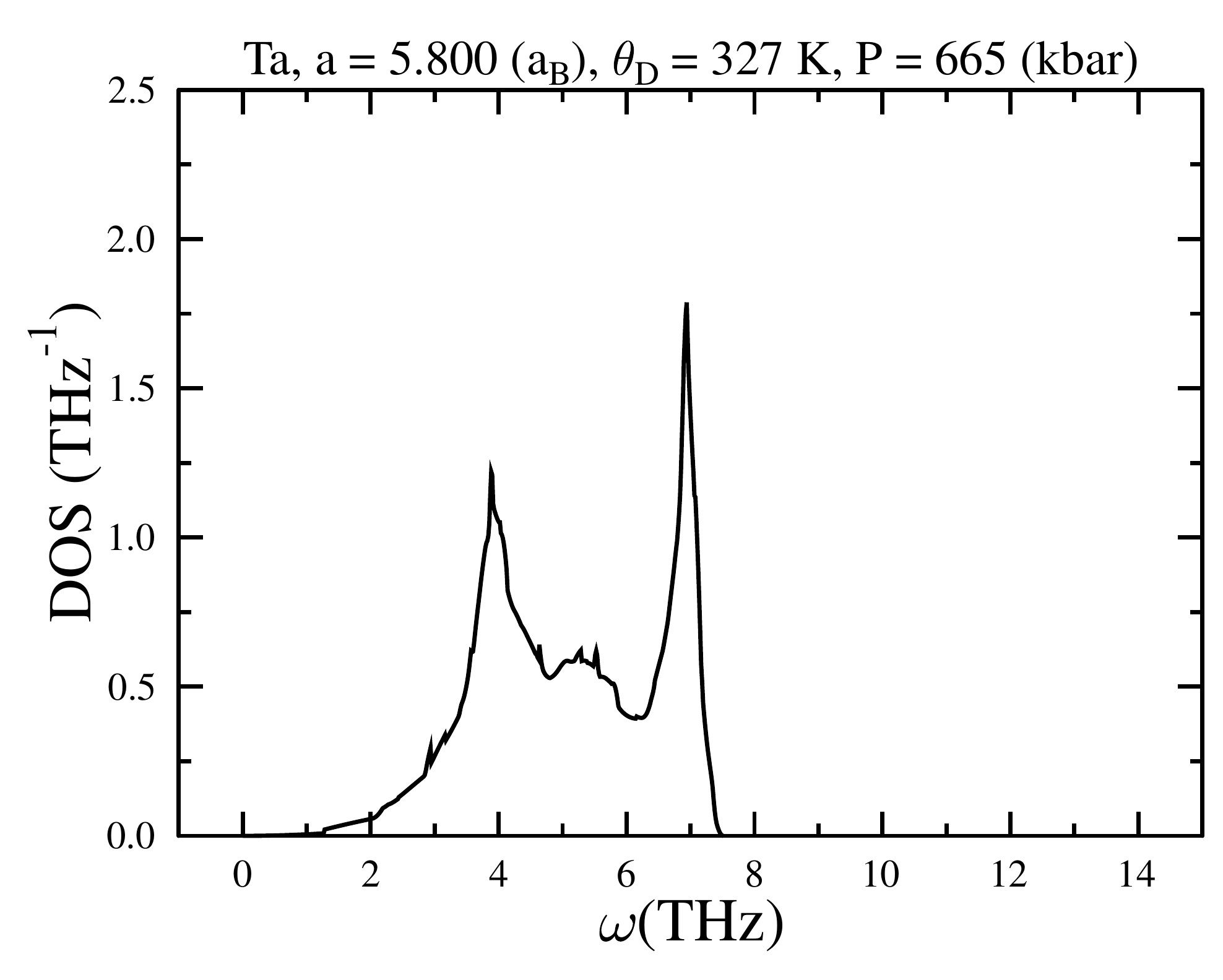}
\includegraphics[width=0.31\textwidth]{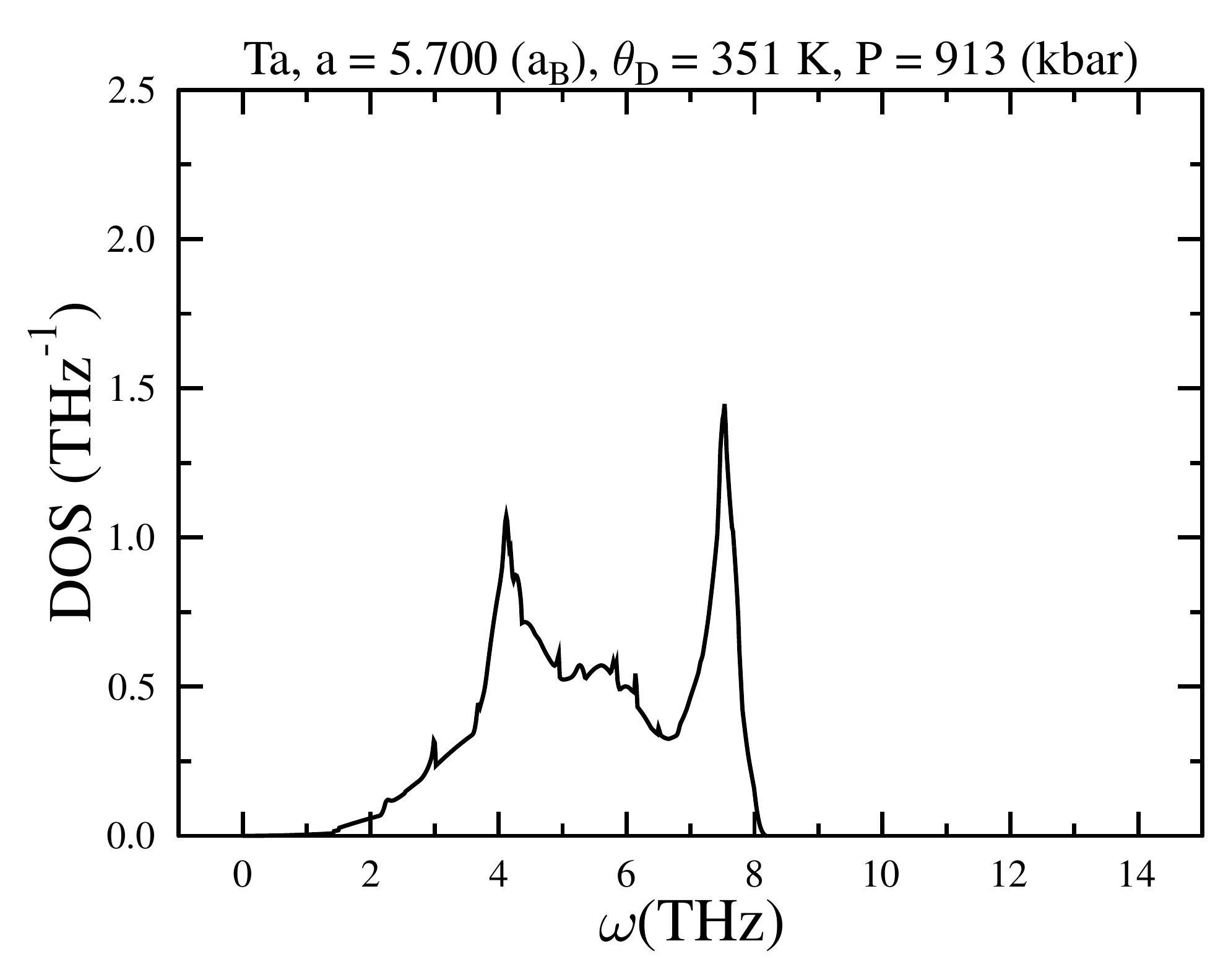}
\includegraphics[width=0.31\textwidth]{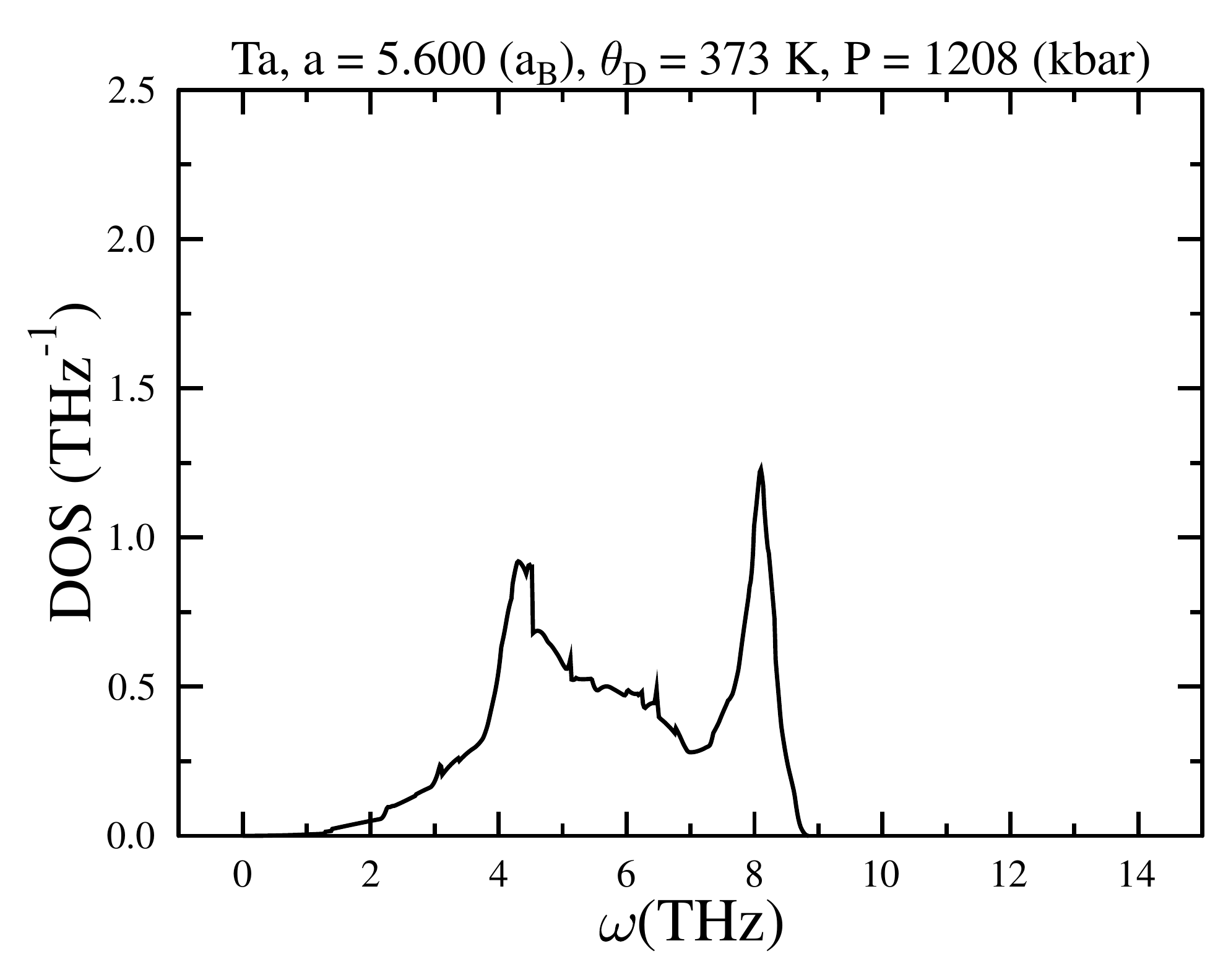}
\end{center}
\caption{Phonon DOS for Ta}
\label{ta}
\end{figure}

\newpage
\section{XRD diffraction patterns}
Fig.~\ref{xrd} shows the XRD patterns of our HEA sample.

\begin{figure}[h!]
\begin{center}
\includegraphics[width=0.90\textwidth]{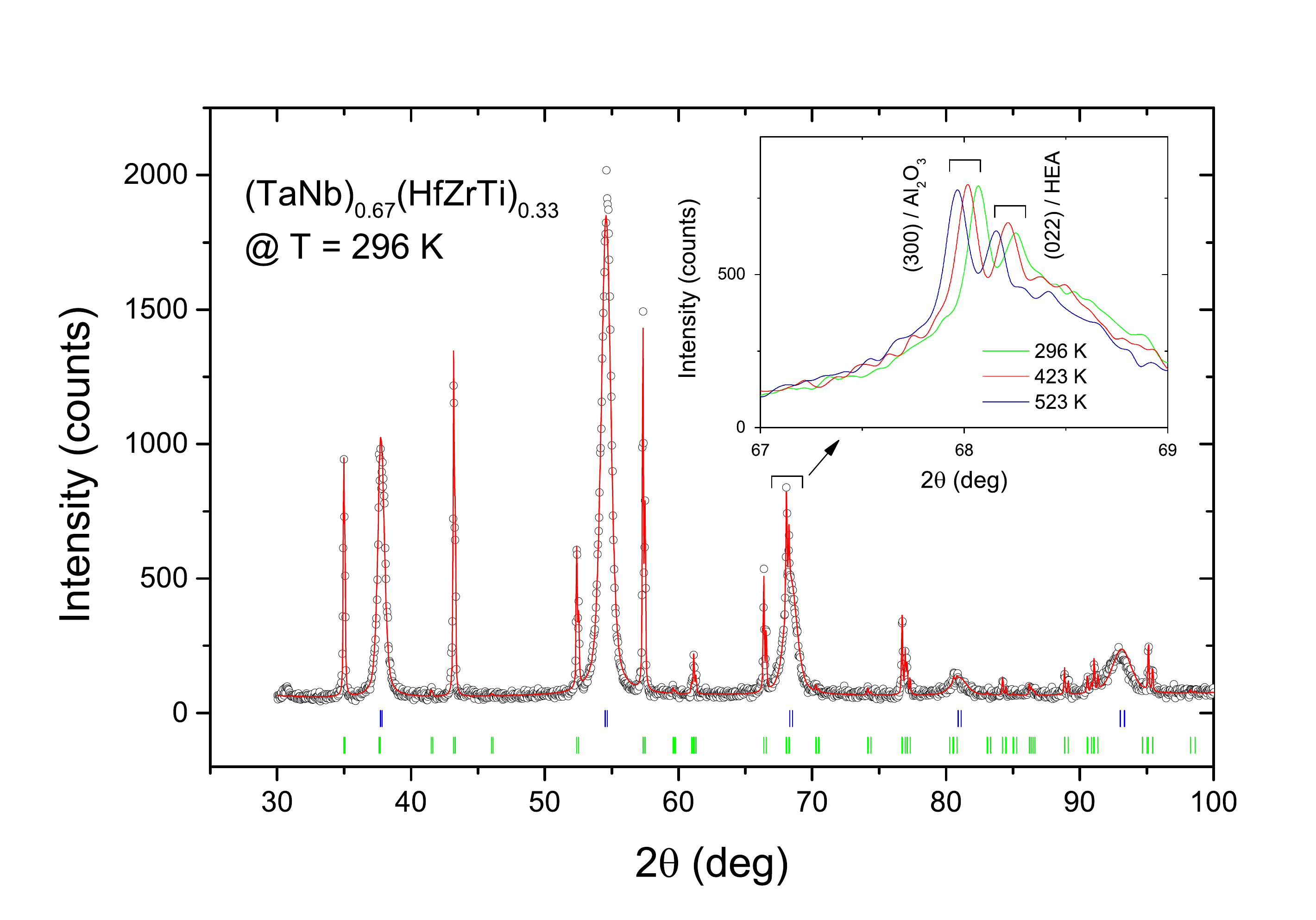}
\end{center}
\caption{Main panel: a room temperature X-ray diffraction pattern for (TaNb)$_{0.67}$(HfZrTi)$_{0.33}$. Experimental data points are represented by open circles and a LeBail profile fit is shown by a solid red line. Tick marks are the Bragg positions of a cubic $Im$-$3m$ (TaNb)$_{0.67}$(HfZrTi)$_{0.33}$ phase (blue) and a trigonal $R$-3c corundum holder (green). Inset shows a clear shift of (022) reflection of HEA and (300) of Al$_2$O$_3$ towards lower angle (larger $d$-spacing) as temperature is increased.}
\label{xrd}
\end{figure}

\newpage

\section{Matrix elements as a function of pressure}
Fig.~\ref{matrix} shows the evolution of the matrix elements, denoted as $\beta_{l\rightarrow l+1}$ ($l = 0, 1, 2$), and the ratio of a product of the partial densities of states $n_ln_{l+1}$ to the total DOS at the Fermi level $N(E_F)$ as a function of pressure. Matrix elements are defined as:
\begin{equation}
\beta_{l\rightarrow l+1} = \left|\int_0^{R_{\mathsf{MT}}}\!\!r^2R_l\frac{dV}{dr}R_{l+1} \right|^2\!
\end{equation}
and together with the ratio of densities of states enter the formula for the McMillan-Hopfield parameters:
\begin{equation}\label{eq:etas}
\eta_i =\!\sum_l \frac{(2l + 2)\,n_l(E_F)\,
n_{l+1}(E_F)}{(2l+1)(2l+3)N(E_F)} \beta_{l\rightarrow l+1}.
\end{equation}
The increase in matrix elements is responsible for the increase in the McMillan-Hopfield parameters, despite the drop in $n_ln_{l+1}/N(E_F)$ term for $p-d$ (1-2) and $d-f$ (2-3) channels.

\begin{figure}[h!]
\begin{center}
\includegraphics[width=0.95\textwidth]{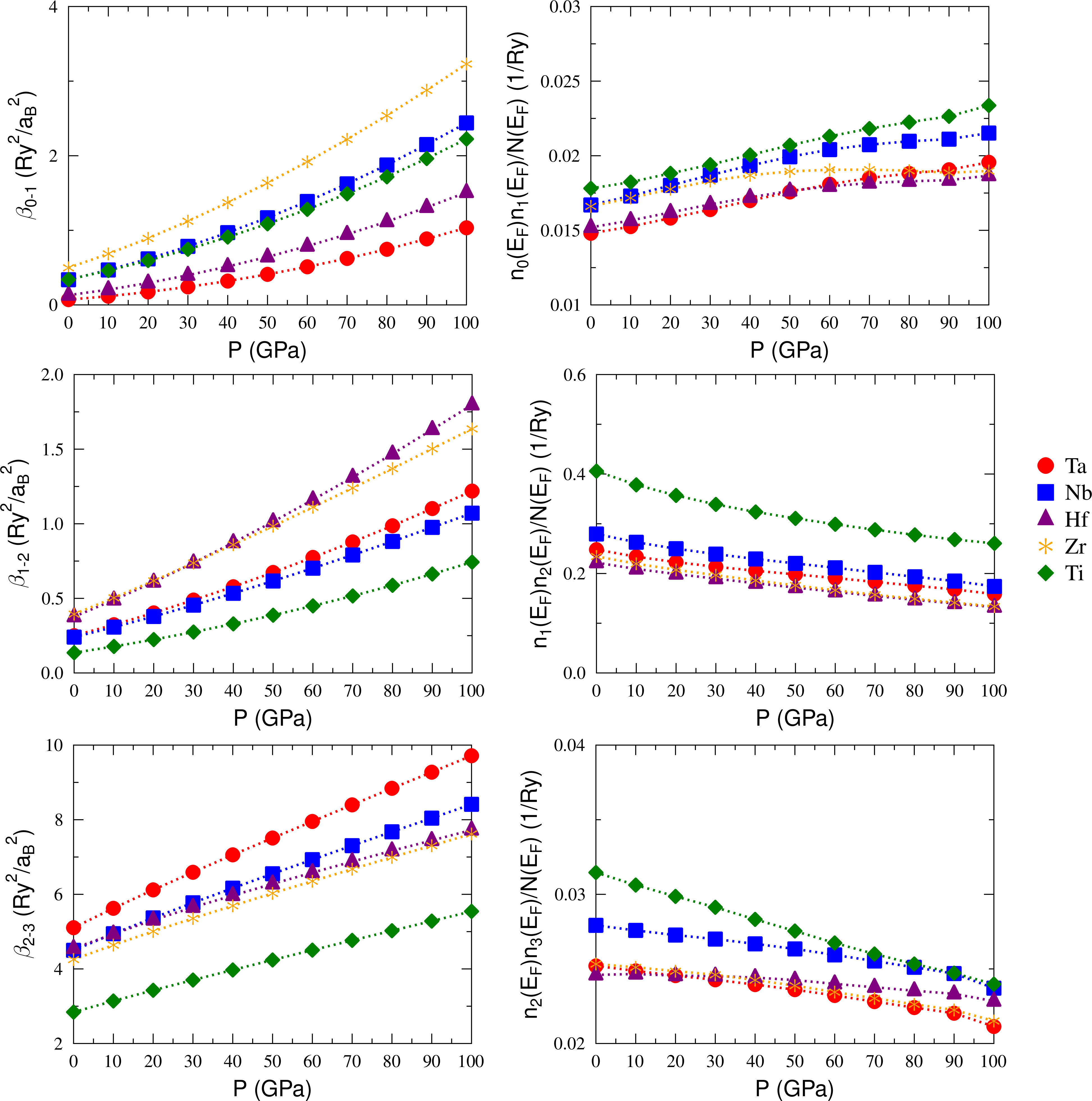}
\end{center}
\caption{Change in the integrals $\beta_{l\rightarrow l+1}$ and in the ratio of a product of the partial densities of states $n_ln_{l+1}$ to the total DOS at the Fermi level $N(E_F)$ in HEA under external pressure. 
The increase in matrix elements is responsible for the increase in the McMillan-Hopfield parameters. }
\label{matrix}
\end{figure}

\end{widetext}

\end{document}